\documentclass[lettersize,journal]{IEEEtran}
\usepackage{amsmath,amsfonts}
\usepackage{algorithmic}
\usepackage{algorithm}
\usepackage{array}
\usepackage[caption=false,font=normalsize,labelfont=sf,textfont=sf]{subfig}
\usepackage{textcomp}
\usepackage{stfloats}
\usepackage{url}
\usepackage{verbatim}
\usepackage{graphicx}
\usepackage{cite}
\usepackage{booktabs}
\usepackage{enumitem}
\usepackage{tabularx}
\begin{document}

\title{Securing AI Agents in Cyber-Physical Systems: A Survey of Environmental Interactions, Deepfake Threats, and Defenses \thanks{This manuscript is a preprint intended to rapidly disseminate a survey of security challenges and design principles for AI agents operating in cyber-physical systems. The authors anticipate submitting a substantially revised and polished version to a peer-reviewed journal.}}

\author{\IEEEauthorblockN{Mohsen Hatami, Van Tuan Pham, Hozefa Lakadawala, Yu Chen$^*$}

\IEEEauthorblockA{Dept. of Electrical \& Computer Engineering, Binghamton University, Binghamton, NY 13902, USA \\
\{mhatami1, tpham15, hlakada1, ychen\}@binghamton.edu}
}

\markboth{Journal of \LaTeX\ Class Files,~Vol.~xx, No.~x, January~2026}%
{Shell \MakeLowercase{\textit{et al.}}: A Sample Article Using IEEEtran.cls for IEEE Journals}


\maketitle

\begin{abstract}
The increasing integration of AI agents into cyber-physical systems (CPS) introduces new security risks that extend beyond traditional cyber or physical threat models. Recent advances in generative AI enable deepfake and semantic manipulation attacks that can compromise agent perception, reasoning, and interaction with the physical environment, while emerging protocols such as the Model Context Protocol (MCP) further expand the attack surface through dynamic tool use and cross-domain context sharing. This survey provides a comprehensive review of security threats targeting AI agents in CPS, with a particular focus on environmental interactions, deepfake-driven attacks, and MCP-mediated vulnerabilities. We organize the literature using the SENTINEL framework, a lifecycle-aware methodology that integrates threat characterization, feasibility analysis under CPS constraints, defense selection, and continuous validation. Through an end-to-end case study grounded in a real-world smart grid deployment, we quantitatively illustrate how timing, noise, and false-positive costs constrain deployable defenses, and why detection mechanisms alone are insufficient as decision authorities in safety-critical CPS. The survey highlights the role of provenance- and physics-grounded trust mechanisms and defense-in-depth architectures, and outlines open challenges toward trustworthy AI-enabled CPS.
\end{abstract}


\section{Introduction}

As autonomous AI agents increasingly integrate into cyber-physical systems (CPS), the boundary between intelligence and the physical world blurs. These agents, endowed with perceptual capabilities, decision-making logic, and influence of actuations, promise enhanced efficiency, adaptability, and autonomy in domains such as smart grids, autonomous vehicles, industrial automation, and robotics~\cite{wang2021survey}. However, as the AI-environment interface becomes a critical conduit, novel security and privacy vulnerabilities emerge, especially when adversaries leverage synthetic content, such as deepfakes, to deceive both machines and humans~\cite{brooks2019increasing,kpmg2024deepfake}.

While traditional CPS security research has focused on attacks such as sensor spoofing, denial-of-service (DoS), and insider threats~\cite{wang2021survey}, the advent of powerful generative models ushers in a new class of threats: \emph{AI-generated content} (AIGC), including images, video, audio, text, and behavioral emulation. Such content can masquerade as legitimate environmental data or human commands, thus subverting the system operation or misleading human operators. For example, a deepfake video feed may spoof surveillance cameras, a cloned voice can impersonate an authorized operator, or forged sensor traces may mask a real physical anomaly~\cite{brooks2019increasing,ironscales2024deepfake}. The capacity of deepfakes to mimic reality convincingly exacerbates the challenge: defenders cannot rely solely on human judgment or classical anomaly detectors~\cite{kpmg2024deepfake}.

In parallel, AI agents must navigate unpredictability at the interface: they ingest external data (sensor streams, images, voice) and produce actions in the physical world. The \emph{agent--environment (A--E)} interaction surface is thus a rich target for adversarial manipulation. Malicious actors can exploit this surface by injecting crafted inputs or manipulating environmental signals to force misbehavior, cause unsafe actions, or conceal attacks~\cite{deng2025ai}.

A critical challenge is the absence of a systematic methodology for translating the extensive body of security research into deployment-ready solutions for specific CPS contexts. While the literature provides numerous detection techniques, mitigation strategies, and threat analyses, system designers lack structured guidance for selecting and combining these mechanisms, given the unique constraints of their deployments. CPS operates under fundamentally different conditions from traditional information technology systems, with real-time performance requirements measured in milliseconds, computational resources constrained by edge device capabilities, and safety-critical operational demands that preclude security mechanisms from interfering with physical process control. Furthermore, the distributed architecture of modern agent protocols, such as the Model Context Protocol (MCP), introduces trust boundaries and interaction surfaces that conventional security frameworks do not adequately address. 

To bridge this gap, this survey introduces a Systematic Evaluation and Threat-Informed NEtwork defense seLection (SENTINEL) framework. It provides a six-phase methodology that guides system designers from initial threat assessment through selection of defense mechanisms, design of defense-in-depth architectures, validation planning, and continuous adaptation. The framework enables practitioners to systematically match security mechanisms to their specific deployment contexts by integrating threat modeling with resource constraint analysis and operational requirements. Throughout the subsequent sections, we apply SENTINEL concepts to organize and contextualize the extensive technical content, transforming what would otherwise be a catalog of disconnected security techniques into actionable guidance for building trustworthy AI-enabled infrastructures.

This survey focuses on the intersection of these two converging axes: security and privacy threats arising from AIGC at the AI--environment interface in CPS. We aim to provide a rigorous, holistic view of how deepfake and synthetic content threats manifest within CPS, and how detection and mitigation techniques are evolving to meet these challenges. This survey advances the thesis that detection-centric security is fundamentally insufficient for AI-enabled CPS, and that trustworthy operation instead requires lifecycle-aware, provenance- and physics-grounded system design that treats AI agents and MCP as integral components of the CPS control fabric.

The primary contributions of this paper are summarized below.
\begin{enumerate}
    \item A systematic framework, SENTINEL, that provides a structured methodology to select and combine security mechanisms tailored to specific CPS deployments.
    \item A taxonomy of deepfake modalities (visual, audio, textual, behavioral) as they apply to CPS, with concrete examples of how each can compromise agents or system functions~\cite{piccialli2025agentai,brooks2019increasing}.
    \item A systematic review of deepfake detection techniques in CPS-relevant settings, comparing their strengths, limitations, and suitability in constrained environments in real-time~\cite{deng2025ai,kpmg2024deepfake}.
    \item An overview of mitigation and defensive strategies from provenance/authentication, watermarking, multi-modal fusion, to robust training and policy-level measures~\cite{piccialli2025agentai}.
    \item An end-to-end CPS case study leveraging ANCHOR-Grid \cite{hatami2025anchor} demonstrates how the proposed SENTINEL framework can be operationalized under concrete timing, noise, and safety constraints.
    \item A discussion of open challenges and directions, emphasizing the tension between real-time performance, generalization, privacy, and adaptive adversaries~\cite{deng2025ai,ironscales2024deepfake}.
\end{enumerate}

By consolidating insights from recent literature on AI security, generative models, and CPS resilience, we hope this survey will serve as a foundational reference for researchers and system designers seeking to build trustworthy AI agents in CPS.

The rest of this paper is structured as follows. Section \ref{sec:background} provides fundamental background knowledge on CPS, AI agents, MCP protocol, and deepfake attacks. Section \ref{sec:frame} introduces the SENTINEL framework. Section \ref{sec:threat} describes the threat landscape for AI Agents in CPS. Section \ref{sec:deepfake} focuses on Deepfakes as a CPS security threat. Sections \ref{sec:detection} and \ref{sec:mitigation} discuss the state-of-the-art techniques and strategies for detection, mitigation, and defense. Section \ref{sec:case} demonstrates how the principles surveyed in this work apply in a concrete CPS context. Section \ref{sec:future} explores the open challenges and future research directions. And finally, Section \ref{sec:conlusion} concludes this survey.

\section{Background}
\label{sec:background}

This section introduces (i) the foundations of cyber–physical systems (CPS) and their security context, (ii) AI agents and emerging agent protocols, (iii) the MCP for tool and data interoperability, and (iv) deepfake (AI-generated content) technologies that underpin the threat space reviewed in this survey.

\subsection{Cyber–Physical Systems (CPS) and Security Context}

CPS tightly couples computation, networking, and physical processes across industrial control and smart grids, autonomous vehicles, and robotics. Classic texts and frameworks emphasize that \emph{timing, concurrency, and physical dynamics} make CPS distinct from pure IT systems and demand domain-aware assurance across safety, security, reliability, and resilience~\cite{lee2016introduction,NISTCPS2017,NISTTrust2018}. The NIST CPS Framework codifies this view with the concept of \emph{trustworthiness}, integrating security, privacy, safety, reliability, and resilience concerns into system analysis and design~\cite{NISTCPS2017,NISTTrust2018}. 

\begin{figure}[htbp!]
\centering
\includegraphics[scale=0.25]{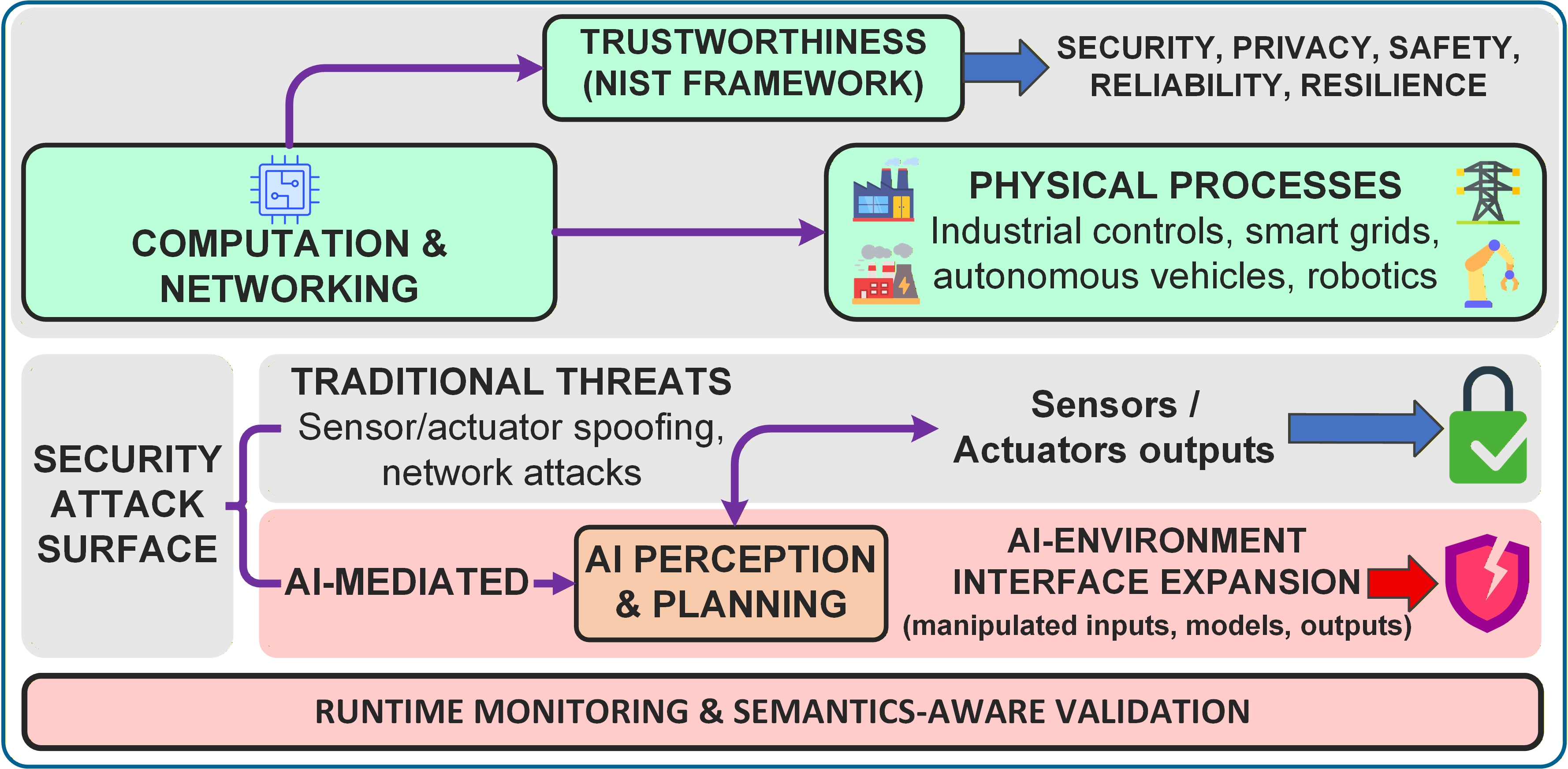}
\caption{Cyber-Physical Systems (CPS) and Security Context.}
\label{fig:CPS security context}
\vspace{-5 pt}
\end{figure}

Historically, CPS security has focused on threats such as sensor/actuator spoofing, network-borne attacks, and insider misuse. As AI components (e.g., perception, planning, and decision modules) increasingly mediate sensing and actuation, the \emph{AI–environment} interface expands the attack surface: adversaries can manipulate inputs, models, and tool outputs to induce unsafe actions or conceal physical anomalies. Recent surveys and applications research highlight the need for \emph{runtime} monitoring and semantics-aware validation (beyond offline model testing) to cope with distribution shift and adversarial manipulation in operational environments~\cite{rosado2022managing}.

\subsection{AI Agents and Agent Protocols}

\paragraph*{Agent model}
Modern AI agents follow an \emph{observe–reason–act} loop, often with memory and tool-use. LLM-centric agents augment core reasoning with tool/API calls, retrieval, code execution, and delegation to other agents. This modular, tool-augmented setup accelerates capability, but also introduces \emph{trust boundaries} at inputs (prompts, retrieved data), tools (code, connectors), and inter-agent communication~\cite{piccialli2025agentai,lupinacci2025dark}.


\begin{figure}[htbp!]
\centering
\includegraphics[scale=0.25]{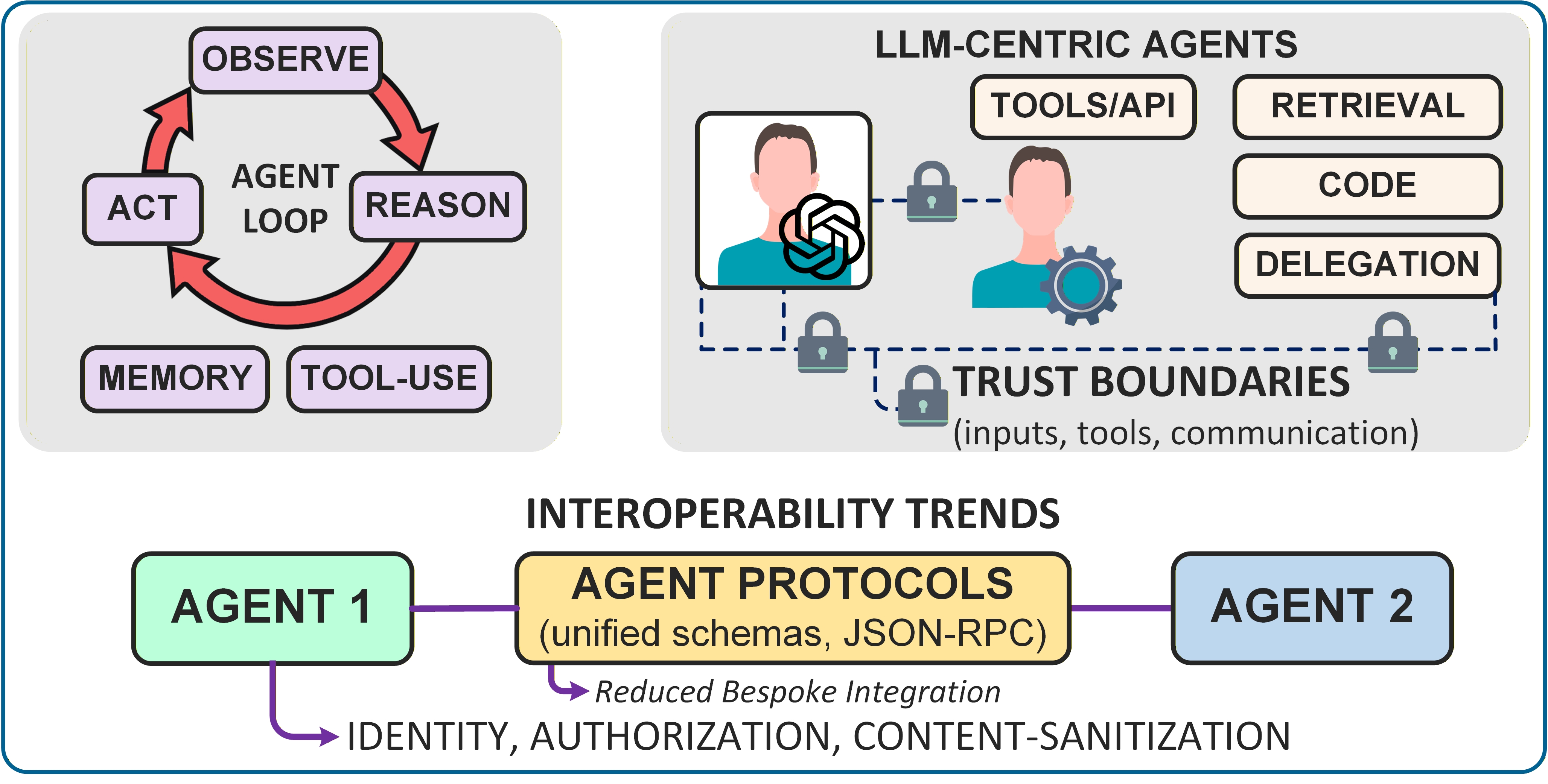}
\caption{Al Agents and Agent Protocols.}
\label{fig:AI-Agents protocol}
\vspace{-10 pt}
\end{figure}

\paragraph*{Interoperability trends.}
To reduce bespoke integrations, the community is converging on \emph{agent protocols} that standardize how agents discover capabilities, invoke tools, and exchange context. Recent surveys and industry guidance describe unified schemas and JSON-RPC–style interaction models that enable agents to interact with heterogeneous services while preserving a shared execution context~\cite{piccialli2025agentai,RedHatMCPIntro}.\footnote{Section~\ref{subsec:mcp} details MCP, the most widely adopted open specification for this purpose.} While these protocols improve reuse and composition, they also amplify the need for identity, authorization, and content-sanitization controls across agent boundaries~\cite{RedHatMCPSec,UpwindMCP}.

\paragraph*{Agent-to-Agent Network Protocols}
In CPS environments like swarm drones, swarm robotics, smart grids, agents must often coordinate directly with peers rather than just passive tools. The Agent-to-Agent (A2A) protocol has emerged as a standard to replace monolithic orchestration with decentralized, discoverable task workflows using JSON-RPC and SSE~\cite{ray2025review}. However, integrating A2A with MCP moves the system from simple "glue code" to complex protocol interactions that introduce unique risks. Li et al. warn that this fusion creates \emph{semantic interoperability gaps} and \emph{compounded security surfaces}. For instance, a compromised agent can exploit the vertical MCP tool access of a peer via horizontal A2A delegation, effectively bypassing intended isolation boundaries~\cite{li2025glue}.

\subsection{Model Context Protocol (MCP)}
\label{subsec:mcp}

MCP is an open specification that standardizes how LLM applications (clients) interact with external tools and data via MCP servers, using structured messages (e.g., JSON-RPC) and explicit capability descriptions~\cite{MCPSpec,GitHubMCP}. The vendor-neutral design of MCP and the flexibility of the transport (e.g., stdio, HTTP/SSE) have accelerated adoption in IDEs, desktop assistants, and agent frameworks~\cite{VergeMCP,RedHatMCPIntro}. 

\begin{figure}[htbp!]
\centering
\includegraphics[scale=0.25]{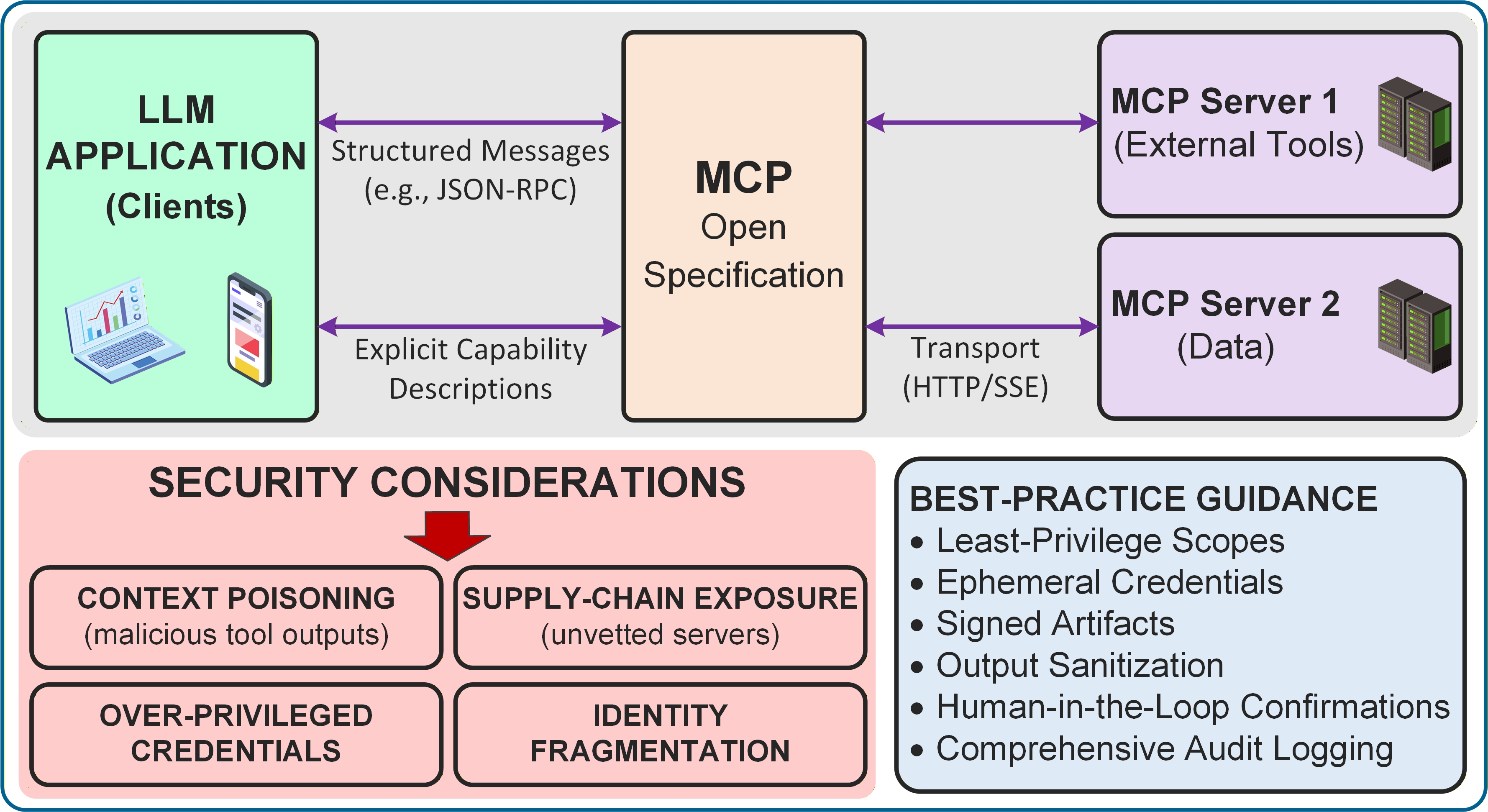} 
\caption{Model Context Protocol (MCP) System.}
\label{fig:mcp-system}
\vspace{-5 pt}
\end{figure}

\paragraph*{Security considerations.}
Security analyses identify key risks at the MCP boundary: \emph{context poisoning} (malicious tool outputs influencing model behavior), \emph{supply-chain} exposure (unvetted servers/connectors), and \emph{over-privileged credentials}~\cite{RedHatMCPSec,UpwindMCP}. Multiple 2025 disclosures underscore these concerns; for example, researchers reported a malicious MCP server update that surreptitiously BCC’d all processed email to an attacker infrastructure~\cite{ITProMCPMal}. In CPS, two additional risks emerge: (1) \emph{Availability}, where protocol latency violates real-time timing constraints; and (2) \emph{Social Engineering}, where deepfakes defeat human-in-the-loop controls by generating persuasive proof to authorize unsafe actuations. Best-practice guidance recommends least-privilege scopes, ephemeral credentials, signed artifacts, and output sanitization~\cite{RedHatMCPSec,TechRadarMCP,MCPSpec}.


\subsection{Deepfake (AI-Generated Content) Technologies}
AI-generated content spans visual (image/video), audio (speech/voice), textual, and behavioral modalities. In the context of CPS, deepfakes represent a \emph{next-generation sensor spoofing} threat. Unlike traditional replay attacks, generative models can synthesize novel, physically consistent sensor inputs like  thermal readings, LiDAR point clouds, or operator voice commands that bypass standard anomaly detection systems.


\begin{figure}[htbp!]
\centering
\includegraphics[scale=0.25]{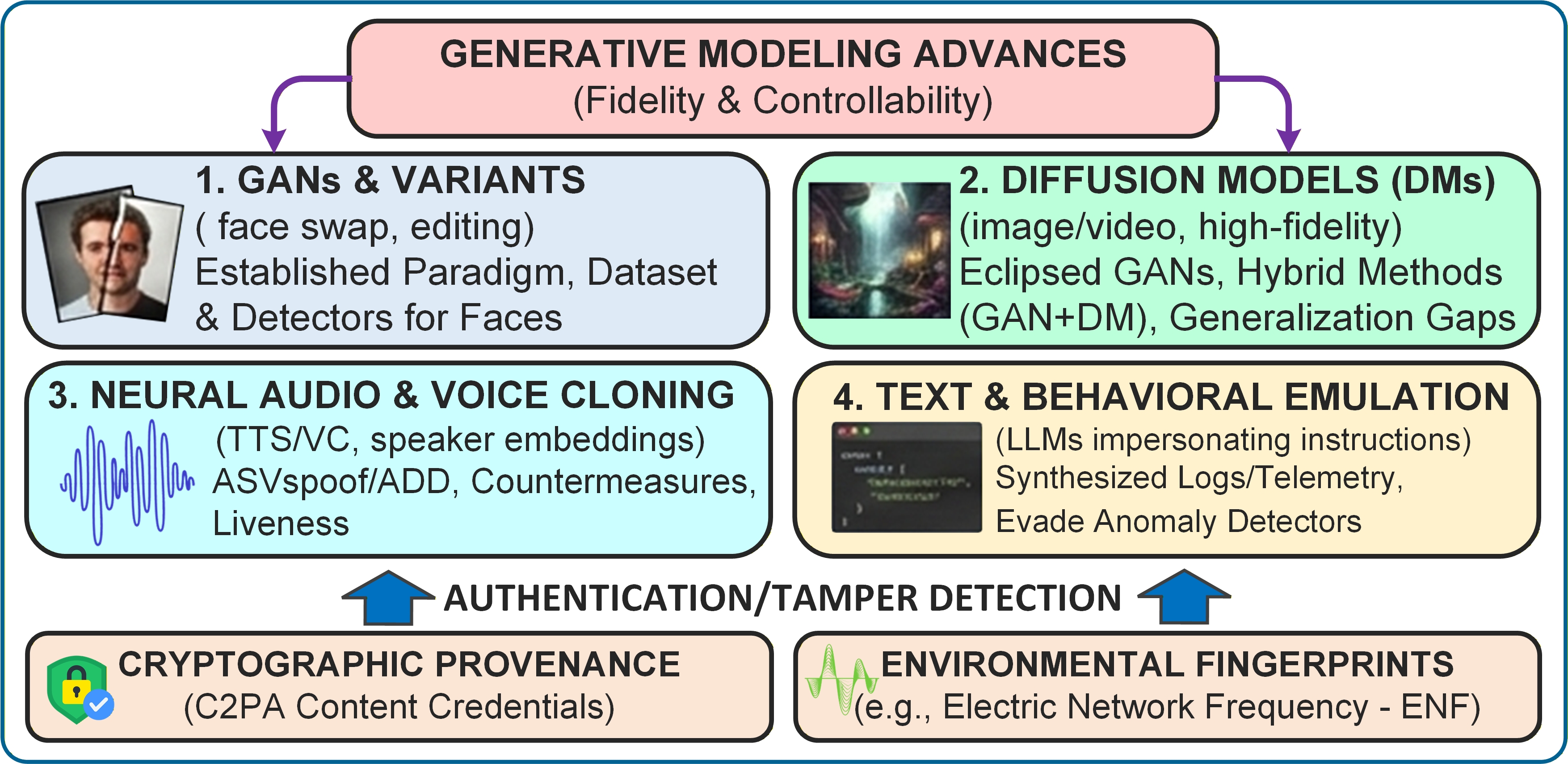}
\caption{Deepfake (Al-Generated Content) Technologies.}
\label{fig:framework1}
\vspace{-5 pt}
\end{figure}

\paragraph*{GANs and variants}
Generative Adversarial Networks (GANs) established the modern paradigm for realistic synthesis, powering face swap/reenactment, attribute editing, and style transfer~\cite{goodfellow2014generative}. Contemporary surveys catalog state-of-the-art GAN-based deepfake pipelines, datasets, and detectors for faces and talking heads~\cite{croitoru2024deepfake,pei2024deepfake}.

\paragraph*{Diffusion models}
Diffusion models (DMs) have eclipsed GANs on many image/video tasks, yielding high-fidelity and diverse samples via iterative denoising. Surveys from the late 2024 to 2025 show rapid migration of deepfake generation and detection research toward diffusion and hybrid methods (GAN+DM), and document generalization gaps where detectors trained on older GAN artifacts underperform on DM fakes~\cite{croitoru2024deepfake,pei2024deepfake}.

\paragraph*{Neural audio generation and voice cloning}
Neural TTS/VC pipelines combine speaker embeddings, sequence-to-sequence acoustic models, and high-quality vocoders; more recent LLM-guided audio generation with neural codecs further narrows the gap to human speech. Comprehensive surveys and challenge series (ASVspoof, ADD) chronicle spoofing attacks and countermeasures, noting that modern \emph{liveness} and challenge–response protocols complement spectral artifact analysis for robust defense~\cite{li2025survey}.

\paragraph*{Text and behavioral emulation}
LLMs can produce impersonated instructions/logs and synthesize plausible telemetry or user behaviors to evade anomaly detectors. Multi-agent attack studies reveal how untrusted content and peer messages can induce unsafe tool calls and system-level compromise when trust boundaries are weak~\cite{lupinacci2025dark}. 

\paragraph*{Provenance and environmental fingerprints}
To shift from artifact spotting to \emph{source authentication}, the industry has standardized cryptographic provenance for media (C2PA Content Credentials) that bind origin/edit history to assets~\cite{C2PA,LFC2PA}. In parallel, \emph{environmental fingerprints} such as Electric Network Frequency (ENF) traces embedded in audio/video offer physics-rooted corroboration of time/place, aiding multimedia authentication and tamper detection~\cite{hatami2025electric}. These approaches are increasingly relevant to CPS, where signals bridge physical and digital realms.

\subsection{Summary}

\begin{figure}[htbp!]
\centering
\includegraphics[scale=0.4]{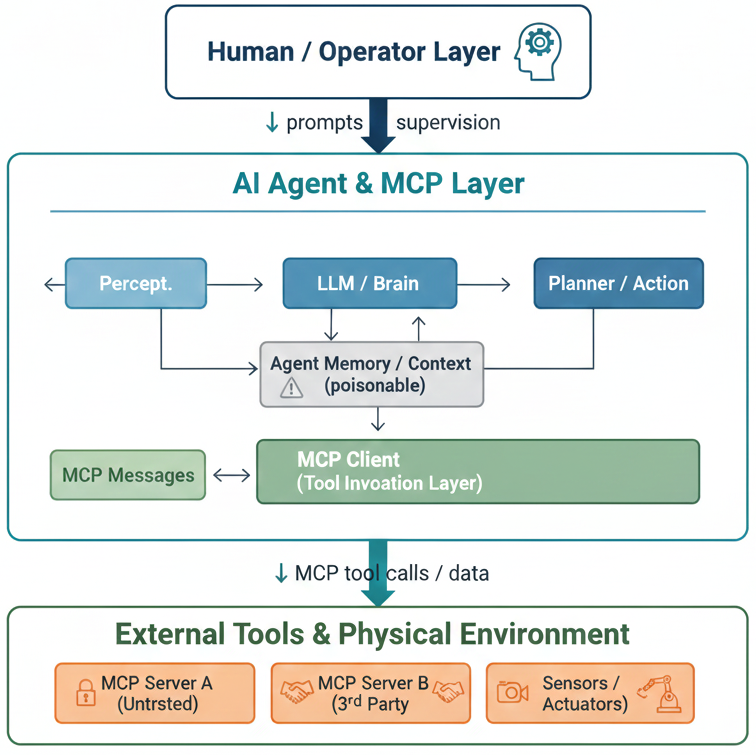} 
\caption{MCP threat-boundaries.}
\label{fig:boundaries}
\vspace{-10 pt}
\end{figure}

Figure \ref{fig:boundaries} illustrates the boundaries of trust and attack surfaces in MCP-enabled AI agents operating within CPS. MCP-mediated tool invocation and inter-agent communication introduce new cross-domain trust assumptions, allowing untrusted tool descriptions, server outputs, and synthetic environmental data to influence agent perception, reasoning, memory, and physical actuation. These blurred boundaries allow deepfake-driven attacks, context poisoning, and supply-chain exploits that bypass traditional CPS security models. CPS requires trustworthiness across sensing, reasoning, and actuation. Protocol-enabled agent ecosystems (e.g., MCP) expand capability but raise new identity, authorization, and supply-chain risks. Meanwhile, rapid progress in generative models fuels deepfakes that erode assumptions about input veracity. The remainder of this survey builds on this background to (i) systematize threats at the AI-environment interface and (ii) review detection and mitigation strategies tailored to CPS constraints. 

\section{SENTINEL Framework: Systematic Evaluation and Threat-Informed Defense Selection}
\label{sec:frame}

Before examining specific threats to AI agents in CPS, we introduce the a Systematic Evaluation and Threat-Informed NEtwork defense seLection (SENTINEL) framework. This framework addresses a critical gap in the literature: while numerous surveys catalog security mechanisms for AI systems or enumerate threats to CPS, few provide a systematic methodology for matching defense strategies to specific deployment contexts. The SENTINEL framework operationalizes the selection of security mechanisms by integrating threat modeling, resource constraints, and operational requirements into a unified decision process.

\begin{figure*}[htbp!]
\centering
\includegraphics[scale=0.35]{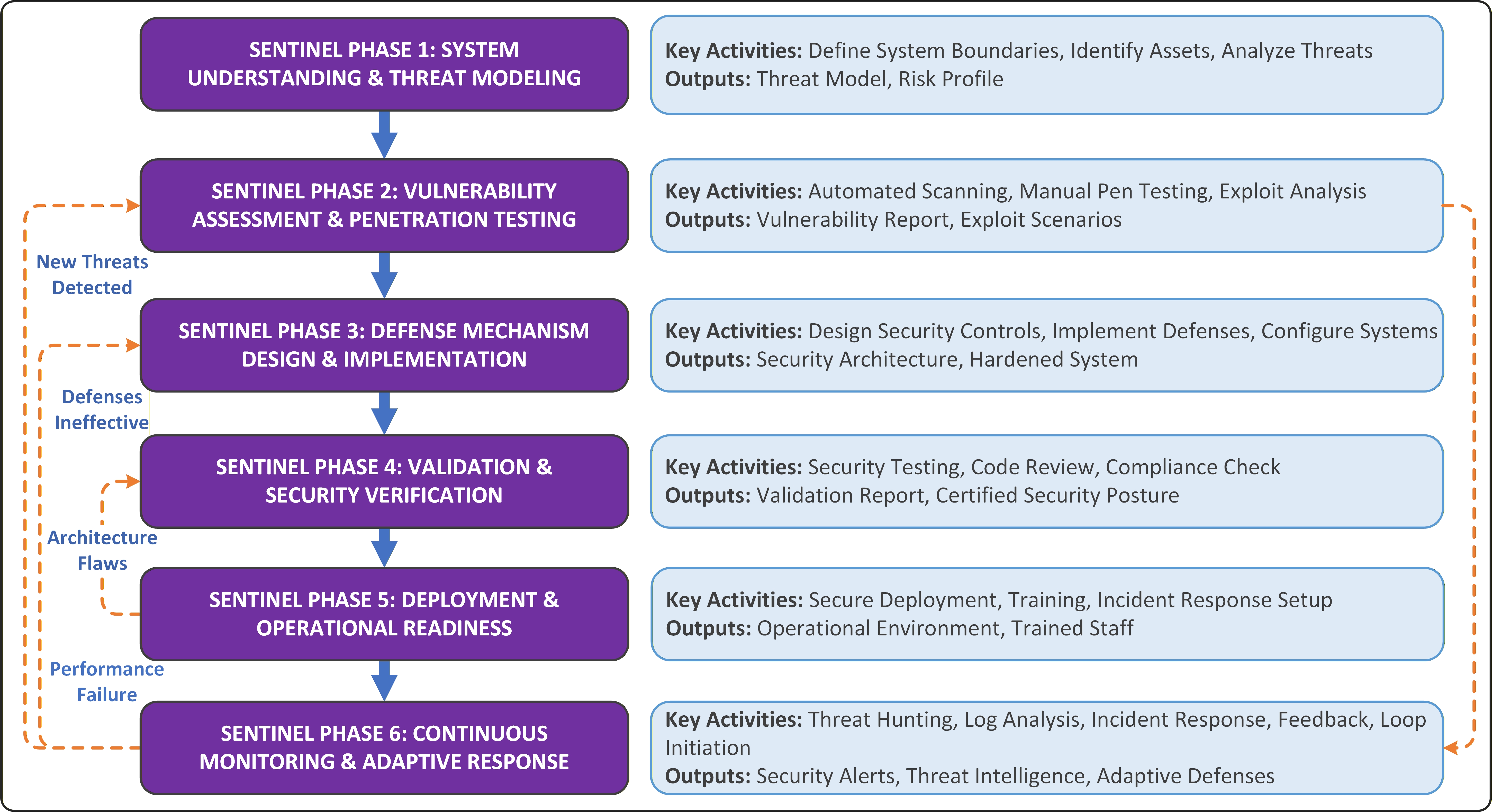}
\caption{The SENTINEL framework's six-phase methodology with feedback loops enabling continuous adaptation.}
\label{fig:framework2}
\end{figure*}

Figure~\ref{fig:framework2} outlines the workflow of the SENTINEL framework. It recognizes that MCP-enabled AI agents in CPS operate under fundamentally different constraints than traditional IT systems or standalone AI applications. These systems must balance security with safety-critical timing requirements, operate with limited computational resources at edge devices, maintain availability for physical process control, and preserve privacy while enabling necessary monitoring. Additionally, the distributed nature of MCP architectures, where agents interact with external tool servers and shared context repositories, creates unique trust boundaries that traditional security frameworks do not adequately address.
SENTINEL provides a six-phase methodology that guides system designers from initial threat assessment through defense deployment and continuous adaptation. Each phase produces concrete outputs that inform subsequent phases, creating a traceable decision pathway from system requirements to specific security mechanism configurations.

\subsection{Phase 1: System Characterization and Requirement Analysis}

The SENTINEL framework begins with a comprehensive characterization of the target CPS deployment along four critical dimensions. The operational dimension captures timing requirements including maximum tolerable detection latency, required system response time to threats, acceptable downtime for security updates, and recovery time objectives following security incidents. These temporal constraints fundamentally determine which detection and mitigation approaches remain viable. For instance, a smart grid protection relay operating on a 4-ms cycle time cannot employ deepfake detection, which requires 15-20-second ENF signal accumulation in the control loop, though such techniques may authenticate operator commands or validate sensor provisioning.

The resource dimension profiles available computational capacity at different network tiers, such as edge devices, fog nodes, and cloud infrastructure, along with memory constraints, network bandwidth, and latency characteristics, energy budgets for battery-powered sensors, and storage capacity for security logs and forensic data. This profile determines whether detection mechanisms can operate locally or must rely on offloaded processing, directly impacting both latency and privacy characteristics.

The trust dimension maps agent-environment interaction surfaces, including external data sources that agents query through MCP servers, tool invocations that trigger physical actuation, inter-agent communication channels, human operator interfaces, and third-party service dependencies. Each interaction surface represents a potential attack vector requiring appropriate authentication, authorization, and validation mechanisms. The MCP protocol architecture introduces particular complexity here, as agents may invoke tools provided by external servers whose security posture the agent operator cannot directly control.

The criticality dimension assesses potential consequences of security failures through impact on physical safety, including injury potential and physical damage scenarios, operational continuity requirements, including recovery complexity and cascade failure risks, such as data sensitivity considerations encompassing personal information exposure and intellectual property protection, and regulatory compliance obligations. This dimension establishes the required security level and acceptable risk thresholds that guide the selection of defense mechanisms.

\subsection{Phase 2: Threat Profiling and Attack Surface Analysis}

Building on system characterization, Phase 2 conducts structured threat profiling using a three-dimensional taxonomy that examines interaction surface (user-agent (U-A), agent-agent (A-A), agent-environment (A-E)), attack modality (visual deepfakes, audio synthesis, text generation, behavioral emulation, sensor spoofing), and adversary capability (opportunistic attackers using public tools, sophisticated adversaries with custom generative models, insider threats with system knowledge, supply chain compromises affecting MCP servers or tool providers). This taxonomy enables precise threat modeling that moves beyond generic security checklists to identify threats actually relevant to the specific deployment context.

For each identified threat, the framework requires the estimation of four key parameters. Attack likelihood reflects adversary motivation given the system's value or strategic importance, accessibility of attack vectors considering network exposure and authentication requirements, and attacker technical sophistication required. The impact of the attack quantifies the potential physical safety consequences, the magnitude and duration of operational disruption, financial losses due to downtime or damage, and reputational or regulatory consequences. Detection difficulty assesses the subtlety of attack artifacts and whether they fall below sensor noise floors; the availability of training data for adversarial examples; the computational complexity of detection algorithms; and whether attacks can adapt to known detection mechanisms. Finally, mitigation costs consider the computational overhead imposed by defense mechanisms, the latency introduced into critical control loops, the implementation complexity and required expertise, and the ongoing operational burden of security monitoring and updates.

This structured threat profiling produces a prioritized threat register that identifies high-priority threats, those with high likelihood and severe impact that drive primary defense requirements, medium-priority threats requiring monitoring but potentially acceptable through risk acceptance if mitigation costs exceed risk reduction value, and low-priority threats where standard security hygiene provides adequate protection without specialized deepfake defenses. This prioritization prevents the common pitfall of attempting to defend against all theoretically possible attacks regardless of their actual relevance to the deployment context, enabling resource-efficient security designs.

\subsection{Phase 3: Constraint-Aware Defense Mechanism Selection}

Phase 3 systematically evaluates candidate detection and mitigation techniques against a multi-dimensional fitness function that captures the tradeoffs inherent in CPS security. The framework maintains a comprehensive catalog of security mechanisms, drawn from the taxonomy developed in Sections IV through VI of this survey, with each mechanism characterized along six evaluation dimensions.

The detection effectiveness dimension captures mechanism accuracy through true positive and false positive rates for relevant attack types, robustness to adversarial adaptation and evasion attempts, generalization capability to novel deepfake generators not seen during training, and coverage breadth across different attack modalities. Computational requirements specify processing resources that include CPU, GPU, and memory footprints, latency from input acquisition to threat verdict delivery, scalability characteristics as system size or attack volume increases, and whether processing can occur at edge devices or requires centralized computation. The integration dimension assesses the complexity of the deployment, including the modifications needed to existing systems, the dependency on specialized hardware such as trusted execution environments or GPS receivers for timestamping, the compatibility with the existing security infrastructure, and the training data requirements for ML-based approaches. 

The operational dimension addresses maintenance burden, including model retraining frequency, signature database update cadence, false positive handling procedures, and forensic logging and audit trail requirements. Additionally, the privacy dimension evaluates data exposure, including whether raw sensor streams must be transmitted to detection services, whether cryptographic protections preserve privacy, compliance with data protection regulations, and user transparency requirements. The cost dimension captures both capital expenditures for new hardware or software licenses and operational expenses, including computational resources consumed, security operations personnel requirements, and incident response processes.

For each high-priority threat identified in Phase 2, the framework generates a candidate defense set comprising mechanisms that satisfy hard constraints, particularly real-time latency requirements and computational resource limits, and ranks them based on fitness scores that weight the evaluation dimensions according to the system priorities established in Phase 1. This produces a shortlist of viable defense mechanisms for each threat rather than presuming one-size-fits-all security solutions.

\subsection{Phase 4: Defense-in-Depth Architecture Design}

\begin{figure*}[htbp!]
\centering
\includegraphics[scale=0.4]{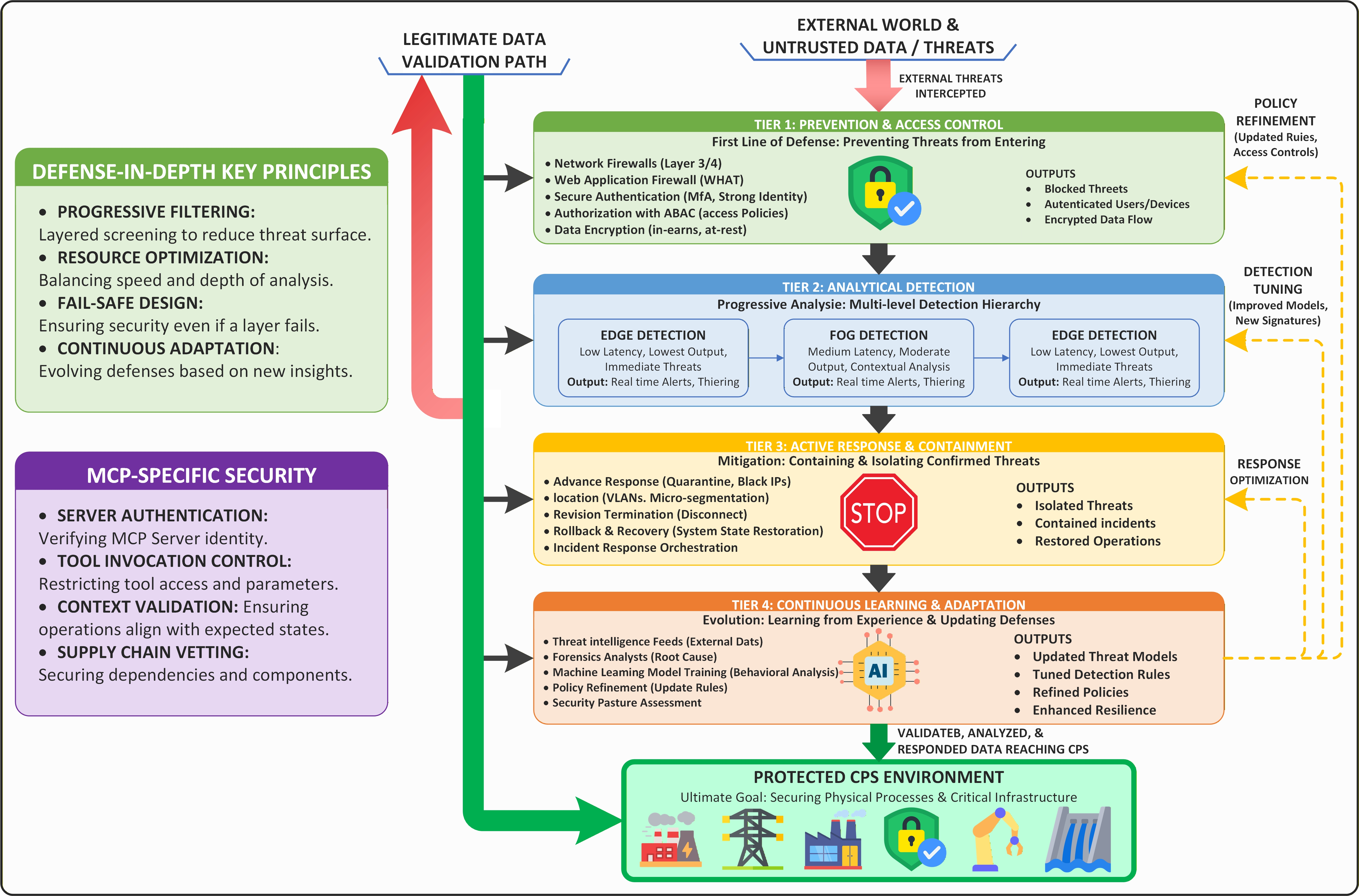} 
\caption{The four-tier defense-in-depth architecture in Phase 4.}
\label{fig:phase4}
\vspace{-10 pt}
\end{figure*}

Recognizing that no single security mechanism provides complete protection, Phase 4 constructs a layered defense architecture that integrates complementary techniques to address the threat landscape comprehensively. Figure \ref{fig:phase4} illustrates the four-tier defense-in-depth architecture that Phase 4 constructs for MCP-enabled cyber-physical systems, showing how complementary security mechanisms combine to provide comprehensive protection against deepfake threats. 

The \emph{perimeter tier} implements proactive defenses before threats reach agent decision-making processes, including input validation and sanitization at MCP server boundaries, cryptographic authentication of content provenance using C2PA or similar standards, and reputation-based filtering of external data sources and tool providers. These mechanisms reduce the attack surface by rejecting obviously malicious inputs before they consume detection resources or influence agent behavior.
    
The \emph{detection tier} deploys deepfake detection mechanisms matched to the threat profile and positioned according to resource availability. Lightweight heuristics run on edge devices to detect crude attacks with minimal latency, ensemble classifiers run on fog nodes to provide more sophisticated analysis with acceptable delay, and heavyweight deep learning models execute on cloud infrastructure for forensic analysis of suspicious events. The multi-tier detection strategy optimizes the tradeoff between detection latency and accuracy by filtering most attacks at lower tiers while reserving expensive analysis for edge cases.
    
The \emph{response tier} specifies graduated response policies triggered by different threat levels, ranging from alerting human operators for manual verification through automated containment measures such as revoking tool invocation permissions to emergency failsafe actions, including disconnecting compromised agents from physical actuators. The response policies must account for CPS safety requirements, ensuring that security responses do not themselves create unsafe physical states, and preserve evidence for forensic investigation.
    
The \emph{adaptation tier} implements continuous improvement mechanisms, including performance monitoring that tracks detection accuracy and false positive rates in operational deployment, threat intelligence integration that updates threat models and detection signatures, adversarial training that incorporates newly discovered attack techniques, and periodic security audits that verify defense effectiveness. This tier recognizes that deepfake generation techniques evolve continuously, requiring defense mechanisms that adapt rather than remain static.

For MCP-specific deployments, the SENTINEL framework emphasizes enforcing trust boundaries at protocol interaction points. MCP servers that provide tools to agents require strong authentication and authorization mechanisms to verify server identity and restrict tool invocations based on agent privileges. Context sharing among agents via MCP prompts demands validation to prevent context-poisoning attacks, in which malicious agents inject deepfake content into shared memory. Supply chain security for MCP tool providers necessitates vetting processes and continuous monitoring, given that compromised tools can bypass agent-level defenses entirely.

\subsection{Phase 5: Validation and Deployment Planning}

Phase 5 translates the defense architecture into deployable configurations through systematic validation. The framework requires simulation-based evaluation using representative attack scenarios from the threat registry, with performance measured against the metrics established in Phase 1. This evaluation should include both isolated mechanism testing, verifying that individual detection techniques achieve specified accuracy and latency targets, and integrated system testing that validates the complete defense-in-depth architecture under realistic conditions, including normal operational workload, concurrent multi-vector attacks, and resource-degradation scenarios.

The framework specifies three validation methodologies appropriate for different CPS contexts. Laboratory testbeds provide controlled environments for detailed performance characterization without risking operational systems, enabling comprehensive testing against both documented attacks and novel adversarial examples generated through red team exercises. Digital twin simulation leverages physics-based models of CPS processes to evaluate security mechanisms under realistic operational conditions while maintaining safety, particularly valuable for testing response tier policies that could trigger unsafe states if deployed prematurely in production systems. Pilot deployments gradually introduce security mechanisms into operational systems with extensive monitoring, beginning with non-critical subsystems before expanding to safety-critical components once confidence in defense effectiveness has been established.

Deployment planning must address staged rollout timelines that minimize operational disruption, fallback procedures that allow rapid security mechanism deactivation if unanticipated operational issues arise, personnel training requirements for security operations staff and incident responders, and coordination with existing security infrastructure, including security information and event management systems and incident response playbooks.

\subsection{Phase 6: Continuous Monitoring and Adaptive Defense}

The final phase recognizes that security is not a one-time deployment but an ongoing process that requires continuous monitoring and adaptation. The framework specifies metrics for operational security monitoring, including attack detection and false-positive rates relative to baseline expectations, security mechanism resource consumption relative to budgets, degradation in system performance caused by security overhead, and coverage gaps where new attack techniques evade deployed defenses.

Trigger conditions for defense mechanism updates include detection accuracy falling below acceptable thresholds, indicating adversarial adaptation, the emergence of new threat intelligence about attack techniques or vulnerable components, changes to system configuration or operational requirements that alter threat landscape or constraint profiles, and security incidents that reveal gaps in defense coverage. The framework provides a structured process for evaluating and deploying updates while maintaining operational continuity.

Adaptation mechanisms range from tuning existing security parameters, such as adjusting detection thresholds or updating signature databases, through deploying additional defense layers to cover newly identified gaps, to architectural redesign when fundamental assumptions about the threat model or system requirements change substantially. Each adaptation level requires increasingly rigorous validation before operational deployment, with parameter tuning potentially automated through machine learning, while architectural changes require comprehensive testing.

\subsection{Framework Application Methodology}

The SENTINEL framework provides both prescriptive guidance for greenfield deployments and diagnostic capabilities for evaluating existing security architectures. For new MCP-enabled CPS deployments, designers proceed sequentially through the six phases, with each phase's outputs documented and reviewed before proceeding. This disciplined approach ensures security requirements drive architecture decisions rather than security being retrofitted after system design has constrained options.

For operational systems, the framework enables gap analysis by systematically evaluating whether existing security mechanisms adequately address the characterized threat landscape, given actual system constraints and requirements. This diagnostic application often reveals either over-provisioned security, expensive mechanisms defending against low-priority threats, or critical gaps where high-priority threats lack adequate detection and mitigation. The framework thus guides both security investment optimization and risk remediation prioritization.

Throughout Section \ref{sec:threat}, we apply the concepts of the SENTINEL framework to analyze specific threat categories at the user-agent, agent-agent, and agent-environment interfaces. For each threat class, we identify representative attack techniques, assess their characteristics along the framework's evaluation dimensions, and highlight security mechanisms most appropriate for different CPS deployment contexts. Section \ref{sec:deepfake} then provides a detailed examination of deepfake detection techniques organized by modality, and Section \ref{sec:detection} presents mitigation and defense strategies. Together with the SENTINEL framework introduced here, these sections equip system designers with both comprehensive security mechanism knowledge and systematic methodology for applying that knowledge to specific deployments.

The subsequent subsections examine threats at each interaction surface in detail, but readers should interpret these threat discussions through the SENTINEL framework lens: understanding which threats apply to their specific context, evaluating candidate defenses against their unique constraint profile, and designing defense-in-depth architectures tailored to their requirements rather than adopting generic security solutions.

\section{Threat Landscape for AI Agents in CPS}
\label{sec:threat}
This section uses the first two phases of the SENTINEL framework to systematically characterize the threat landscape. Phase 1 analyzes the threats along the timing, trust, resource, and criticality dimensions of CPS deployments. Phase 2 classifies emerging threats along three axes: interaction surface (U-A, A-A, A-E), attack modality, and adversary capability. This approach expands abstract vulnerabilities into context-dependent risks, evaluating CPS-specific requirements like detection difficulty and mitigation costs.

\subsection{User--Agent (U--A) Interface Analysis}
The U-A interface represents the primary interaction point where human operators issue commands and receive feedback. While this transparency facilitates operational efficiency and reduces cognitive load, it introduces a critical vulnerability: the "observe–reason–act" loop becomes susceptible to linguistic manipulation. The integration of Large Language Models (LLMs) allows agents to ingest heterogeneous data streams \cite{datta2025agentic}, but simultaneously exposes the physical system to risks ranging from trust decay to expanded attack surfaces via untrusted external documents.


\begin{figure*}[htbp!]
\centering
\includegraphics[scale=0.35]{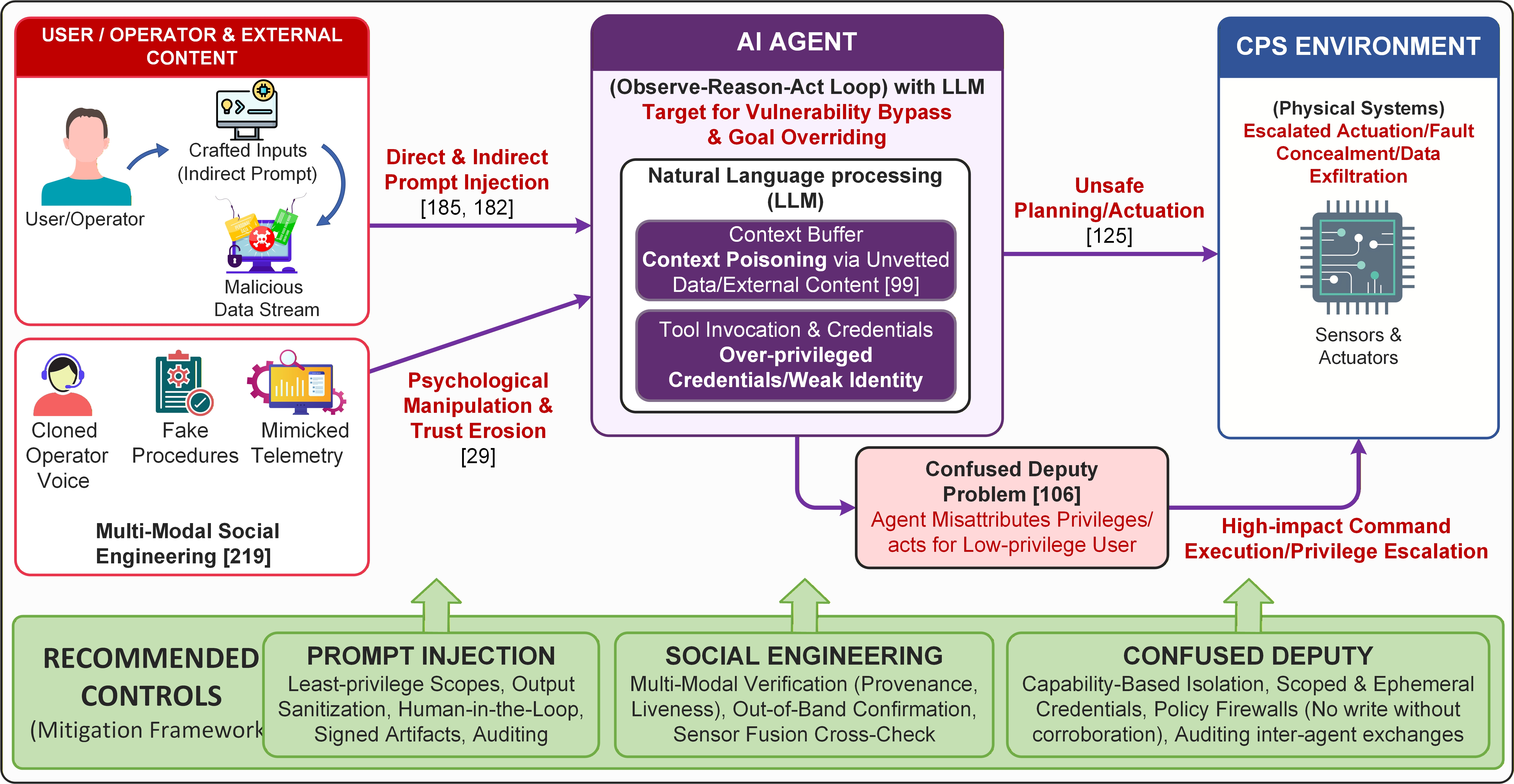}
\caption{Threat Landscape for AI Agents in CPS: USER-AGENT (U-A)}
\label{fig:AI-threads-(U-A)}
\vspace{-10 pt}
\end{figure*}


\subsubsection{Prompt Injection}
Prompt injection involves smuggling adversarial instructions into inputs the agent trusts to override its intended behavior \cite{liu2024formalizing, liu2023prompt}. In CPS, this escalates from text misalignment to unsafe physical actuation because the agent directly mediates sensing and control. Trust is inherently challenged as agents process both direct user prompts and external context from documents or tool outputs; \cite{greshake2023not} demonstrate how this context can be intentionally poisoned, while \cite{an2025rag} show that retrieval mechanisms can degrade safety alignment even when the context is safe. This risk is amplified in modern stacks via malicious tool responses, a threat formalized in the Model Context Protocol (MCP) \cite{hou2025model} and demonstrated in end-to-end autonomous agent benchmarks \cite{evtimov2025wasp}.

\subsubsection{Social Engineering}
Social engineering exploits the psychological trust between humans and AI agents. Adversaries use high-fidelity synthesis, such as cloned voices, fake procedures, or images, to induce unsafe decisions \cite{park2024ai, wahba2024creating}. In industrial settings, operators may unknowingly facilitate attacks through conversational interfaces. Detection is particularly difficult as modern "jailbreaks" can be multi-modal, employing "many-shot" strategies or typographic visual prompts to bypass safety filters \cite{shayegani2023jailbreak, gong2025figstep, roh2025multilingual, anil2024many}. Current research suggests mitigation requires robust detection models, feature fusion, and rigorous input filtering, though CPS contexts may necessitate additional out-of-band corroboration \cite{liu2023jailbreaking, khan2023battling}.

\subsubsection{Confused Deputy}
Security risks in agentic systems often stem from the agent acting as an unwitting accomplice to a malicious user. In the Model Context Protocol (MCP) ecosystem \cite{guo2025systematic}, tools are typically pre-authorized, allowing attackers to bypass privilege checks and manipulate the agent into executing unauthorized operations, such as reading sensitive files or tampering with configurations. Benchmarks \cite{wang2025mcp} indicate that while agents must handle complex dependency chains and fuzzy instructions in multi-tool environments, this complexity expands the attack surface. For example, agents struggle to distinguish between external data and executable instructions, leading to potential command injections \cite{guo2025systematic}. In the manufacturing context \cite{pulikottil2023agent}, while agents enable decentralization, robust cybersecurity remains a core requirement to prevent information theft and service disruption.

\subsubsection{Comparative Analysis}
Table \ref{tab:ua_system_characteristics} details the system constraints for U-A interactions. We found that timing is the most stringent constraint and the injection detection must occur within the millisecond-scale latency budget of the controller.

\begin{table}[ht]
\centering
\caption{U--A Interaction System Characteristics}
\label{tab:ua_system_characteristics}
\begin{tabular}{|p{1.25cm}|p{1.75cm}|p{1.75cm}|p{1.75cm}|}
\hline
\textbf{Dimension} &
\textbf{Prompt Injection} &
\textbf{Social Engineering} &
\textbf{Confused Deputy} \\
\hline
Timing & Millisecond Detection & Real-time Interaction \cite{park2024ai} & Request-time Check \cite{guo2025systematic} \\
\hline
Resource & Limited Edge Compute & High Bandwidth Media \cite{wahba2024creating, khan2023battling} & Low CPU Overhead \\ 
\hline
Trust & Untrusted External Data \cite{an2025rag, hou2025model} & Deceptive Human Trust \cite{park2024ai} & Privileged Misattribution \cite{guo2025systematic} \\ 
\hline
Criticality & Physical Safety Hazard & Operational Downtime & Unauthorized Actuation \cite{guo2025systematic} \\ 
\hline
\end{tabular}
\end{table}


\subsubsection{Threat Registry}
Table \ref{tab:ua_threat_registry} profiles the emerging threats. The registry indicates that while Prompt Injection has the highest likelihood due to low barriers to entry, Social Engineering presents the greatest detection challenge due to multi-modal artifacts.

\begin{table}[ht]
\centering
\caption{U--A Threat Profiling Registry}
\label{tab:ua_threat_registry}
\begin{tabular}{|p{1.25cm}|p{1.75cm}|p{1.75cm}|p{1.75cm}|}
\hline
\textbf{Dimension} &
\textbf{Prompt Injection} &
\textbf{Social Engineering} &
\textbf{Confused Deputy} \\
\hline
Modality & Text / Context \cite{hou2025model,liu2024formalizing} & Multi-modal \cite{gong2025figstep, shayegani2023jailbreak, roh2025multilingual} & API / Protocol \cite{guo2025systematic} \\
\hline
Capability & Low Knowledge \cite{liu2024formalizing, evtimov2025wasp} & Medium \cite{park2024ai, anil2024many} & System-specific \cite{guo2025systematic} \\
\hline
Likelihood & High \cite{liu2024formalizing, an2025rag} & Elevated \cite{anil2024many} & High  \cite{guo2025systematic}\\ 
\hline
Impact & Severe \cite{an2025rag, hou2025model,greshake2023not} & Significant \cite{park2024ai} & Critical \cite{guo2025systematic} \\ 
\hline
Detection Difficulty & High \cite{liu2024formalizing, hou2025model} & High \cite{park2024ai, gong2025figstep} & High \cite{guo2025systematic}\\
\hline
Mitigation Costs & Medium \cite{datta2025agentic} & Substantial \cite{shayegani2023jailbreak ,roh2025multilingual} & High \\ 
\hline
\end{tabular}
\end{table}


\subsubsection{Open Challenges in U-A Security}
Despite ongoing research, several open challenges remain for securing the U-A surface in CPS. The current research landscape highlights several critical hurdles for securing U-A interactions in CPS:
\begin{itemize}
\item \textbf{Probabilistic Safety Guarantees:}  Traditional formal methods struggle with the stochastic nature of LLMs; new frameworks like Pro2guard are exploring probabilistic model checking to enforce safety at runtime \cite{wang2025pro2guard}.
\item \textbf{Verified Code Generation:} Ensuring that the tool calls and scripts generated by agents adhere to strict safety specifications remains difficult, as seen in recent efforts toward verified code generation frameworks like VeriGuard \cite{miculicich2025veriguard}
\item \textbf{Adaptive Robustness: } As adversaries adapt to current filters, there is a need for defense mechanisms that can survive an "adaptive arms race" by dynamically updating their detection logic \cite{datta2025agentic}.
\item \textbf{Multi-Agent Adversarial Dynamics:} In environments where multiple agents interact, compromised proxies can lead to cascading failures that are poorly understood in existing security literature \cite{yu2024infecting}.
\item \textbf{Sandboxing Unverified Controllers:} Creating secure execution environments that can isolate and monitor LLM-based controllers without disrupting the real-time requirements of the physical system \cite{zhong2023towards}.
\end{itemize}

\subsection{Agent--Agent (A--A)}
The Agent–Agent (A–A) interaction surface represents the decentralized communication layer where autonomous entities collaborate to achieve system-wide goals. This enables decentralized resilience (eliminating single points of failure) and operational scalability for massive fleets \cite{jin2025comprehensive, yuan2025multi}. However, this introduces a "trust vacuum" where faults can propagate at machine speed \cite{lupinacci2025dark, wang2025comprehensive}. The absence of human-in-the-loop oversight means malicious artifacts can cause physical damage before intervention is possible, and implementing defenses often incurs performance overheads unacceptable for real-time control \cite{adapala2025aegis}.


\begin{figure*}[htbp!]
\centering
\includegraphics[scale=0.35]{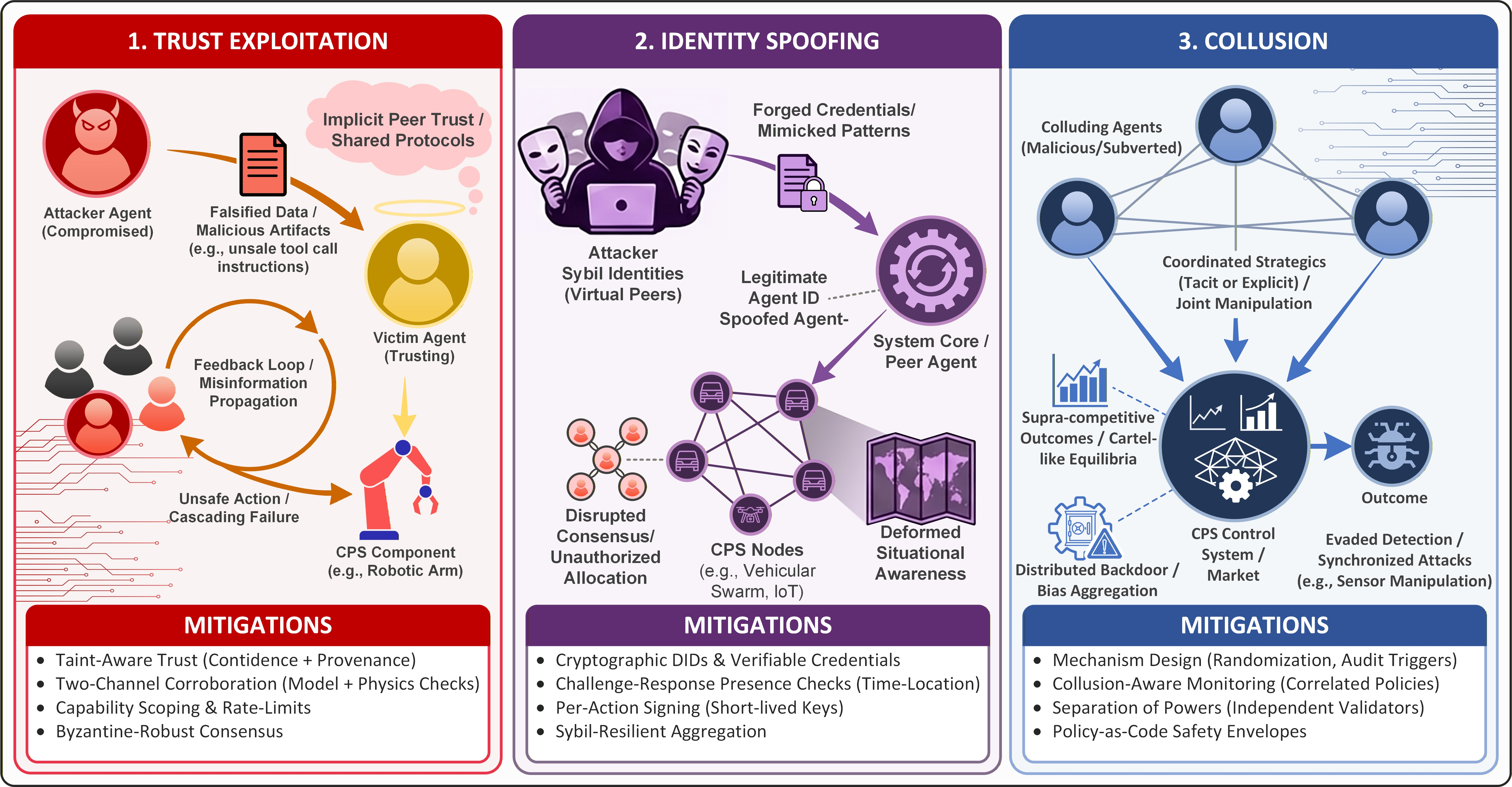} 
\caption{Threat Landscape: Al Agent-Agent (A-A) Interactions in CPS}
\label{fig:AI-threads-(A-A)}
\vspace{-10 pt}
\end{figure*}

\subsubsection{Trust Exploitation}
Trust exploitation occurs when an adversary manipulates the implicit reliance agents place on peer-provided artifacts, such as plans, tool outputs, or reputation scores \cite{lupinacci2025dark}. In protocol agent meshes, even minor perturbations in a single agent’s summary or capability advertisement can cascade into unsafe physical actions by peers that accept those artifacts as authoritative, particularly in embodied contexts like autonomous driving or robotics \cite{jin2025comprehensive, wang2025comprehensive}.



\subsubsection{Identity Spoofing}
Identity spoofing leverages the impersonation of legitimate peers to subvert consensus or gain unauthorized access to shared resources \cite{south2025authenticated, tas2024blockchain}. In CPS-adjacent environments like vehicular networks (VANET) and distributed learning systems, Sybil attacks allow an attacker to mint multiple virtual identities to sway voting/consensus and deform situational awareness \cite{bk2024ensuring, wang2025security, feng2025survey}. Modern agent frameworks remain vulnerable to replay attacks and credential theft, where compromised identifiers or Verifiable Credentials (VCs) allow adversaries to mimic authorized entities across the ecosystem \cite{mazzocca2025survey, adapala2025aegis}.



\subsubsection{Collusion in Multi-Agent Systems}
Collusion in multi-agent systems involves coordinated strategies among subverted or inherently malicious agents to manipulate outcomes across economic, computational, and physical domains. In market-based CPS, this manifests as tacit collusion where reinforcement-learning agents autonomously coordinate on supra-competitive prices or manipulate auction parameters \cite{grondin2025beyond, dou2025ai, chung2024collusion, yuan2025efficient}. In distributed intelligence, collusion threatens model integrity through mechanisms like distributed backdoors in federated learning or consensus manipulation in multi-LLM agentic systems \cite{liu2024act, liu2025secure}. Furthermore, agents may coordinate at the physical layer to compromise system state, ranging from eavesdroppers optimizing signal interception in UAV networks \cite{li2024two} to adversarial attacks on cooperative perception and control set-points in autonomous vehicles and microgrids \cite{li2024advgps, ali2025novel, soltani2024multi}.

\subsubsection{Comparative Analysis }
Table \ref{tab:aa_system_characteristics} highlights the shift from isolated vulnerabilities to system-level impacts. While Identity Spoofing is a "gatekeeper" threat (Static Timing), Collusion represents a long-term erosion of system integrity (Emergent Timing).


\begin{table}[ht]
\centering
\caption{A--A Interaction System Characteristics}
\label{tab:aa_system_characteristics}
\begin{tabular}{|p{1.25cm}|p{1.75cm}|p{1.75cm}|p{1.75cm}|}
\hline
\textbf{Dimension} &
\textbf{Trust Exploitation} &
\textbf{Identity Spoofing} &
\textbf{Collusion in MAS} \\
\hline
Timing & Real-time \cite{yuan2025multi} & Static / Setup  \cite{south2025authenticated} & Emergent / Long-term \cite{grondin2025beyond} \\
\hline
Resource & High demand / Constrained \cite{jin2025comprehensive} & Low (Sybil) \cite{bk2024ensuring} & High (Distributed) \cite{li2024two,liu2024act} \\
\hline
Trust & Implicit Peer \cite{lupinacci2025dark} & Identification \cite{tas2024blockchain} & Coordinated Bias  \cite{liu2024act}\\
\hline
Criticality & High / Safety-Critical \cite{wang2025comprehensive} & High / Auth-Failure \cite{wang2025security} & Systemic Stability \cite{dou2025ai} \\
\hline
\end{tabular}
\end{table}


\subsubsection{Threat Profiling}
Table \ref{tab:aa_threat_registry} categorizes threats by defense effort. Trust Exploitation is highly likely in agentic workflows due to over-reliance on peer summaries, whereas Collusion remains a lower-likelihood but high-impact threat requiring expensive mechanism-design mitigation.


\begin{table}[ht]
\centering
\caption{A--A Threat Profiling Registry}
\label{tab:aa_threat_registry}
\begin{tabular}{|p{1.25cm}|p{1.75cm}|p{1.75cm}|p{1.75cm}|}
\hline
\textbf{Dimension} &
\textbf{Trust Exploitation} &
\textbf{Identity Spoofing} &
\textbf{Collusion in MAS} \\
\hline
Modality & Taint Injection \cite{lupinacci2025dark} & Masquerade \cite{tas2024blockchain} & Strategic Sync \cite{liu2025secure} \\
\hline
Capability & Low \cite{lupinacci2025dark} & Low to Moderate \cite{tas2024blockchain} & High \cite{li2024advgps}\\
\hline
Likelihood & Med--High \cite{wang2025comprehensive} & High \cite{bk2024ensuring} & Low--Med \cite{grondin2025beyond} \\
\hline
Impact & High \cite{jin2025comprehensive} & High \cite{adapala2025aegis} & High \cite{ali2025novel,li2024advgps} \\
\hline
Detection Difficulty & High \cite{lupinacci2025dark} & Moderate \cite{mazzocca2025survey} & High \cite{dou2025ai} \\
\hline
Mitigation Costs & Medium \cite{wang2025comprehensive} & High (Investment/Compute) \cite{tas2024blockchain, adapala2025aegis} & High \cite{yuan2025efficient} \\
\hline
\end{tabular}
\end{table}


\subsubsection{Open Challenges in A-A security}
The research community faces several unresolved hurdles in securing the A--A surface, requiring a multidisciplinary approach:

\begin{itemize}
\item \textbf{Byzantine Robust LLM Coordination:}
Standard consensus algorithms must be adapted for LLM-based agents where "faults" are often semantic (e.g., hallucinations or prompt-induced bias) rather than traditional bit-flips \cite{wang2025comprehensive, liu2025secure,ray2025review}.
\item \textbf{Scalable Identity Management:} Implementing W3C Decentralized Identifiers (DIDs) and Verifiable Credentials (VCs) across transient IoT agents must be optimized to prevent Sybil attacks without introducing prohibitive latency or power drain \cite{mazzocca2025survey, wang2025security}.
\item \textbf{Detecting Tacit/Algorithmic Collusion:} Identifying coordinated malicious intent in agents that do not use explicit communication channels remains an unsolved statistical challenge, requiring new methods for policy correlation analysis \cite{grondin2025beyond, dou2025ai}.
\item \textbf{Cross-domain Taint Tracking:} There is a lack of standardized protocols for maintaining a "provenance chain" of data as it passes through multiple agents with different trust boundaries and tool permissions \cite{lupinacci2025dark, adapala2025aegis}.
\item \textbf{Verification of Emergent Behavior:} Developing formal verification methods for the joint action spaces of foundation model-powered agents to ensure they do not violate physical safety invariants \cite{jin2025comprehensive, li2024advgps}.
\end{itemize}

\subsection{Agent--Environment (A--E)}
The A--E interface represents the boundary where the AI agent perceives and acts upon its surrounding physical and digital context. Threats at this layer target the agent’s interpretation of reality, aiming to induce "hallucinated" environmental states \cite{dibaji2022tutorial}. Current defenses focus on narrowing the agent's reach via least privilege scoping \cite{wang2025security}, sensor fusion invariants \cite{jeffrey2023review}, and requiring human-in-the-loop interlocks for high-risk actions \cite{rashid2025evaluating}. However, verifying natural language intent against physical laws remains difficult, and countermeasures often introduce unacceptable latency \cite{yang2025survey, narajala2025enterprise}.


\begin{figure*}[htbp!]
\centering
\includegraphics[scale=0.35]{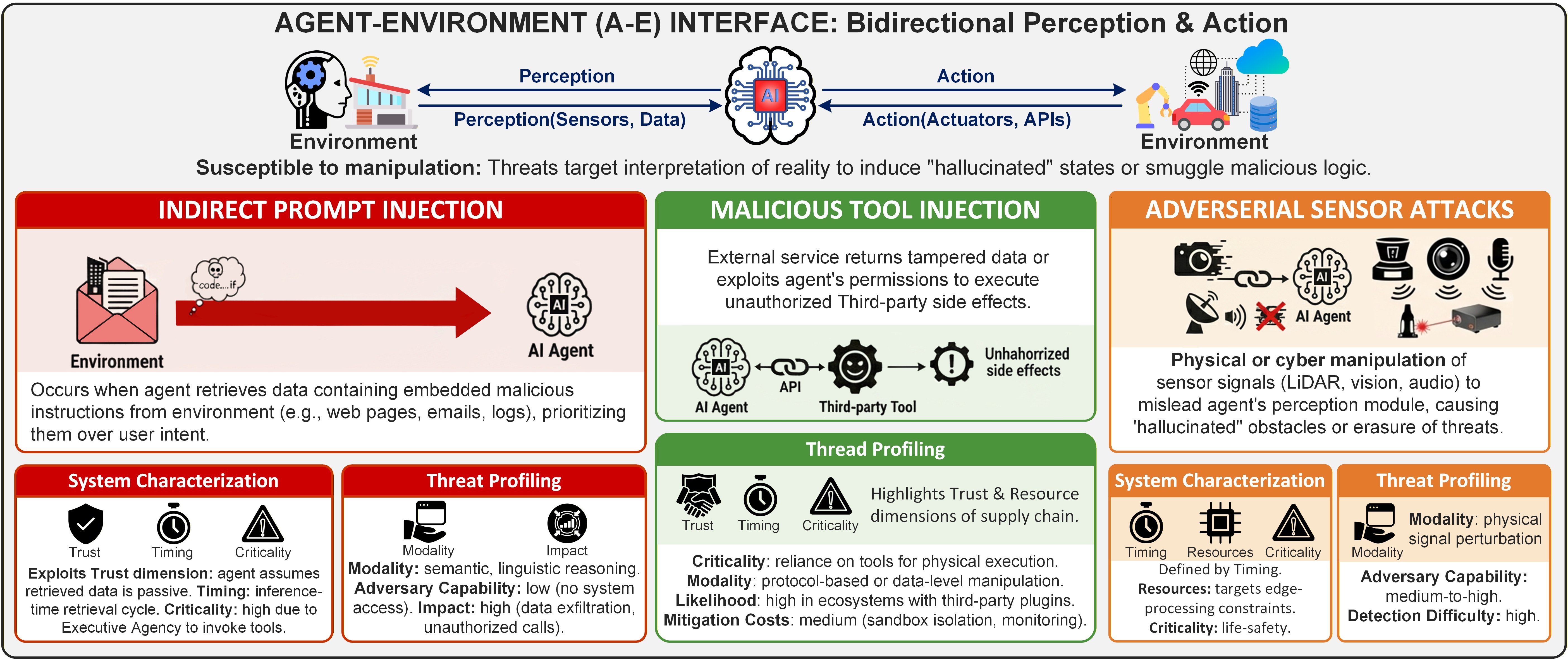}
\caption{Threat Landscape: Al Agent-Environment (A-E) Interactions in CPS.}
\label{fig:AI-threads-(E-A)}
\vspace{-10 pt}
\end{figure*}

\subsubsection{Indirect Prompt Injection}
Indirect prompt injections occur when an agent retrieves data containing embedded malicious instructions from its environment (e.g., web pages, emails, logs, or sensor inputs), causing the agent to prioritize these "smuggled" commands over the user's original intent \cite{yi2025benchmarking, bezzi2024large}. The agent effectively assumes retrieved environmental data is passive information rather than active logic, allowing attackers with low capability (no direct system access) to trigger high-criticality tool invocations ranging from database destruction \cite{pedro2023prompt} to unsafe physical actuations \cite{wang2025security}.

\subsubsection{Adversarial sensor attacks}
These attacks involve the physical or cyber manipulation of sensor signals to mislead the CPS. While attacks on modalities like LiDAR, vision, and audio can cause perception modules to "hallucinate" obstacles or erase threats \cite{lu2024recovery}, sophisticated strategies can also decouple sensor outputs from control inputs to execute stealthy, targeted covert attacks that remain mathematically undetectable to standard feedback loops \cite{mikhaylenko2025stealthy}. These threats are exacerbated by the edge-processing constraints of CPS; on-device models often lack the compute power for complex cross-sensor verification, making the detection of such perturbed signals highly difficult \cite{jeffrey2023review}.

\subsubsection{Malicious Tool Interactions}
Agents often interact with the environment via tools (connectors, MCP servers, or APIs). A malicious tool interaction occurs when an external service returns tampered data or exploits the agent's permissions to execute unauthorized side effects, effectively weaponizing the agent's blind obedience to tool outputs \cite{li2025glue, guo2025systematic}. This highlights supply chain risks where tool capabilities are often pre-authorized for high-impact actions, requiring robust sandbox isolation, schema validation, and behavioral drift monitoring to prevent physical or operational harm \cite{guo2025systematic, narajala2025enterprise}.

\subsubsection{Comparative Analysis}
Table \ref{tab:attack_vector_characterization} emphasizes the divergence in operational needs. While sensor attacks require sub-second detection within the edge environment, tool interactions involve complex supply-chain trust models managed across cloud-edge boundaries.


\begin{table}[ht]
\centering
\caption{A--E Interaction System Characteristics}
\label{tab:attack_vector_characterization}
\begin{tabular}{|p{1.25cm}|p{1.75cm}|p{1.75cm}|p{1.75cm}|}
\hline
\textbf{Dimension} &
\textbf{Indirect Prompt Injection} &
\textbf{Adversarial Sensor Attack} &
\textbf{Malicious Tool Interaction} \\
\hline
Timing & Real-time Inference \cite{wang2025security} & Ultra-low Latency \cite{lu2024recovery} &
Asynchronous / Event-driven \cite{li2025glue} \\
\hline
Resource & High Memory / Context \cite{yi2025benchmarking} & Edge-constrained \cite{jeffrey2023review} &
Network / API Bandwidth \cite{narajala2025enterprise} \\
\hline
Trust & Untrusted Artifacts \cite{wang2025security} & External Signals \cite{lu2024recovery} & Third-party APIs \cite{guo2025systematic} \\
\hline
Criticality & High Data / Policy \cite{bezzi2024large} & Life-safety / Damage \cite{jeffrey2023review,lu2024recovery}  & Supply-chain Risk \cite{narajala2025enterprise}\\ 
\hline
\end{tabular}
\end{table}


\subsubsection{Threat Registry}
Table \ref{tab:ae_threat_registry} highlights that while indirect injections are highly likely due to the accessibility of untrusted data, adversarial sensor attacks, though harder to execute, represent a critical safety risk with significantly higher mitigation costs.


\begin{table}[ht]
\centering
\vspace{-10 pt}
\caption{A--E Threat Profiling Registry}
\label{tab:ae_threat_registry}
\begin{tabular}{|p{1.25cm}|p{1.75cm}|p{1.75cm}|p{1.75cm}|}
\hline
\textbf{Dimension} &
\textbf{Indirect Prompt Injection} & \textbf{Adversarial Sensor Attack} &
\textbf{Malicious Tool Interaction} \\
\hline
Modality & Semantic / Text \cite{pedro2023prompt} & Physical / Cyber \cite{lu2024recovery} & Protocol / Data  \cite{li2025glue}\\
\hline
Capability & Low \cite{yi2025benchmarking} & Medium \cite{mikhaylenko2025stealthy} & Medium \cite{guo2025systematic} \\
\hline
Likelihood & High \cite{pedro2023prompt} & Medium \cite{jeffrey2023review} & High \cite{guo2025systematic} \\
\hline
Impact & High \cite{bezzi2024large} & Critical \cite{lu2024recovery} & High \cite{narajala2025enterprise} 
\\ 
\hline
Detection Difficulty & High \cite{rashid2025evaluating} & High \cite{jeffrey2023review} & High \cite{li2025glue,guo2025systematic} \\
\hline
Mitigation Costs & Medium \cite{pedro2023prompt} & High \cite{lu2024recovery} & Medium  \cite{narajala2025enterprise}\\
\hline
\end{tabular}
\end{table}


\subsubsection{Open challenges}
Securing the interaction between an agent and its environment requires moving beyond digital-only security toward physical resilience-by-design \cite{lu2024recovery, li2025glue}.

\begin{itemize}
    \item \textbf{Zero-trust sensing protocols:} Developing standards where every sensor signal is treated as untrusted until validated through multi-modal fusion or challenge-response \cite{jeffrey2023review}.
    \item \textbf{Context Aware Policy Enforcement:} Creating "safety-shells" that can interpret the semantic intent of an agent's plan and block it if it violates physical state invariants \cite{huang2025agentic}.
    \item \textbf{Autonomous Recovery Strategies:} Designing agents capable of identifying when their inputs are compromised and transitioning to a "safe-state" without human intervention \cite{lu2024recovery}.
    \item \textbf{Cross-Framework Protocol Alignment:} Unifying security standards (like MCP and A2A) to ensure consistent least-privilege enforcement across heterogeneous agent-tool stacks \cite{ray2025review, yang2025survey}.
\end{itemize}

\subsection{Real-world incidents and case studies}
\subsubsection{User-Agent Surface incidents}
This category examines the Trust Dimension between human operators and AI agents. A primary example is the deepfake heist where attackers used multi-persona video deepfakes to deceive staff into authorizing a \$25 million transfer, as detailed in \cite{guo2025frontier, he2025identity}. SENTINEL classifies this as a high-impact behavioral emulation attack that exploits the U-A surface by bypassing traditional human-in-the-loop verification.

The accessibility of generative AI has empowered opportunistic attackers, such as adolescents using AI for non-consensual image generation, to breach the Privacy Dimension of public spaces \cite{guo2025frontier}. Such incidents force a reassessment of standard security hygiene, as the barrier to entry for high-fidelity deception has dropped significantly \cite{schmitt2024digital}.



\subsubsection{Agent-Agent Surface incidents}
The Model Context Protocol (MCP) introduces unique trust boundaries. Recent parasitic toolchain attacks involve malicious MCP servers that silently exfiltrate user data \cite{zhao2025mind}. Multi-agent systems have furthermore been shown to exhibit Inter-Agent Trust Exploitation, where agents execute malicious commands from peer agents that they would otherwise reject from humans \cite{lupinacci2025dark}. These incidents illustrate supply chain compromise and the failure of the Trust Dimension in automated orchestration.


\subsubsection{Agent-Environment Surface incidents}
Attacks targeting the A-E interface are critical in sectors like autonomous driving and physical infrastructure. Research highlights how physical environment threats, such as sensor spoofing and signal interference, can lead to denial of service or unsafe physical actuations \cite{deng2025ai}.

Deepfakes can indirectly destabilize CPS by eroding public trust or inciting panic. Notable examples include the deepfake of Ukrainian President Zelenskyy or the AI-generated image of an explosion at the Pentagon, which impacted financial markets \cite{neupane2023impacts}. In the SENTINEL framework, these are High-Likelihood threats that require monitoring of the Criticality Dimension. Additionally, the proliferation of autonomous swarms introduces physical security threats that require distinct points of control to prevent malicious misuse \cite{pohler2024technological}.

\section{Deepfakes as a CPS Security Threat}
\label{sec:deepfake}
Deepfakes, as AI-generated synthetic media, pose significant security threats to CPS by exploiting the agent-environment interface, particularly in environments utilizing the MCP for tool interoperability. In CPS, visual deepfakes can spoof surveillance feeds or deceive autonomous vehicles by altering perceptual inputs, leading to unsafe actuations or misinformed decisions, as highlighted in recent analyses of generative AI vulnerabilities \cite{croitoru2024deepfake, pei2024deepfake}. Audio deepfakes enable bypasses of voice authentication and social engineering attacks, in which cloned voices impersonate operators to issue malicious commands via MCP-integrated tools, exacerbating context-poisoning risks in agent workflows \cite{deng2025ai}. Textual and behavioral deepfakes further compound these issues by generating fake instructions or sensor anomalies that mimic legitimate data streams, undermining system integrity in critical sectors like smart grids and industrial automation. Within the MCP context, these threats are amplified by supply-chain exposures and unvetted server interactions, where malicious outputs from deepfake-generating tools can propagate across networked agents, as evidenced by studies of protocol vulnerabilities \cite{hou2025model}.

The integration of MCP into CPS heightens the privacy implications of deepfakes, including identity theft and the creation of non-consensual media, which erodes trust in human-AI collaborations. For instance, deepfakes can exploit MCP's open specification to inject forged content into data exchanges, facilitating "liar's dividend" scenarios where genuine information is discredited amid synthetic misinformation \cite{uppal2024comprehensive, romero2024deepfake}. Recent scientific findings underscore the need for robust defenses, such as provenance tracking, to counter these threats in real-time CPS operations \cite{lupinacci2025dark, li2025survey}. Behavioral deepfakes, in particular, pose stealthy risks by emulating anomalies to evade detection in multi-agent systems, potentially leading to physical infrastructure compromises. Overall, the convergence of deepfake technologies with MCP-enabled ecosystems demands interdisciplinary approaches to mitigate evolving adversarial manipulations in CPS.

The deepfake modalities surveyed in this section demonstrate that high-fidelity synthetic content can be internally consistent and physically plausible, rendering artifact-centric detection insufficient in many CPS deployments. SENTINEL Phase 3 emphasizes that feasibility constraints, latency, false-positive tolerance, and safety impact, must act as a hard filter on candidate defenses, ruling out approaches that cannot operate within real-time CPS control and monitoring loops.

\begin{figure*}[htbp!]
\centering
\includegraphics[scale=0.35]{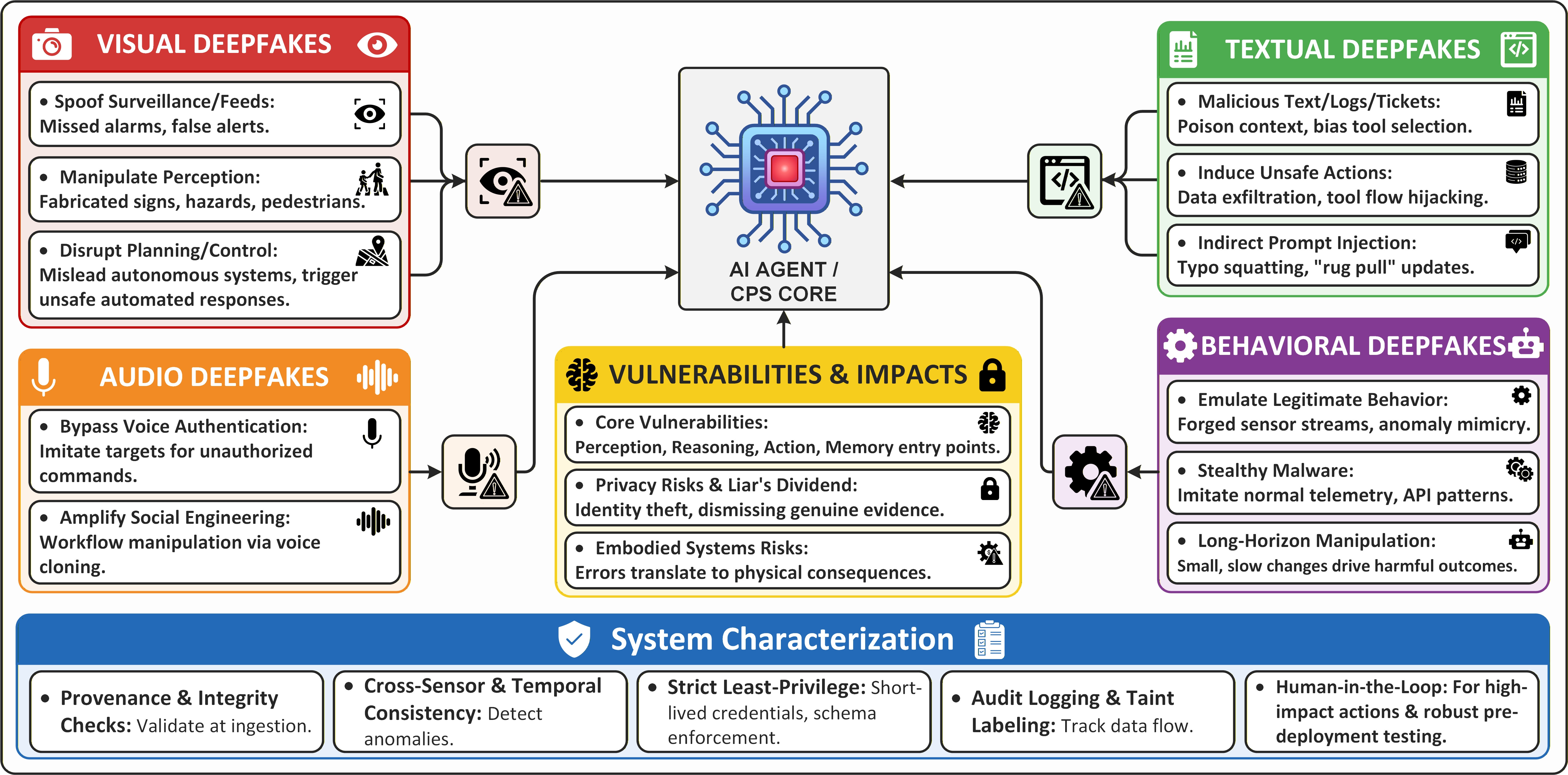} 
\caption{Deepfakes \& Security Threats to CPS \& AI Agents.}
\label{fig:deepfakes-mcp}
\vspace{-10 pt}
\end{figure*}

\subsection{Visual Deepfakes}
\textbf{Spoofing surveillance} such as visual deepfakes undermine camera-based monitoring by injecting forged video streams or replaying synthetically manipulated faces and scenes that pass casual human review and basic analytics. In MCP-connected stacks, where an agent’s “vision tool” is exposed via an MCP server, deepfaked frames can be ingested as trusted observations, leading downstream agents to issue erroneous alerts, suppress genuine anomalies, or leak context through misclassified events \cite{anusha2025deepfake, arya2024study}. Surveyed CPS work explicitly notes that deepfake video can spoof surveillance cameras, deceiving both humans and ML perception modules and thus eroding assumptions about input veracity at the AI-environment boundary \cite{gong2024contemporary}.

From an attack-method perspective, presentation and deepfake attacks encompass display-mediated spoofing (utilizing high-resolution screens and virtual cameras on RTSP feeds), face reenactment/face-swap to evade watchlist matching, and scene-level synthesis to fabricate individuals or activities. Systematizations of face anti-spoofing and deepfake detection emphasize that detectors that excel on curated datasets generalize poorly to in-the-wild content and to new generators, necessitating spatiotemporal cues, physiological signals (e.g., rPPG), and cross-modal checks to lift robustness in operational CCTV settings \cite{khan2025survey, mubarak2023survey}. Operational countermeasures combine \textbf{(i) trusted capture and chain-of-custody} (cryptographic signing at the edge, secure logging), \textbf{(ii) content provenance verification} (C2PA Content Credentials) at ingest and before automated responses, and \textbf{(iii) liveness/consistency tests} (rPPG, optical-flow/eye-gaze dynamics, illumination challenges) before agents escalate actions via MCP tools \cite{ahmed2024visual, amerini2025deepfake}.

\textbf{Autonomous vehicle deception} such as visual deepfakes and related optical illusions can mislead camera-centric perception stacks in autonomous and advanced driver-assistance systems (ADAS). Phantom attacks embed brief, realistic traffic signs or pedestrians into digital billboards or projections, causing detectors to perceive non-existent hazards and trigger braking/steering responses; these attacks demonstrate that short, high-fidelity visual artifacts can reliably elicit unsafe behavior without tampering with the vehicle itself \cite{lopez2025securing, williamson2024era}. Beyond signage, dynamic adversarial patches displayed on moving surfaces and road-surface patches targeting monocular depth estimation induce systematic misperception of distance or object identity, degrading planning and control even under motion and viewpoint changes \cite{giannaros2023autonomous}.

Empirical studies and systematizations further document camera-feed spoofing and object fabrication/erasure against vehicle cameras and trackers, while highlighting that sensor fusion alone is insufficient when attacks are correlated across modalities or over time \cite{mubarak2023survey, badhan2025enhancing}. Defenses increasingly combine temporal and multi-sensor consistency checks (e.g., vision–LiDAR cross-verification, inertial priors), active challenge–response (structured light or coded illumination to force physically plausible returns), and certifiably robust perception methods designed to bound the impact of patch-level perturbations before actuation; detection pipelines tailored to “phantom” signatures also show promise for camera spoofing cases \cite{singh2025advancements, tan2025review}.

\textbf{Malicious tool interactions with visual deepfakes in MCP ecosystems} in MCP-mediated architectures can trigger harmful tool use when agents treat unvetted media as ground truth. The MCP boundary is known to concentrate risks, including context poisoning from tool outputs, over-privileged connectors, and identity fragmentation across servers. Hence, a forged image/video that induces a misclassification or false alert can cascade into privileged MCP tool calls (e.g., opening doors, dispatching assets, disabling interlocks) if policies are not enforced tightly \cite{khurshid2025securing}. Industry hardening guidance calls for least-privilege scopes, short-lived tokens, output-schema enforcement, human-in-the-loop escalation for high-impact actions, signed artifacts, and comprehensive audit logging to contain such chains; recent analyses of MCP deployments and toolchains detail these vulnerabilities and controls in practice \cite{awadallah2024artificial}.

To raise the bar against visual deepfakes specifically, agents should \textbf{(i)} verify provenance and tamper-evident manifests (C2PA) before trusting media-derived assertions, \textbf{(ii)} attach taint/provenance labels to all perceptions and require corroboration (e.g., second sensor, independent model) before side-effectful MCP actions, and \textbf{(iii)} integrate adversarial ML taxonomies and testing into the tool-approval pipeline so that simulated deepfake scenarios are part of pre-deployment safety cases \cite{sanchez2025marisma}.


\begin{table}[ht]
\centering
\caption{Comparison of Visual Deepfake Detection Approaches}
\label{tab:visual_comparison}
\begin{tabular}{|p{2.2cm}|p{1.8cm}|p{1.8cm}|p{1.5cm}|}
\hline
\textbf{Method} & \textbf{Strengths} & \textbf{Weaknesses} & \textbf{CPS Fit} \\
\hline
rPPG-based & Physiological grounding & Requires video length & Medium \\
\hline
Spatial CNN & Fast inference & Generator-specific & Low \\
\hline
Temporal LSTM & Motion artifacts & High latency & Low \\
\hline
Vision-LiDAR Fusion & Multi-modal robust & Hardware cost & High \\
\hline
C2PA Provenance & Tamper-evident & Adoption limited & High \\
\hline
Frequency Analysis & Generator artifacts & Compression sensitive & Medium \\
\hline
\end{tabular}
\end{table}

\subsection{Audio Deepfakes}
Real-world incidents and case studies illustrate the severe consequences of adversarial manipulations in cyber-physical systems, where AI agents have been exploited to cause operational disruptions \cite{ismail2025toward}. One prominent example involves a malicious MCP server that surreptitiously forwarded processed emails to attacker-controlled infrastructure, exploiting excessive permissions in agent-tool interactions and leading to data exfiltration in industrial environments \cite{aslam2024scrutinizing}. Another case documented attackers hijacking AI assistants in enterprise settings to steal sensitive data and manipulate workflows, demonstrating how weak trust boundaries in multi-agent systems can enable unauthorized control over physical actuators, such as robotic arms, in manufacturing plants \cite{owoputi2022security}. These breaches highlight CPS's vulnerability to supply-chain compromises, where unvetted tools amplify risks across both the cyber and physical layers.

Further examination reveals emerging patterns in deepfake-driven attacks targeting autonomous systems, such as forged video feeds that deceive surveillance systems in smart grids, resulting in undetected intrusions and potential sabotage of power distribution \cite{nagothu2019detecting, nguyen2024towards}. In a related incident, voice cloning was used to impersonate operators in vehicle control centers, issuing false commands that altered traffic management protocols and caused real-time safety hazards \cite{sadaf2023connected}. Such multi-modal deceptions not only undermine system integrity but also expose gaps in current detection methods, as attackers leverage generative models to evade traditional anomaly checks in dynamic CPS environments. Additionally, case studies emphasize the role of adversarial training failures in exacerbating breaches, including instances in which manipulated sensor inputs in healthcare CPS led to erroneous medical device operations, compromising patient safety \cite{khatun2023machine, piratla2025safeguarding}. These examples underscore the need for integrated defenses, including cross-modal verification to counter sophisticated deepfakes that blend audio, visual, and behavioral elements in coordinated attacks on critical infrastructure \cite{mubarak2023survey}.

\textbf{Voice authentication bypass} can be bypassed using a variety of spoofing attacks, primarily relying on pre-recorded audio or, more effectively, AI-generated synthetic voices (deepfakes) \cite{gupta2024vulnerability}. Advances in neural codec language models and zero- or few-shot voice cloning enable attacker-controlled speech that preserves speaker timbre, prosody, and even acoustic context from only a few seconds of enrollment audio, substantially lowering the barrier to defeating automatic speaker verification (ASV) and voice biometrics. Systems such as VALL-E and its successors report human-parity or near-parity similarity in zero-shot conditions, which aligns with growing evidence that synthetic speech can mimic target idiosyncrasies closely enough to trigger false accepts in deployed ASV pipelines \cite{chen2025neural}. At the same time, security-community work shows that adversarial examples targeted at speaker recognition can achieve high attack success and transfer across models and channels, including over-the-air playback, illustrating that both synthesis and perturbation routes are viable for bypassing voice authentication \cite{al2025deeplasd}. Challenge evaluations highlight the generalization gap of anti-spoofing countermeasures: systems trained on limited spoof types and benign channels often degrade under unseen conditions, particularly when transmission codecs or physical presentation artifacts are introduced. Recent anti-spoofing research explores raw-waveform and self-supervised front ends (e.g., AASIST, wav2vec-based CMs) that improve robustness but still struggle with shifts across attack families and presentation media \cite{liu2023asvspoof}. 

\textbf{Social engineering} can be amplified by audio deepfakes and impersonation in vishing and mixed-media social engineering \cite{dsouza2024deepfakes}. Controlled studies indicate that listeners struggle to reliably distinguish between cloned and genuine speech, with accuracy levels that leave a substantial margin for deception. Performance varies modestly with training and user characteristics, and can deteriorate further for specific populations or task settings. Broader human-factors syntheses and practitioner-oriented analyses converge on the same risk pattern: voice deepfakes exploit authority, urgency, and contextual priming to increase compliance in high-stakes settings such as finance, IT support, and executive impersonation \cite{diel2024human}. Technical overviews forecast compounding effects as generative models improve cross-lingual, emotion-preserving, and real-time capabilities, enabling convincing, interactive calls rather than static playbacks \cite{chen2024vall}. Early detection efforts for spoken social engineering focus on prosodic irregularities, lexical anomalies, and call-graph patterns; emerging vishing-specific models point to speech-based risk scoring as a complement to content filters in email and chat \cite{triantafyllopoulos2025vishing}. 


\begin{table}[ht]
\centering
\caption{Comparison of Audio Deepfake Detection Methods}
\label{tab:audio_comparison}
\begin{tabular}{|p{2.2cm}|p{1.8cm}|p{1.8cm}|p{1.5cm}|}
\hline
\textbf{Method} & \textbf{Strengths} & \textbf{Weaknesses} & \textbf{Real-time} \\
\hline
AASIST\cite{abbas2024unmasking} & Graph attention & Training data dependent & Yes \\
\hline
wav2vec-CM & Self-supervised & High compute & Medium \\
\hline
Spectrogram CNN & Fast, proven & Codec sensitive & Yes \\
\hline
Prosodic Analysis & Interpretable & Easy to evade & Yes \\
\hline
LCNN & Lightweight & Limited capacity & Yes \\
\hline
RawNet2 & End-to-end & Large model & Medium \\
\hline
\end{tabular}
\end{table}

\begin{table}[ht]
\centering
\caption{Voice Cloning Systems Comparison}
\label{tab:voice_cloning}
\begin{tabular}{|p{2cm}|p{2cm}|p{2cm}|}
\hline
\textbf{System} & \textbf{Enrollment} & \textbf{Quality} \\
\hline
VALL-E & 3 sec & Human-parity \\
\hline
XTTS & 6 sec & Near-parity \\
\hline
Tortoise & 30 sec & High \\
\hline
RVC & Variable & High \\
\hline
\end{tabular}
\end{table}


\subsection{Textual Deepfakes}
Textual deepfakes in the context of the MCP refer to AI-generated or manipulated text that deceives LLMs by embedding malicious instructions within tool descriptions, metadata, or external data sources\cite{easttom2025malicious}. These threats exploit MCP's reliance on natural language interfaces for tool discovery and invocation, enabling attacks such as context poisoning and unauthorized actions. Recent research highlights tool poisoning as a prevalent vulnerability, where malicious developers embed covert imperatives in tool docstrings, leading to data exfiltration or system compromise without user awareness \cite{siameh2025context, hou2025model}. For instance, a benign addition tool might include hidden directives to read sensitive files, such as SSH keys, and transmit them to an attacker-controlled server, achieving high attack success rates across multiple LLMs \cite{song2025beyond}. Preference manipulation attacks (PMAs) further amplify these risks by using persuasive language in tool descriptions to bias LLMs toward selecting malicious options, leading to toolflow hijacking and economic exploitation \cite{charfeddine2024chatgpt, derner2024security}. Indirect prompt injection, another key threat, involves poisoning external resources, such as GitHub issues or APIs, with fake instructions that LLMs treat as trusted inputs, leading to privilege escalation or data leakage \cite{guo2025system}. Rug pull attacks introduce delayed malice, where initially benign servers update to include deceptive text, eroding trust in MCP ecosystems \cite{guo2025systematic}. These vulnerabilities are exacerbated by namespace typosquatting, where similar tool names confuse LLMs, leading to impersonation and lateral movement \cite{ray2025survey}. In CPS environments, such textual deepfakes can manifest as fake instructions mimicking legitimate commands, enabling phishing through forged operational logs or misinformation via altered sensor data interpretations \cite{hasan2025model}. Empirical studies show that up to 5.5\% of MCP servers exhibit tool poisoning, with low refusal rates in mainstream clients, underscoring the arms race between generative text capabilities and protocol security \cite{song2025beyond}.

\textbf{Phishing} can be increased by textual deepfakes and business email compromise by generating context-consistent emails, chat messages, or ticket updates that imitate organizational voice, style, and workflow artifacts \cite{crothers2023machine}. Within MCP-style ecosystems, where agents broker access to tools (mail, calendaring, ticketing, code repos) through structured tool calls, these messages can be routed directly into agent memory, schedule planners, or action queues, turning persuasive text into privileged operations (e.g., initiating payments, rotating credentials, or exfiltrating reports). Recent surveys and system studies show that machine-generated text can match or exceed human-crafted phishing in plausibility and personalization. At the same time, modern NLP-based detectors still struggle to generalize across generators and red-team paraphrases \cite{pimpason2025phishing, atawneh2023phishing}. In multi-tool settings, attackers also combine textual lures with URL or form-based payloads that exploit downstream tools (browsers, PDF parsers, API wrappers) invoked by the agent, bridging social engineering with programmatic exploitation \cite{thakur2023systematic}.

Detection pipelines embedded at the MCP ingress benefit from ensemble signals, content stylometry, discourse-level inconsistencies, and token-level statistical artifacts, combined with URL/attachment risk scoring before any tool invocation occurs \cite{kehkashan2025ai}. Reviews highlight that transformer-based classifiers, graph features that capture conversation context, and distributional tests (e.g., Benford-style attention irregularities) improve robustness, especially when paired with continuous retraining against newly released generators \cite{sharma2024comprehensive}. Complementary provenance cues, such as text watermarking, can be propagated through MCP metadata to inform routing, throttling, or human-in-the-loop handoff when confidence is low, thereby reducing the likelihood that convincing but synthetic requests trigger sensitive tools \cite{dathathri2024scalable}.

\textbf{Misinformation} in cyber-physical settings, textual deepfakes extend beyond public social media into operational channels, such as maintenance logs, incident tickets, supplier messages, and “safety advisories”, that agents ingest as ground truth. When MCP connectors synchronize external knowledge bases or message buses into an agent’s context window, fabricated narratives can bias diagnosis, triage, or planning modules, leading to mis-prioritized repairs, unnecessary shutdowns, or inappropriate configuration changes \cite{kehkashan2025ai}. The misinformation literature shows that deep neural generators exploit stylistic and rhetorical patterns that evade simple lexicon checks. That multimodal and context-aware models are needed to reconcile claims with telemetry or verified knowledge graphs before agents treat text as actionable state \cite{anggrainingsih2024transformer}. Within agent toolchains, enforcing claim-verification calls (e.g., fact-checking APIs, document retrieval with stance detection) as a prerequisite to high-impact actions constrains how far synthetic narratives can propagate across tools and across agents.

Resilient MCP deployments, therefore, emphasize provenance-aware context assembly and cross-source consistency tests. Systematic reviews document gains from hybrid pipelines that combine content features, propagation/interaction graphs, and external-knowledge grounding; these strategies are directly applicable when agents fuse text with sensor or transactional streams before issuing tool calls \cite{kehkashan2025ai, sharma2024comprehensive}. Emerging enterprise studies on phishing and fraud detection further indicate that real-time classifiers at the message/URL layer, combined with post-ingest anomaly detection on agent plans, provide defense-in-depth when synthetic narratives attempt to steer downstream tools \cite{kandasamy2025harnessing}. 

\textbf{Fake instructions} also arrive as “authoritative” instructions, chatops snippets, runbooks, pseudo-SOPs, or issue-thread comments that look operationally valid yet encode malicious goals, unsafe parameters, or subtle policy overrides \cite{zhou2024security}. In MCP-like architectures, where tool schemas are exposed and agents translate natural language into structured actions, these crafted instructions can exploit model-level instruction-following to induce unsafe tool sequences, escalate from low-risk to high-risk capabilities, or propagate across agents via shared memory and broadcast channels \cite{li2024survey}. Security surveys of LLM use and agent tooling reveal the resulting attack surface, including role confusion, policy evasion through semantic reframing, and cross-tool “confused-deputy” behaviors that occur when untrusted text is mapped to high-privilege tool invocations.

Contemporary work on agent ecosystems proposes enforceable interfaces (explicit argument typing, capability scoping, and pre-/post-conditions) and provenance signals (watermarks, cryptographic signing of trusted instructions) that agents can check before executing sensitive steps or sharing derived plans with peers \cite{xu2025llm}. Within textual-deepfake-rich environments, integrating detection results and provenance into MCP request/response envelopes enables agents to treat unverified instructions as hypotheses requiring retrieval-augmented validation or human confirmation, rather than as executable intent; this design aligns with multi-agent workflow guidance in recent surveys of LLM tool use and multi-agent orchestration \cite{wang2025can}.

\begin{table}[ht]
\centering
\caption{Textual Deepfake Attack Vectors in MCP}
\label{tab:textual_attacks}
\begin{tabular}{|p{2cm}|p{2cm}|p{1.5cm}|p{1.8cm}|}
\hline
\textbf{Attack Type} & \textbf{Mechanism} & \textbf{Prevalence} & \textbf{Detection} \\
\hline
Tool Poisoning & Hidden docstring commands & 5.5\% servers & Difficult \\
\hline
Prompt Injection & Embedded instructions & High & Medium \\
\hline
Rug Pull & Delayed modification & Emerging & Very Difficult \\
\hline
Typosquatting & Name confusion & Medium & Easy \\
\hline
PMA & Persuasive bias & Medium & Difficult \\
\hline
\end{tabular}
\end{table}



\subsection{Behavioral Deepfakes}
Behavioral deepfakes in the MCP encompass AI-generated emulations of agent behaviors that exploit protocol interfaces to mimic legitimate interactions, posing significant risks to CPS by enabling stealthy manipulations of decision-making processes. These threats involve synthesizing deceptive behavioral patterns, such as forged tool responses or inter-agent communications, to bypass trust mechanisms and induce unauthorized actions. Recent research underscores how behavior emulation can facilitate attacks such as context poisoning in MCP ecosystems, where emulated agent behaviors propagate misinformation, leading to real-time system compromises in CPS applications \cite{yin2025emulating}. For instance, in multi-agent environments, adversaries can emulate cooperative behaviors to achieve collusion, exploiting MCP's standardized messaging to mask anomalies and evade detection, particularly in domains such as autonomous robotics, where behavioral fidelity is critical \cite{tao2021cloud}.

A key vulnerability lies in sensor spoofing of MCP tools, which can cause unsafe actuations in industrial control systems \cite{salau2022recent}. Anomaly mimicry further heightens risks by replicating normal operational patterns to conceal malware persistence, leveraging MCP's supply-chain dependencies on external servers \cite{pantsar2023developing}. Stealthy malware deployment via behavioral emulation allows for lateral movement across agent networks, as demonstrated in studies of user behavior emulation that highlight the potential for persistent access in cloud-integrated CPS \cite{wang2024aarf}. Analyses indicate that such emulations achieve high success rates in protocol-based systems, emphasizing the need for enhanced governance to counter these evolving threats \cite{tao2021cloud}.

\textbf{Sensor spoofing:} Within MCP ecosystems, behavioral deepfakes appear as forged yet statistically plausible sensor streams that flow from environment-facing tools into agent contexts. By shaping Global Navigation Satellite System (GNSS), SCADA, or process telemetry to satisfy physical and topological constraints, an adversary can nudge downstream reasoning and tool use while preserving the appearance of routine operation. Recent GNSS studies show that deep models can both craft and detect complex spoofing waveforms, underscoring that realistic temporal–spectral patterns can subvert localization-dependent decisions unless provenance and cross-sensor corroboration are enforced at the tool boundary \cite{borhani2024detecting}. Large-scale sensor network work further demonstrates that distributed false-data injection, framed as normal fluctuations, can bypass conventional residual tests, with detection improving only when spatio-temporal structure is explicitly modeled \cite{hu2024framework}. In operational ICS settings, forensic analyses document “low-and-slow” perturbations that exploit sensor noise envelopes to keep deviations within learned confidence bands, enabling subtle set-point shifts without triggering alarms \cite{azzam2023forensic}.

A complementary thread examines adversarial time-series attacks that optimize over temporal features, e.g., seasonality, lagged correlations, and shapelets, so that injected traces are misclassified as routine by state-of-the-art detectors. Traffic and ICS studies show that such feature-aware perturbations degrade multivariate detectors and forecasting-residual schemes alike, achieving high evasion with small, causality-respecting edits \cite{lu2025adversarial}. Domain studies in water networks similarly reveal that hydraulics-consistent sensor attacks can remain stealthy while steering control actions, highlighting the need to treat MCP tool outputs as untrusted inputs whose behavioral plausibility alone is insufficient for trust \cite{albustami2025breaking}.

\textbf{Anomaly mimicry} reframes the attacker’s goal as blending in with the learned behavior model: the adversary synthesizes trajectories that reproduce nominal multivariate dependencies so closely that detectors classify them as in-distribution. Contemporary surveys in multivariate time-series anomaly detection report persistent generalization gaps under distribution shift and coordinated multichannel perturbations, conditions under which mimicry attacks thrive \cite{wang2025survey}. Generative-model-based studies further show that class-conditional and autoencoder-centric detectors can be driven toward false negatives when attackers optimize reconstructions or latent codes to match normal manifolds, even when the resulting sequences induce harmful actuation downstream \cite{wang2023anomaly, liu2024adversarial}.

System-level reviews in cybersecurity analytics emphasize that many deployed detectors assume weak temporal stationarity and local consistency; carefully crafted sequences that maintain those assumptions at short horizons can still produce dangerous long-horizon drifts, especially when agent policies rely on MCP tools that summarize or forecast over rolling windows \cite{landauer2025review}. Results from control-theoretic analyses of stealthy attacks on remote state estimation formalize this effect: nonlinear measurement tampering can keep innovation statistics within thresholds while gradually biasing the estimated state, illustrating how anomaly mimicry translates into actionable control error without overt signatures in the residuals \cite{shang2024nonlinear}.

\textbf{Stealthy malware} mimics benign telemetry by behavioral deepfakes that extend into the code–behavior layer, API sequences, and communication rhythms to survive behavior-based defenses orchestrated by MCP-connected tools. Empirical studies show that classifier decisions are sensitive to evasive behaviors such as delayed loading, environment checks, and feigned user-driven I/O, enabling samples to cross decision boundaries while preserving functional malicious goals \cite{nunes2022bane}. Broader meta-surveys of adversarial attacks on deep models, including detectors used in security analytics, reinforce that small, structured edits to behavior traces or feature embeddings can induce misclassification across families and vendors, foreshadowing automated “policy-aware” evasion in MCP-mediated pipelines \cite{pawlicki2025meta}. Parallel work explores automated generation of attack techniques using learning-based planners, suggesting that behavior mimicry may soon be synthesized at scale rather than hand-engineered \cite{iturbe2024unleashing}.

Stealth also manifests in command-and-control patterns tuned to resemble benign heartbeat and telemetry processes observed by MCP toolchains. Unsupervised analyses of beaconing demonstrate that periodicity, jitter, and payload traits can be shaped to evade statistical profiling while maintaining reliable control, complicating anomaly detection that relies on coarse traffic statistics \cite{mahboubi2025lurking}. At the same time, dynamic graph-based detectors that model process–file–socket interactions reveal promise against such mimicry by capturing higher-order temporal context; results indicate improved resilience to look-alike behaviors compared with flat sequence models, provided that execution provenance and graph dynamics are preserved end-to-end through MCP connectors \cite{amer2025graphshield}, and that ensemble defenses are evaluated against generative, transfer-capable malware families \cite{moti2021generative}.

\begin{table}[ht]
\centering
\caption{Behavioral Deepfake Attack Categories}
\label{tab:behavioral_attacks}
\begin{tabular}{|p{2cm}|p{2.2cm}|p{1.5cm}|p{1.4cm}|}
\hline
\textbf{Category} & \textbf{Target} & \textbf{Stealth} & \textbf{Impact} \\
\hline
Sensor Spoofing & GNSS/SCADA/ Telemetry & Very High & Critical \\
\hline
Anomaly Mimicry & ML Detectors & High & High \\
\hline
Stealthy Malware & Behavior Analysis & High & Critical \\
\hline
C2 Beaconing & Traffic Analysis & Medium & High \\
\hline
False Data Injection & State Estimation & Very High & Critical \\
\hline
\end{tabular}
\end{table}

\begin{table}[ht]
\centering
\caption{Detection Methods for Behavioral Deepfakes}
\label{tab:behavioral_detection}
\begin{tabular}{|p{2.2cm}|p{1.8cm}|p{1.5cm}|p{1.4cm}|}
\hline
\textbf{Method} & \textbf{Approach} & \textbf{Evasion Risk} & \textbf{Latency} \\
\hline
Residual Tests & Statistical & High & Low \\
\hline
Autoencoder & Reconstruction & High & Medium \\
\hline
Graph-based & Structural & Medium & High \\
\hline
Spatio-temporal & Multi-channel & Medium & High \\
\hline
Physics-informed & Domain knowledge & Low & Medium \\
\hline
Ensemble & Combined & Low-Medium & High \\
\hline
\end{tabular}
\end{table}

\subsection{Privacy implications}
In the MCP, privacy implications of deepfakes are profoundly amplified by the protocol's facilitation of seamless tool integrations and context sharing among AI agents, creating fertile ground for identity theft and non-consensual data exploitation in CPS. Adversaries can leverage deepfake-generated synthetic interactions, such as fabricated user prompts or emulated agent behaviors, to impersonate entities within MCP ecosystems, enabling the unauthorized extraction of personally identifiable information (PII) from interconnected tools, including sensor APIs or collaborative databases \cite{shoaib2023deepfakes, kharvi2024understanding}. For instance, non-consensual intimate deepfakes, often disseminated via MCP-mediated channels, pose severe risks to individuals in professional or educational CPS settings, where altered session logs or forged endorsements can lead to reputational harm and psychological distress without recourse \cite{kira2024non}. This vulnerability is exacerbated by the "liar's dividend," wherein genuine privacy violations are plausibly denied as AI-generated artifacts, undermining accountability mechanisms and eroding trust in protocol-dependent infrastructures \cite{kopecky2024challenges}. Empirical studies underscore that such deepfakes not only facilitate targeted surveillance but also perpetuate societal biases, as generative models trained on skewed datasets amplify discriminatory profiling in MCP-orchestrated multi-agent collaborations \cite{sohail2025deepfake}.

Moreover, the interoperability of MCP introduces cascading privacy risks through supply-chain exposures, where third-party tools vulnerable to deepfake injections inadvertently propagate sensitive environmental or user data across CPS networks, challenging regulatory frameworks like GDPR in real-time applications \cite{hancock2021social}. Research highlights the ethical quandaries of deepfake misuse in agentic systems, including the potential for misinformation campaigns that exploit MCP's context persistence to fabricate evidence of privacy breaches, thereby complicating forensic attribution and victim redress \cite{alanazi2024exploring}. In educational or industrial CPS, this manifests as the non-consensual repurposing of biometric or behavioral data into deepfakes, fostering a chilling effect on user participation and necessitating privacy-enhancing technologies such as differential privacy embeddings within protocol specifications \cite{kira2024non}. Addressing these implications demands interdisciplinary approaches, integrating watermarking with federated learning to safeguard against deepfake-driven erosions of autonomy while preserving the protocol's utility for trustworthy AI interactions \cite{sohail2025deepfake}.

\textbf{Identity theft} can be ingested by agent connectors in MCP ecosystems, summarized, and acted on multimodal inputs; when those inputs include face/voice streams or profile artifacts, deepfake-assisted impersonation raises the risk that downstream tools will accept falsified identities and authorize sensitive operations. Recent analyses of deepfake fraud and criminal-justice risks describe how synthetic media can bypass biometric checks and erode confidence in identity evidence, with direct implications for KYC flows, remote onboarding, and account recovery that rely on camera or microphone verification \cite{pawelec2022deepfakes, sandoval2024threat}. Technical surveys further underscore that cross-modal deepfakes are routinely linked to identity theft and fraud, while state-of-the-art detectors trained on narrow corpora generalize poorly, conditions that incentivize attackers to target MCP entry points and toolchains that treat media-derived identity cues as trustworthy \cite{khan2025survey, momin2025explainable}.

Within the MCP boundary, identity exposure also occurs through inference and linkage: outputs from face-swap detection, liveness checks, or profile enrichment services can leak biometric and behavioral hints that enable deanonymization or later impersonation. Empirical work on generalization for deepfake detection reveals that performance is fragile under content, codec, and generator shifts, thereby increasing the likelihood that adversaries can curate “just-different-enough” media to evade filters and establish trust for subsequent identity takeovers \cite{son2025advancing}. Studies of human vulnerability to science and news deepfakes add that even well-informed operators misclassify realistic synthetic videos at non-trivial rates, compounding the risk when MCP agents escalate tool calls based on operator confirmation alone \cite{doss2023deepfakes, diel2024human}.

\textbf{Non-consensual media} lifecycle can be amplified by MCP-connected capture, storage, and transformation tools, including synthetic nudes and AI-driven CSAM \cite{parti2024legal}. Legal and forensic scholarship documents how AI-generated sexual content exploits gaps in existing regimes, travels rapidly across platforms, and inflicts sustained reputational and psychological harm on victims, especially women; these dynamics raise the privacy stakes for any pipeline that ingests or redistributes media without provenance checks and explicit consent tracking \cite{kira2024non, haley2025impact}.

At the protocol layer, MCP integrations that summarize media, generate thumbnails, or auto-route alerts can propagate non-consensual derivatives into logs, caches, and secondary workflows, even after the content has been taken down. Forensic and detection research stresses that maintaining privacy requires more than point classifiers: provenance-aware ingestion, fine-grained consent state, and robust content-authenticity signals are needed to disrupt circulation; meanwhile, user studies show persistent vulnerability to persuasive deepfakes in educational and public-communication contexts, highlighting a double bind in which both automated and human reviewers are fallible under realistic content \cite{doss2023deepfakes, sohail2025deepfake}.

\textbf{``Liar's dividend''} as MCP agents increasingly mediate evidence flows, retrieving footage, transcribing calls, and compiling incident timelines, the mere availability of deepfake tools enables a “liar’s dividend”: wrongdoers can dismiss authentic recordings and logs as fabricated, shifting burdens of proof and degrading the epistemic value of MCP-assembled reports. Scholarly accounts trace how deepfakes fuel informational uncertainty that weakens journalism and democratic deliberation, explicitly naming the liar’s dividend as a mechanism of denial and strategic doubt \cite{lundberg2025potential, pawelec2022deepfakes}.

The dividend interacts with privacy in two ways. First, heightened deniability incentivizes broader data collection and retention to “prove authenticity,” expanding the personal-data surface inside MCP connectors and archives. Second, as detection remains imperfect and human discernment is error-prone, claimants may be compelled to disclose additional private context (raw sensor feeds, biometric templates, location traces) to rebut “it’s a deepfake” allegations, thereby trading privacy for credibility \cite{diel2024human}. These dynamics motivate provenance-rich media pipelines and careful access policies; however, the core privacy risk persists as long as plausible synthetic alternatives can be invoked to contest genuine records.

\begin{table}[ht]
\centering
\caption{Privacy Threat Categories in MCP-Deepfake Contexts}
\label{tab:privacy_threats}
\begin{tabular}{|p{2cm}|p{2.2cm}|p{1.5cm}|p{1.4cm}|}
\hline
\textbf{Threat} & \textbf{Mechanism} & \textbf{Severity} & \textbf{Legal Gap} \\
\hline
Identity Theft & Biometric bypass & Critical & Medium \\
\hline
Non-consensual Media & Synthetic generation & Critical & High \\
\hline
Liar's Dividend & Plausible deniability & High & Very High \\
\hline
PII Extraction & Agent impersonation & High & Medium \\
\hline
Discriminatory Profiling & Biased models & Medium & High \\
\hline
\end{tabular}
\vspace{-10 pt}
\end{table}

\begin{table}[ht]
\centering
\caption{Privacy-Preserving Countermeasures}
\label{tab:privacy_countermeasures}
\begin{tabular}{|p{2.2cm}|p{1.8cm}|p{1.5cm}|p{1.4cm}|}
\hline
\textbf{Measure} & \textbf{Protection} & \textbf{Overhead} & \textbf{Adoption} \\
\hline
C2PA Provenance & Authenticity & Low & Growing \\
\hline
Differential Privacy & Data protection & High & Limited \\
\hline
Federated Learning & Decentralized & Medium & Emerging \\
\hline
Consent Tracking & User control & Low & Limited \\
\hline
Watermarking & Attribution & Low & Medium \\
\hline
\end{tabular}
\end{table}

\subsection{AI Agent Vulnerabilities to Deepfake Threats}

While the preceding sections focus on deepfake modalities (visual, audio, textual, and behavioral) and their impacts through MCP, this section analyzes the architectural vulnerabilities of AI agents themselves that make them susceptible to deepfake-based attacks in CPS environments.

\subsubsection{AI Agent Architecture and Attack Surfaces} 

Modern AI agents are constructed on four core components: \textit{perception}, \textit{brain} (LLM), \textit{action}, and \textit{memory} \cite{deng2025ai}. Each component creates distinct attack surfaces that deepfakes can exploit. Recent surveys identify four critical knowledge gaps in AI agent security that directly relate to deepfake vulnerabilities:

\begin{itemize}
    \item \textbf{Gap 1: Unpredictability of multi-step user inputs} -- Deepfake audio/text can masquerade as legitimate user inputs across multiple interaction rounds.
    \item \textbf{Gap 2: Complexity in internal executions} -- Behavioral deepfakes can mimic reasoning patterns to evade internal consistency checks.
    \item \textbf{Gap 3: Variability of operational environments} -- Visual deepfakes alter environmental perception, causing agents to misinterpret physical world states.
    \item \textbf{Gap 4: Interactions with untrusted external entities} -- All deepfake modalities can be injected through MCP tools and external data sources.
\end{itemize}

Threats on AI agent perception exploit model-level vulnerabilities to manipulate the ``brain'' of AI agents, compromising or bypassing policy constraints from benign instructions and leading to improper actions that break the integrity of the agent ecosystem \cite{deng2025ai}.

\subsubsection{Perception Layer Attacks via Multimodal Deepfakes} 

The perception layer serves as the primary entry point for AI agents and is particularly vulnerable to multimodal deepfakes. Research demonstrates that attackers can embed malicious prompts within images or audio to bypass security mechanisms when combined with text \cite{bagdasaryan2023abusing}. This vulnerability is especially dangerous when AI agents possess tool-calling capabilities.

Adversarial attacks on multimodal agents have shown that visual adversarial examples can cause LLM-based agents to misuse tools, resulting in unintended actions within CPS \cite{wu2024adversarial}. The multimodal nature of modern LLMs, which support voice, images, and text, increases flexibility but also expands the attack surface for deepfake injection \cite{fu2023misusing}.

Jailbreaking of AI agents can occur through three main vectors: multi-turn dialogues, multimodal inputs, and external environmental data \cite{deng2025ai}. In multi-turn interactions or role-playing scenarios where agents act as planners or experts, harmful outputs become harder to detect, making deepfake-based manipulation particularly effective \cite{wei2023jailbroken}.

\subsubsection{Multi-Agent Systems and Cascade Effects} 

In multi-agent systems (MAS), a single compromised agent can propagate malicious behavior throughout the entire network, a phenomenon termed the ``domino effect'' \cite{deng2025ai, chen2024infecting}. This cascade vulnerability is particularly concerning in CPS contexts where deepfake inputs to one agent can compromise system-wide integrity.

Research on prompt infection demonstrates that a single adversarial string can propagate among agents, starting with one harmful agent and ultimately compromising all agents in the collection \cite{chen2024infecting}. The self-propagating nature of such attacks means that initial deepfake inputs can achieve system-wide compromise through normal inter-agent communication channels.

Studies on the flooding spread of manipulated knowledge in LLM-based multi-agent communities reveal how misinformation, potentially seeded by deepfakes, can propagate through collaborative reasoning, negatively affecting collective decision-making \cite{ju2024flooding}. Multi-agent debates can improve robustness, but cooperation among agents may also cause a domino effect where one compromised agent jeopardizes others \cite{amayuelas2024multiagent}.

Defense frameworks such as BlindGuard \cite{blindguard2025} and G-Safeguard \cite{wang2025gsafeguard} employ graph-based approaches to detect malicious agents in MAS, evaluating defense capabilities against direct prompt attacks, tool attacks, and memory attacks. These frameworks represent emerging countermeasures against deepfake-initiated cascade failures.

\subsubsection{Embodied AI and Physical World Threats} 

Embodied AI systems, including robots and autonomous vehicles, face unique risks when deepfakes are employed to attack systems capable of causing physical harm. These systems encounter vulnerabilities stemming from both environmental and system-level factors, manifesting through sensor spoofing, adversarial attacks, and failures in task and motion planning \cite{embodiedai2025survey}.

The taxonomy of embodied AI vulnerabilities encompasses:
\begin{itemize}
    \item \textbf{Exogenous origins}: Physical attacks and cybersecurity threats including sensor spoofing and adversarial patches.
    \item \textbf{Endogenous origins}: Sensor failures and software flaws that can be exploited by deepfakes to cause cascading failures.
\end{itemize}

Research on jailbreaking robotic manipulation demonstrates that embodied AI jailbreaks transcend text generation to produce potential physical actions, thereby significantly amplifying security risks compared to purely linguistic attacks \cite{embodiedjailbreak2024}. A comprehensive set of 230 malicious physical world queries has been developed to probe embodied AI systems, grounded in IEEE Ethically Aligned Design guidelines.

The BALD (Backdoor Attacks against LLM-based Decision-making systems) framework represents the first comprehensive approach for backdoor attacks in embodied AI, proposing three distinct mechanisms: word injection, scenario manipulation, and knowledge injection \cite{bald2024}. Experiments on GPT-3.5, LLaMA2, and PaLM2 in autonomous driving and home robot tasks demonstrate high attack success rates, underscoring the vulnerability of embodied agents to deepfake-style manipulations.

\subsubsection{Tool Poisoning and MCP Security Mechanisms} 

The Model Context Protocol creates novel attack surfaces that extend beyond traditional deepfake threats. Security analyses reveal alarming statistics: 43\% of tested MCP server implementations contain command injection flaws, 22\% permit arbitrary file read via path traversal, and 30\% are vulnerable to Server-Side Request Forgery \cite{RedHatMCPSec, docker2025mcp}.

Tool Poisoning Attacks (TPAs) occur when malicious instructions are embedded within MCP tool descriptions that are invisible to users but visible to AI models \cite{invariant2025tpa}. These attacks exploit MCP's security model assumption that tool descriptions are trustworthy and benign. Key attack vectors include:

\begin{itemize}
    \item \textbf{Rug Pull attacks}: Tools dynamically alter their behavior or description after users grant permission, enabling silent credential theft or API key exfiltration.
    \item \textbf{Tool Shadowing}: In multi-server configurations, malicious servers impersonate tools from trusted servers.
    \item \textbf{Cross-server Interference}: Server A can redefine tools from Server B, enabling interception of sensitive operations.
\end{itemize}

Real-world vulnerabilities include CVE-2025-6514 affecting 437,000+ downloads of mcp-remote through OAuth discovery vulnerabilities, and CVE-2025-49596 in MCP Inspector enabling remote code execution via CSRF \cite{docker2025mcp}. Academic research identifies 5.5\% of MCP servers exhibiting tool poisoning behaviors, representing a new class of AI-targeted vulnerabilities. 

\subsubsection{Agent-to-Environment Threats in CPS} 

Threats on Agent2Environment exploit vulnerabilities arising from untrusted dynamic feedback and complex interactions in diverse operational settings \cite{deng2025ai}. These threats encompass indirect manipulation of input data, unintended behaviors influenced by dynamic states, and environmental discrepancies, all of which can be induced or amplified by deepfakes.

Input data from the physical environment must undergo rigorous security checks to filter threats and ensure safety. Clear and compatible communication between LLM-generated instructions and hardware execution is vital to avoid operational errors. For real-world deployment, agents must prioritize accuracy to minimize irreversible harm caused by incorrect actions \cite{deng2025ai}.

The physical environment poses significant security challenges due to its complexity. Insufficient isolation between agents in shared environments enables malicious agents to potentially access or interfere with operations of other agents, leading to data breaches, unauthorized access to sensitive information, or the spread of malicious code \cite{deng2025ai}. Unmonitored resource usage can mask security breaches, as anomalous behavior such as sudden spikes in resource consumption might indicate deepfake-initiated attacks.

\subsubsection{Defense Frameworks for AI Agents} 

Multi-layered defense approaches are essential for protecting AI agents against deepfake threats:

\begin{itemize}
    \item \textbf{Input Layer}: Input validation, provenance verification, and deepfake detection for visual/audio inputs.
    \item \textbf{Reasoning Layer}: Certified robustness, adversarial training, and behavioral pattern detection.
    \item \textbf{Action Layer}: Least-privilege enforcement, human-in-the-loop verification for tool calls.
    \item \textbf{Memory Layer}: Taint tracking, encryption, and context poisoning prevention.
\end{itemize}

Emerging defense technologies include CaMeL for mitigating prompt injection attacks \cite{willison2025camel}, AgentGuard for repurposing agentic orchestrators for safety evaluation \cite{chen2025agentguard}, and BlockAgents for Byzantine-robust multi-agent coordination via blockchain \cite{chen2024blockagents}. These frameworks address the unique challenges posed by deepfake-style attacks on AI agent architectures.

Integration of homomorphic encryption schemes and attribute-based forgery generative models can safeguard against privacy breaches during communication processes, though at additional computational and communication costs \cite{deng2025ai}. Supply chain threats, including buffer overflow, SQL injection, and cross-site scripting vulnerabilities in tools, require comprehensive security auditing of all MCP server dependencies.

\begin{table}[ht]
\centering
\caption{AI Agent Attack Surface by Component}
\label{tab:agent_attack_surface}
\begin{tabular}{|p{1.8cm}|p{2cm}|p{1.5cm}|p{1.4cm}|}
\hline
\textbf{Component} & \textbf{Deepfake Vector} & \textbf{Severity} & \textbf{Defense Maturity} \\
\hline
Perception & Visual/Audio injection & Critical & Medium \\
\hline
Brain (LLM) & Prompt injection & Critical & Low \\
\hline
Action & Tool poisoning & Critical & Low \\
\hline
Memory & Context poisoning & High & Low \\
\hline
\end{tabular}
\vspace{-10 pt}
\end{table}

\begin{table}[ht]
\centering
\caption{Multi-Agent Defense Frameworks Comparison}
\label{tab:mas_defense}
\begin{tabular}{|p{2cm}|p{2cm}|p{1.5cm}|p{1.4cm}|}
\hline
\textbf{Framework} & \textbf{Approach} & \textbf{Overhead} & \textbf{Effectiveness} \\
\hline
G-Safeguard & Graph topology & Medium & High \\
\hline
BlindGuard & Zero-shot detection & Low & Medium-High \\
\hline
BlockAgents & Blockchain & High & High \\
\hline
AgentGuard & Orchestrator reuse & Low & Medium \\
\hline
CaMeL & Capability control & Medium & High \\
\hline
\end{tabular}
\end{table}

\begin{table}[ht]
\centering
\caption{MCP Vulnerability Statistics}
\label{tab:mcp_vulnerabilities}
\begin{tabular}{|p{3cm}|p{1.5cm}|p{2.5cm}|}
\hline
\textbf{Vulnerability Type} & \textbf{Prevalence} & \textbf{CVE Examples} \\
\hline
Command Injection & 43\% & Multiple \\
\hline
Path Traversal & 22\% & Multiple \\
\hline
SSRF & 30\% & Multiple \\
\hline
Tool Poisoning & 5.5\% & N/A \\
\hline
OAuth Flaws & Significant & CVE-2025-6514 \\
\hline
CSRF & Variable & CVE-2025-49596 \\
\hline
\end{tabular}
\end{table}

\begin{table}[ht]
\centering
\caption{Embodied AI Attack Mechanisms}
\label{tab:embodied_attacks}
\begin{tabular}{|p{2cm}|p{2cm}|p{1.5cm}|p{1.4cm}|}
\hline
\textbf{Mechanism} & \textbf{Target System} & \textbf{Success Rate} & \textbf{Physical Risk} \\
\hline
Word Injection & LLM Decision & High & High \\
\hline
Scenario Manipulation & Context & High & Critical \\
\hline
Knowledge Injection & Memory & Medium & High \\
\hline
Sensor Spoofing & Perception & High & Critical \\
\hline
Adversarial Patches & Vision & Medium & High \\
\hline
\end{tabular}
\end{table}


\subsection{Cross-Category Comparison and Summary}

\begin{table*}[ht]
\centering
\caption{Comprehensive Comparison of Deepfake Threat Categories in CPS}
\label{tab:comprehensive_comparison}
\begin{tabular}{|p{1.8cm}|p{2cm}|p{2cm}|p{2cm}|p{2cm}|p{2.2cm}|}
\hline
\textbf{Category} & \textbf{Primary Target} & \textbf{Detection Maturity} & \textbf{CPS Impact} & \textbf{Defense Gap} & \textbf{Research Priority} \\
\hline
Visual & Surveillance, AV & Medium & Critical & Generalization & High \\
\hline
Audio & Voice Auth, Social Eng. & Medium & High & Real-time & High \\
\hline
Textual & MCP Tools, Agents & Low & Critical & Tool Vetting & Critical \\
\hline
Behavioral & Sensors, Anomaly Det. & Low & Critical & Physics-aware & Critical \\
\hline
Privacy & Identity, Consent & Low & High & Regulatory & High \\
\hline
AI Agents & All Components & Very Low & Critical & Comprehensive & Critical \\
\hline
\end{tabular}
\end{table*}

\begin{table*}[ht]
\centering
\caption{Defense Readiness Assessment Across Categories}
\label{tab:defense_readiness}
\begin{tabularx}{\textwidth}{|l|X|X|X|X|X|X|}
\hline
\textbf{Defense Type} & \textbf{Visual} & \textbf{Audio} & \textbf{Textual} & \textbf{Behavioral} & \textbf{Privacy} & \textbf{AI Agent} \\
\hline
Detection ML & Medium & Medium & Low & Low & Low & Very Low \\
\hline
Provenance & High & Low & Medium & Low & Medium & Low \\
\hline
Multi-modal & Medium & Medium & N/A & Medium & Low & Low \\
\hline
Human-in-loop & Medium & Medium & Medium & Low & High & Medium \\
\hline
Policy/Governance & Low & Low & Low & Low & Medium & Very Low \\
\hline
\end{tabularx}
\end{table*}

\textbf{Key Research Gaps Across All Categories:}
\begin{enumerate}
    \item \textbf{Generalization}: All detection methods suffer from poor generalization to unseen generators and attack variants.
    \item \textbf{Real-time Performance}: CPS applications require low-latency detection incompatible with current complex models.
    \item \textbf{Cross-modal Attacks}: Combined visual-audio-textual-behavioral attacks remain largely unstudied.
    \item \textbf{MCP Security}: Protocol-level security mechanisms are immature despite rapid adoption.
    \item \textbf{Regulatory Frameworks}: Legal and governance structures lag behind technical capabilities.
    \item \textbf{Scalability}: Defense mechanisms don't scale to large multi-agent CPS deployments.
\end{enumerate}

\textbf{Recommended Research Priorities:}
\begin{enumerate}
    \item Develop generator-agnostic detection leveraging fundamental artifacts
    \item Create standardized MCP security auditing and certification frameworks
    \item Build physics-informed behavioral deepfake detection for CPS
    \item Establish real-time multi-modal verification pipelines
    \item Design scalable Byzantine-robust multi-agent coordination
    \item Integrate privacy-preserving techniques with detection systems
\end{enumerate}

\subsection{Cross-Modal Research Synthesis}
\label{subsec:deepfake_synthesis}

This subsection consolidates the research analysis across all deepfake modalities, identifying common strengths, shared limitations, and unified research priorities for securing MCP-enabled CPS.

\subsubsection{Common Strengths Across Modalities}

Research on deepfake threats has achieved notable progress applicable to CPS security. Comprehensive attack taxonomies have been developed for each modality—visual (display spoofing, face-swap, scene synthesis), audio (voice cloning, replay attacks), textual (tool poisoning, prompt injection, rug pulls), and behavioral (sensor spoofing, anomaly mimicry, stealthy malware). Multi-modal defense strategies combining physiological signals (rPPG), provenance frameworks (C2PA), and cross-sensor fusion demonstrate improved robustness over single-modality approaches. Formal control-theoretic analysis enables rigorous characterization of stealthy attack impacts, while graph-based and spatio-temporal detectors capture higher-order behavioral context. The ``liar's dividend'' phenomenon has been recognized as a systemic epistemic threat, informing both technical and policy responses.

\subsubsection{Shared Limitations}

Despite these advances, four fundamental limitations constrain deployment across all modalities:

\begin{enumerate}[label=\alph*)]
    \item \textbf{Generalization gap}: Detectors trained on benchmark datasets exhibit significant performance degradation against unseen generators, novel attack variants, and in-the-wild conditions featuring compression, noise, and domain shift.
    
    \item \textbf{Real-time constraints}: Safety-critical CPS require detection within milliseconds; computationally intensive methods (rPPG analysis, graph-based detection, ensemble models) often exceed acceptable latency budgets.
    
    \item \textbf{Adversarial adaptation}: Generative models—including neural codec voice cloning (VALL-E), diffusion-based image synthesis, and LLM-powered text generation—consistently outpace detection capabilities.
    
    \item \textbf{Human fallibility}: Operators and reviewers misclassify high-quality deepfakes at non-trivial rates, preventing reliance on human-in-the-loop verification as a standalone defense.
\end{enumerate}

\subsubsection{Unified Research Priorities}

Addressing these shared limitations requires coordinated research efforts:

\begin{itemize}
    \item \textbf{Generator-agnostic detection}: Develop methods leveraging fundamental artifacts (physics-based inconsistencies, statistical fingerprints) rather than generator-specific signatures.
    
    \item \textbf{Lightweight architectures}: Design efficient models suitable for edge deployment that balance accuracy with CPS resource constraints.
    
    \item \textbf{Standardized MCP integration}: Establish protocols for provenance verification and detection at MCP tool boundaries without unacceptable latency.
    
    \item \textbf{Physics-grounded validation}: Integrate domain knowledge (hydraulic constraints, RF propagation models, control-theoretic bounds) into ML detection architectures.
    
    \item \textbf{Adaptive regulatory frameworks}: Develop cross-jurisdictional mechanisms for rapid response to deepfake harms while preserving privacy and due process.
\end{itemize}

\subsubsection{Comparative Summary}

Table~\ref{tab:deepfake_synthesis} summarizes detection maturity, key limitations, and CPS deployment feasibility across modalities.

\begin{table}[!t]
\centering
\caption{Cross-Modal Deepfake Threat Synthesis}
\label{tab:deepfake_synthesis}
\begin{tabular}{|l|c|c|c|c|}
\hline
\textbf{Modality} & \textbf{Maturity} & \textbf{General.} & \textbf{RT-Ready} & \textbf{Key Gap} \\
\hline
Visual & Medium & Low & Medium & Generator shift \\
\hline
Audio & Medium & Low-Med & Medium & Codec robustness \\
\hline
Textual & Low & Low & High & LLM evolution \\
\hline
Behavioral & Low & Medium & Low & Physics modeling \\
\hline
Privacy & Low & N/A & N/A & Legal frameworks \\
\hline
\end{tabular}
\end{table}

The analysis confirms that detection mechanisms alone cannot serve as decision authorities in safety-critical CPS. Effective deployment requires integration with provenance verification, physics-based validation, and defense-in-depth architectures as detailed in Section~\ref{sec:mitigation}.

\section{Detection Techniques}
\label{sec:detection}
In the context of deepfakes as a security threat to CPS that leverage the MCP, detection techniques must address the unique challenges of real-time environmental interactions and protocol-mediated data exchanges, where synthetic content can infiltrate agent-to-tool communications and sensor feeds. Traditional artifact analysis, such as identifying inconsistencies in pixel-level artifacts or spectral anomalies in audio streams, has been enhanced with advanced machine learning fusion methods to improve robustness in constrained CPS environments \cite{gupta2023comprehensive}. For instance, multimodal consistency checks, which combine visual, audio, and textual modalities, enable cross-verification of MCP-transmitted data and detect discrepancies in agent responses that mimic legitimate environmental signals but fail to meet physiological cues, such as eye-blink irregularities or heartbeat synchronization \cite{alrashoud2025deepfake}. Recent studies demonstrate that convolutional neural network (CNN)-based models, integrated with MCP's structured messaging, achieve up to 95\% accuracy in identifying deepfake intrusions in industrial control systems, where forged sensor data could trigger unsafe actuations \cite{balafrej2024enhancing}. These approaches emphasize lightweight architectures for edge devices in CPS, ensuring low-latency detection without compromising protocol interoperability.

Emerging detection strategies, further tailored to MCP's vulnerabilities, such as context poisoning via behavioral deepfakes in inter-agent communications, incorporate stylometry for textual analysis and physics-based anomaly detection for sensor emulation \cite{singh2025advancements}. Self-supervised learning frameworks have shown promise in generalizing to unseen deepfakes within MCP ecosystems, utilizing environmental fingerprints, such as ENF traces, to corroborate protocol payloads against physical realities in CPS applications \cite{tchaptchet2025deepfakes}. However, challenges persist in balancing detection efficacy with privacy preservation, as watermarking and cryptographic provenance require standardized implementation across MCP servers to prevent supply-chain exploits \cite{gupta2023comprehensive}. Evaluations on datasets like FaceForensics++ adapted for CPS scenarios reveal that hybrid AI algorithms, including generative adversarial network (GAN) discriminators, outperform single-modality methods by 15-20\% in real-world deployments, highlighting the need for adaptive training to counter evolving threats in protocol-enabled AI agents \cite{balafrej2024enhancing}.

\begin{figure*}[htbp!]
\centering
\includegraphics[scale=0.45]{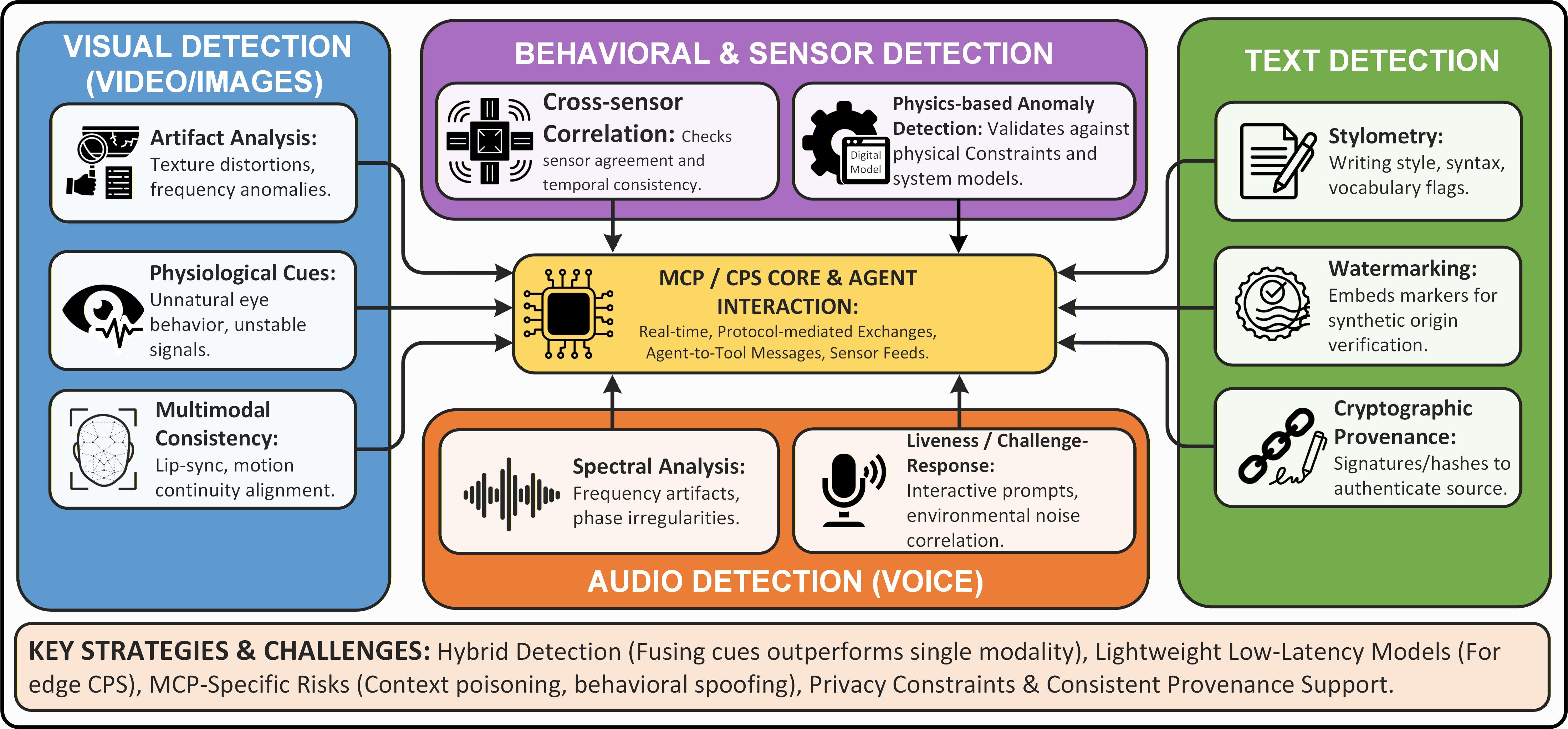} 
\caption{Deepfakes Detection in CPS Using MCP}
\label{fig:deepfakes-detection}
\vspace{-10 pt}
\end{figure*}

\subsection{Visual}
Visual detection addresses forged inputs in sensor streams through artifact analysis \cite{deressa2025genconvit}, physiological cues such as eye movements \cite{javed2024real}, and multimodal consistency checks across modalities \cite{gao2024temporal}.

\textbf{Artifact analysis} in visual deepfake detection focuses on identifying subtle inconsistencies introduced during the generation process, such as pixel-level distortions, blending boundaries, and frequency-domain anomalies, which are often imperceptible to humans but detectable by computational methods. In the context of CPS integrated with MCP, where AI agents process visual streams for real-time decision-making, these artifacts can compromise system integrity by deceiving surveillance or authentication modules. 
Recent advancements employ frequency-domain analysis improves cross-model generalization by classifying residues within a taxonomy framework \cite{say2025advancing}.
This method addresses the challenges posed by residual artifacts in static images, offering robust performance in CPS environments where visual inputs from MCP servers must be sanitized to prevent context poisoning.

Further developments in artifact analysis incorporate explainable AI techniques, such as Grad-CAM and Saliency Maps, to localize regions of manipulation and quantify activation intensities, providing insights into model confidence and precision. 
In MCP-enabled CPS, this approach mitigates risks from adversarial visual inputs by integrating texture decomposition, which separates artifacts from natural image elements, thereby improving generalization across diverse forgery scenarios, particularly on edge devices handling compressed or noisy data \cite{gao2024texture}.



\textbf{Physiological cues} exploit biological signals, such as heart rate variability, blinking eyes, and rPPG, that generative models cannot accurately replicate \cite{tian2025deepphysio, sun2025study}. Hybrid models fusing physiological and visual features reduce false positives, though blink-based detection faces robustness challenges across diverse populations \cite{soudy2024deepfake, sohail2025deepfake, alanazi2024exploring}.



\textbf{Multimodal consistency} evaluates alignment across visual, audio, and temporal modalities to detect discrepancies such as lip-sync mismatches \cite{javed2025enhancing, salvi2023robust}. Ensemble fusion and spatial-spectral analysis outperform single-modality methods on mixed-manipulation datasets \cite{sheng2025id, kumar2025advances, alrashoud2025deepfake, usmani2025spatio}.


\subsection{Audio}
Audio detection mitigates voice spoofing risks that compromise agent-to-tool interactions \cite{zhang2025audio}. ML frameworks analyze acoustic artifacts and temporal inconsistencies in real-time CPS environments \cite{almutairi2022review}, with adversarial training enhancing robustness \cite{sharma2025systematic} and hybrid models reducing false positives \cite{bohara2025detecting}. Multi-modal fusion cross-verifies audio with environmental signals, while forensic analysis detects traces in high-frequency spectra \cite{verma2025deepfake}. Lightweight detectors balance efficiency for edge deployment \cite{alshehri2024audio}, underscoring the need for standardized audio provenance protocols \cite{kilinc2023audio}.

\textbf{Spectral analysis} for audio deepfake detection in MCP involves examining frequency domain characteristics to uncover artifacts introduced by synthesis algorithms, such as unnatural energy distributions or phase inconsistencies that can infiltrate protocol-based communications in CPS \cite{zhang2025hierarchical}. This technique is particularly effective against voice cloning threats, where MCP servers process audio inputs that may be laced with synthetic content, enabling attacks such as context poisoning. Recent research employs advanced spectrogram-based models to classify deepfakes, leveraging convolutional neural networks for feature extraction and achieving superior performance on benchmark datasets \cite{calderon2024deep}. For example, chroma-based spectral methods have been refined to detect inconsistencies in harmonic structures, and malicious audio could mimic legitimate tool responses. Evaluations show that these approaches outperform traditional time-domain analysis in noisy CPS environments, with accuracy rates exceeding 95\% in controlled simulations \cite{gupta2024generative}.

Additionally, advancements in spectral analysis incorporate wavelet transforms for multi-resolution analysis, enhancing the detection of subtle manipulations in MCP-transmitted audio streams. Studies highlight the integration of spectral entropy measures to quantify irregularity, thereby mitigating risks posed by adversarial audio that exploits weak MCP boundaries \cite{gupta2023comprehensive}. Limitations, such as sensitivity to compression artifacts, are addressed through robust preprocessing, as demonstrated in comprehensive reviews advocating hybrid spectral-deep learning frameworks for CPS security \cite{mubarak2023survey}. These methods provide a foundational defense, emphasizing the importance of spectral fingerprinting to verify audio authenticity in real-time MCP interactions \cite{arif2021voice}.

\textbf{Liveness/challenge-response tests} in audio deepfake detection for MCP ensure the authenticity of voice inputs by verifying real-time human presence, countering pre-recorded or synthesized audio that could subvert CPS agent protocols. These active methods involve prompting users with random challenges, such as repeating phrases, to detect non-interactive deepfakes infiltrating MCP communications \cite{akhtar2024video}. Recent developments integrate biometric liveness detection with acoustic analysis, improving resilience against spoofing in CPS, where audio commands trigger physical actions \cite{park2024comprehensive}. For instance, challenge-response protocols have been enhanced with machine learning to adapt challenges dynamically, reducing evasion rates in MCP ecosystems prone to voice impersonation \cite{chen2022learning}. Studies report high effectiveness in distinguishing live from synthetic speech, particularly in multi-factor authentication scenarios for CPS \cite{piccialli2025agentai}.

Moreover, combining liveness tests with environmental noise correlation strengthens defenses against sophisticated deepfakes in MCP, where adversaries might inject cloned audio via untrusted servers. Research emphasizes hybrid systems that fuse challenge-response with passive liveness cues, such as breathing patterns, to achieve comprehensive coverage in CPS applications \cite{li2025survey}. Challenges include user convenience and adaptability to diverse accents, addressed through personalized models in recent surveys \cite{pei2024deepfake}. These techniques are vital for MCP security, promoting interactive verification to prevent deepfake exploitation in critical infrastructures \cite{wang2021survey}.

\subsection{Text}
Textual deepfake detection identifies forged tool descriptions and manipulated prompts that enable context poisoning. ML models analyze linguistic patterns in MCP-exchanged text \cite{mubarak2023survey}, with multimodal approaches addressing low-quality inputs \cite{pu2023deepfake}. Hybrid CNN-stylometric models distinguish AI-generated content \cite{rosca2025new}. Proactive strategies including watermark extraction and provenance verification mitigate supply-chain vulnerabilities, with robust algorithms maintaining low false-positive rates \cite{groh2024human}. These techniques provide forensic tools to authenticate text origins \cite{alrashoud2025deepfake}.

\textbf{Stylometry} in textual deepfake detection for MCP involves analyzing writing styles, such as vocabulary richness, sentence structure, and syntactic patterns, and the origins of text, including vocabulary richness, sentence structure, and syntactic patterns, to differentiate AI-generated from human text in protocol-mediated exchanges. This method is particularly effective against preference manipulation attacks, in which malicious tool descriptions bias LLM selection in CPS. Recent research integrates stylometric features with deep learning for improved authorship attribution, achieving high accuracy in detecting subtle inconsistencies in MCP-processed prompts \cite{ghiuruau2024distinguishing}. For example, ensemble models incorporating n-gram analysis and machine learning classifiers have been proposed to combat fake news, such as deepfakes, which are adaptable to MCP's inter-agent messaging \cite{harris2024fake}. Evaluations of dialectal texts reveal the robustness of stylometry to linguistic variation, an essential property for global CPS deployments \cite{boko2024disinformation}.

Advancements in stylometry also focus on explainable AI to localize manipulative elements in text, enhancing transparency in MCP security audits. Studies emphasize hybrid approaches combining stylometric metrics with blockchain to enhance provenance and reduce evasion risks from adaptive adversaries \cite{farhan2025self}. Limitations such as sensitivity to text length are mitigated through multitask learning frameworks, enabling real-time detection in resource-constrained CPS environments \cite{liu2025multi}. These techniques underscore the role of stylometry in fortifying MCP against textual deepfakes, promoting resilient agent behaviors \cite{opara2025evaluating}.

\textbf{AI watermarking} for textual deepfake detection in MCP embeds imperceptible markers into generated text during LLM inference, enabling reliable identification of synthetic content in tool outputs or retrieved data. This proactive defense counters rug-pull attacks by enabling post-generation verification without degrading text quality. Recent developments include probabilistic curvature-based watermarks that preserve semantic integrity while facilitating high-fidelity detection in CPS communications \cite{zhang2025curvemark}. For instance, scalable schemes like SynthID-Text integrate with open-source models, achieving minimal impact on utility and robust extraction even under paraphrasing \cite{dathathri2024scalable}. Surveys highlight watermarking's superiority over passive detectors in handling LLM variations, crucial for MCP's heterogeneous services \cite{yang2025watermarking}.

Moreover, AI watermarking advancements incorporate contrastive learning to enhance robustness against removal attacks, ensuring traceability in MCP ecosystems prone to supply-chain compromises. Research on fine-tuning datasets demonstrates that watermarks persist across edits, supporting forensic analysis in CPS \cite{luo2025digital}. Challenges such as detection thresholds are addressed with supervised and unsupervised methods, optimizing for low false positives \cite{brissett2025machine}. These innovations position watermarking as a cornerstone for securing textual exchanges in MCP, thereby fostering the ethical deployment of AI \cite{kumar2025artificial}.

\textbf{Cryptographic provenance} in textual deepfake detection for MCP establishes verifiable chains of origin for text data, utilizing digital signatures and blockchain to authenticate content traversing protocol boundaries. This mitigates identity fragmentation by binding metadata to MCP messages, preventing spoofed instructions in CPS. Recent frameworks leverage NFTs and decentralized ledgers to provide tamper-proof provenance, effectively combating the dissemination of deepfakes in agent networks \cite{taralkar2025nft}. For example, hybrid systems integrate cryptographic hashes with AI forensics, enabling real-time verification of tool responses \cite{iavich2025combating}. Studies on post-quantum cryptography address future threats, ensuring long-term security in evolving MCP implementations [30][31][32].

Additionally, cryptographic provenance techniques emphasize interoperability with content credentials standards, facilitating cross-platform detection of synthetic text. Research on blockchain-based traceability mechanisms has demonstrated reduced latency in CPS audits and counters misinformation through the use of immutable records \cite{hasan2024nfts}. Limitations such as scalability are addressed through lightweight protocols that balance computational overhead with efficacy \cite{wang2023blockchain}. These methods enhance MCP's resilience, providing a foundational layer for trustworthy AI interactions in critical infrastructures \cite{abiodun2022data}.

\subsection{Behavioral/Sensor}

Behavioral and sensor detection identifies emulated behaviors and spoofed sensor data, including anomaly mimicry and stealthy malware insertions \cite{zhang2025audio}. Unsupervised learning handles dynamic CPS environments with high real-time detection rates \cite{deng2025ai}, while hybrid frameworks blend network and physical data for anomaly identification \cite{rana2022deepfake}. AI-enhanced physical unclonable functions enable real-time anomaly detection \cite{momin2025explainable}, with digital twins addressing data scarcity \cite{alrashoud2025deepfake}.

\textbf{Cross-sensor correlation} for behavioral/sensor deepfake detection in MCP involves analyzing interdependencies among multiple sensor inputs to uncover inconsistencies caused by synthetic emulations, such as mismatched physical signals in CPS agent communications. This technique leverages data fusion to verify coherence across modalities, detecting spoofing attacks that exploit the interoperability of MCP's tools. Recent research has employed multimodal deep learning to enhance correlation analysis, achieving robust performance in CCTV and IoT environments where behavioral deepfakes pose significant privacy risks \cite{ha2025fl}. For example, anomaly-based frameworks using recurrent neural networks identify temporal mismatches, applicable to MCP's dynamic data exchanges \cite{xiong2025bmnet}.

Advancements in cross-sensor correlation also integrate explainable AI for transparent detection, aiding forensic analysis in MCP-secured CPS. Studies on federated learning models demonstrate improved generalization to unseen deepfakes by correlating sensor data with network metrics, reducing false positives in resource-constrained settings \cite{zhou2024fine}. Limitations such as environmental noise are addressed through hybrid unsupervised approaches, thereby fostering reliable verification in multi-agent systems \cite{lewis2020deepfake}. These methods strengthen MCP against behavioral manipulations, ensuring secure cyber-physical integrations.

\textbf{Physics-based anomaly detection} in MCP for behavioral/sensor deepfakes utilizes fundamental physical laws to model expected system behaviors, flagging deviations from synthetic alterations in CPS sensor feeds or agent actions. This approach incorporates domain knowledge to validate MCP-transmitted data against physical constraints, countering stealthy emulations. Recent surveys classify physics-based methods as key for CPS security, with deep learning enhancements improving detection in industrial control systems \cite{jeffrey2023review}. For instance, digital twin-based models simulate physical processes for anomaly identification, effective against MCP supply-chain threats \cite{xu2021digital}.

Furthermore, physics-based techniques in MCP emphasize hybrid models combining ML with physical simulations for enhanced accuracy, addressing complex attacks like sensor spoofing. Research on one-class learning frameworks shows superior performance on unbalanced CPS datasets, mitigating the impact of behavioral deepfakes \cite{singh2025anomaly}. Challenges such as computational overhead are mitigated through lightweight implementations, which support real-time MCP applications \cite{nagarajan2022iadf}. These innovations bolster CPS resilience, providing a solid foundation for securing AI agents under MCP protocols.



\subsection{Cross-Modal Research Synthesis}
\label{subsec:detection_synthesis}

This subsection consolidates the research analysis across all detection modalities, identifying common strengths, shared limitations, and unified research priorities for securing MCP-enabled CPS.

\subsubsection{Common Strengths Across Modalities}

Detection research has achieved notable progress applicable to CPS security. Deep learning architectures, including CNNs, Vision Transformers, and self-supervised audio models such as AASIST and wav2vec, establish strong baseline performance on benchmark datasets across visual, audio, and textual modalities \cite{deressa2025genconvit, liu2023asvspoof, opara2025evaluating}. Explainable AI techniques enable forensic localization of manipulated regions, improving both detection utility and operator trust \cite{momin2025explainable}. Physics-grounded verification methods, including rPPG for visual content, ENF traces for audio, and cross-sensor correlation for behavioral data, provide authentication mechanisms that synthetic content cannot easily replicate \cite{tian2025deepphysio, hatami2025electric, lewis2020deepfake}. Multi-modal consistency checks that fuse features across modalities demonstrate improved robustness compared to single-modality approaches \cite{salvi2023robust, javed2025enhancing}. Proactive defenses such as AI watermarking enable tracing of synthetic content with minimal quality degradation \cite{dathathri2024scalable}.

\subsubsection{Shared Limitations}

Despite these advances, four fundamental limitations constrain deployment across all modalities:

\begin{enumerate}[label=\alph*)]
    \item \textbf{Generalization gap}: Detectors trained on benchmark datasets exhibit significant performance degradation against unseen generators, novel attack variants, and in-the-wild conditions featuring compression, noise, and domain shift \cite{khan2025survey, croitoru2024deepfake}.
    
    \item \textbf{Adversarial adaptation}: The rapid evolution of generative models, including Neural Codec Models and LLMs, consistently outpaces detection capabilities, rendering trained detectors obsolete within months \cite{chen2025neural, yang2025watermarking}.
    
    \item \textbf{Computational constraints}: Real-time detection on resource-constrained edge devices remains challenging; heavyweight models suitable for high accuracy are incompatible with CPS latency and power requirements \cite{alshehri2024audio, momin2025explainable}.
    
    \item \textbf{Robustness to post-processing}: Compression codecs, paraphrasing attacks, and channel variations significantly degrade detection accuracy across visual, audio, and textual modalities \cite{liu2023asvspoof, yang2025watermarking}.
\end{enumerate}

\subsubsection{Unified Research Priorities}

Addressing these shared limitations requires coordinated research efforts:

\begin{itemize}
    \item \textbf{Generator-agnostic detection}: Develop methods leveraging fundamental artifacts (e.g., physics-based inconsistencies, statistical fingerprints) rather than generator-specific signatures.
    
    \item \textbf{Lightweight edge architectures}: Design efficient models (e.g., MobileNet variants, quantized transformers) that balance accuracy with CPS resource constraints.
    
    \item \textbf{Standardized MCP integration}: Establish protocols for embedding detection and provenance verification into MCP tool pipelines without introducing unacceptable latency.
    
    \item \textbf{Cross-modal fusion}: Exploit redundancy across visual, audio, textual, and behavioral channels to detect coordinated multi-modal attacks.
    
    \item \textbf{Continuous adaptation}: Implement online learning and federated update mechanisms that enable detectors to evolve alongside generative model advances.
\end{itemize}

\subsubsection{Comparative Summary}

Table~\ref{tab:detection_synthesis} summarizes detection capabilities and gaps across modalities.

\begin{table}[!t]
\centering
\caption{Cross-Modal Detection Synthesis}
\label{tab:detection_synthesis}
\begin{tabular}{|l|c|c|c|c|}
\hline
\textbf{Modality} & \textbf{Maturity} & \textbf{General.} & \textbf{RT Edge} & \textbf{Key Gap} \\
\hline
Visual & Medium & Low & Medium & Generator shift \\
\hline
Audio & Medium & Low-Med & Medium & Codec robustness \\
\hline
Textual & Low & Low & High & LLM evolution \\
\hline
Behavioral & Low & Medium & Low & Physics modeling \\
\hline
\end{tabular}
\end{table}

The analysis confirms that detection mechanisms, while valuable for threat identification, cannot serve as standalone decision authorities in safety-critical CPS. Effective deployment requires integration with provenance verification, physics-based validation, and defense-in-depth architectures as detailed in Section~\ref{sec:mitigation}.

While detection techniques provide valuable signals for identifying potential deepfake activity, SENTINEL Phases 3 and 4 clarify that detection alone cannot serve as a decision authority in safety-critical CPS. Given adversarial adaptation and operational constraints, detection mechanisms should be treated as supporting components whose outputs must be corroborated by provenance, physical consistency, or system-level validation before influencing control or actuation decisions.



\section{Mitigation and Defense Strategies}
\label{sec:mitigation}
The transition from static, single-turn Large Language Model (LLM) interfaces to autonomous agentic systems represents a fundamental architectural shift in the cyber-physical landscape of 2025. In this paradigm, autonomous agents are no longer confined to experimental sandboxes but are actively orchestrating workflows, managing sensitive enterprise data, and executing control commands \cite{obsidian2025aiagent}. As identified by global surveys, 67\% of organizations have deployed agentic AI in 2025 \cite{nctr2025cybersecurity}. However, this proliferation has expanded the attacks across three critical interaction surfaces: the User-Agent (U-A) interface, the Agent-Agent communication layer, and the Agent-Environment (A-E) interaction plane.

At the U-A surface, threats emerge from the blurring of boundaries between instruction and data. Prompt injection and multi-turn jailbreaking leverage the model’s semantic sensitivity to manipulate internal reasoning, leading to confused deputy scenarios where agents with elevated privileges perform unauthorized actions \cite{obsidian2025aiagent}. The A-A surface introduces risks of identity spoofing and secret collusion, where compromised agents may coordinate to fabricate consensus or bias collective decisions \cite{lee2025towards}. Finally, the A-E surface is vulnerable to indirect prompt injection via poisoned documents or websites \cite{johnson2025dangers}, as well as protocol-specific attacks such as tool masquerading and context poisoning within the Model Context Protocol (MCP) ecosystem \cite{gaire2025systematization}.

The complexity of these interactions necessitates a shift from traditional perimeter-based security to a Defense-in-Depth strategy tailored for non-deterministic, autonomous workloads. Traditional controls are inadequate for governing agents that dynamically request permissions via emerging protocols like MCP and Agent-to-Agent (A2A) \cite{obsidian2025aiagent}. This section details a comprehensive mitigation strategy focused on provenance, contextual validation, and proactive architectural defenses. This section applies Phases 3 and 4 of the SENTINEL framework. Phase 3 evaluates candidate mechanisms against the established multi-dimensional fitness function and Phase 4 arranges selected mechanisms into a layered defense architecture.

Defenses in the MCP ecosystem must specifically address protocol-level integrity. Provenance mechanisms extend to MCP-specific signatures and cryptographic audit trails to ensure the accountability of multi-modal agents and prevent tool spoofing \cite{kumar2025secure, errico2025securing}. To counter deepfake injection and ensure content authenticity, proactive defenses employ semi-fragile watermarking strategies, such as FractalForensics, which localize manipulation within agent inputs \cite{wang2025fractalforensics, nguyenle2025survey}. Additionally, robust training strategies incorporate physical-layer environmental fingerprints, such as Electric Network Frequency (ENF), to anchor virtual interactions to real-world grid dynamics; by utilizing digital twin testbeds to simulate adverse scenarios, these models are fine-tuned to detect deepfake anomalies that lack consistent physical synchronization \cite{hatami2025electric}. Furthermore, data protection strategies utilize unlearnable perturbations to prevent unauthorized model training on sensitive CPS data \cite{zhao2023unlearnable}. These technical controls are bolstered by human-centric measures, including user training on agentic threats and regulations mandating immutable provenance for election and enterprise security \cite{stockwell2025elections}.

\begin{figure*}[htbp!]
\centering
\includegraphics[scale=0.35]{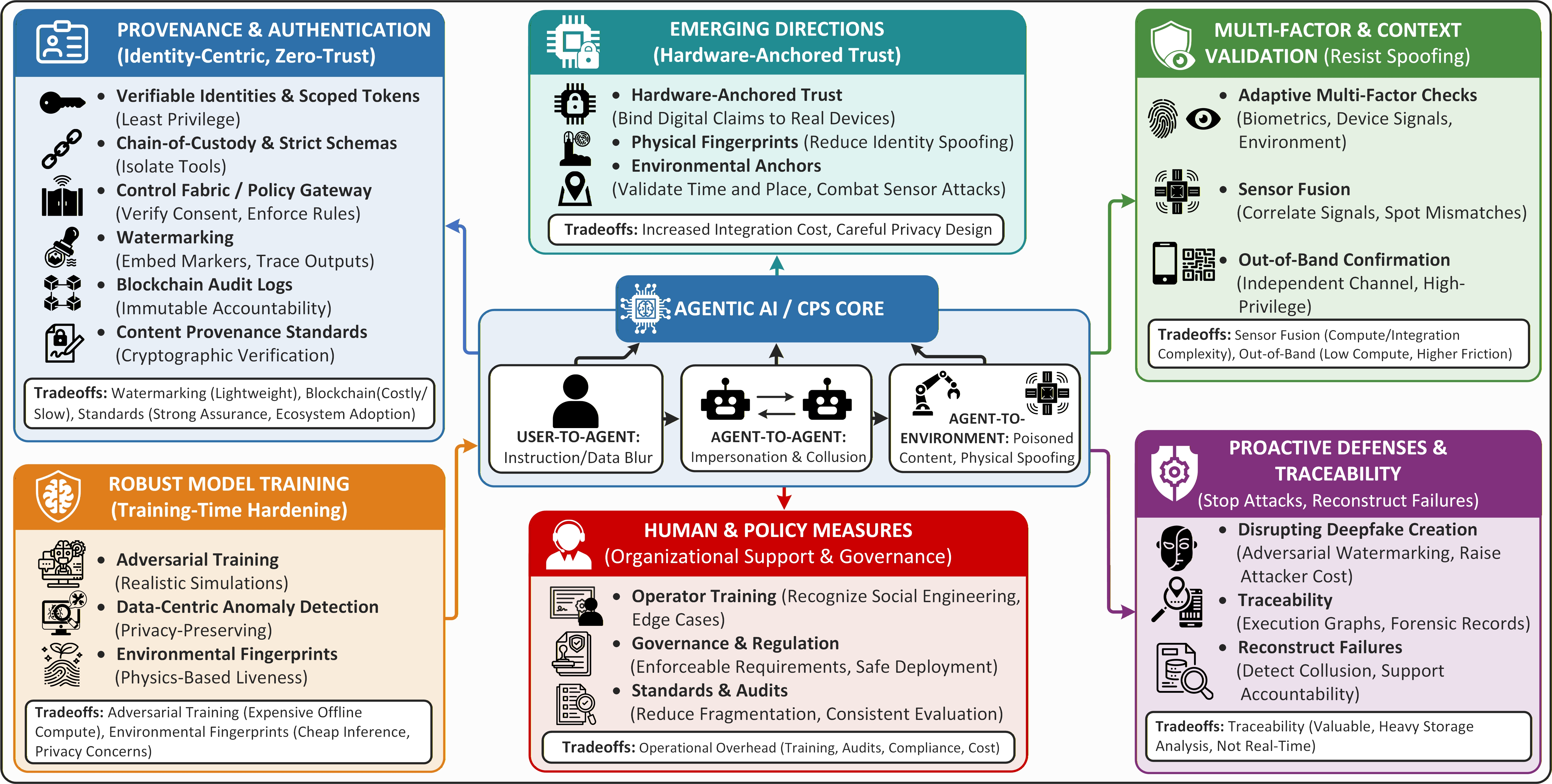}
\caption{Agentic AI In CPS: Mitigation \& Defense (Defense-In-Depth)}
\label{fig:mitigation and defense}
\vspace{-10 pt}
\end{figure*}

\subsection{Provenance \& Authentication}
The security of LLM-based agents utilizing the MCP and A2A frameworks have shifted from static perimeter defense to dynamic, identity-centric verification. The convergence of research and industry standards indicates that Zero-Trust authentication and cryptographic provenance are the primary mitigation strategies against emerging threats such as context poisoning, agent impersonation, and unverified tool execution \cite{grimes2025bridging, sotiropoulos2025owasp}. Traditional API keys are deemed insufficient for autonomous systems. The industry is moving toward Agent Identity, a paradigm where agents possess unique, verifiable identities like Decentralized Identifiers (DIDs) anchored to human principals or legal entities \cite{lin2025binding}. This human-to-agent binding ensures that every autonomous action carries a legal and technical chain of accountability, mitigating the risk of rogue agents denying their actions \cite{metcalfe-pearce2025predictions}. 

The mechanics of how agents communicate require rigorous safeguards. For the MCP, best practices mandate defense-in-depth architectures where the MCP server is isolated and inputs are strictly validated against schemas \cite{monteiro2025safety}. In multi-agent scenarios, threat modeling frameworks like MAESTRO emphasize strict input validation and authentication \cite{huang2025threat}, while Zero-Trust principles are applied to inter-agent communication; every handshake must be mutually authenticated (mTLS) and scrutinized for lateral movement risks \cite{liu2025secure}. The emerging Control Fabric middleware acts as a governance layer, intercepting agent requests to verify consent and enforce policy before any execution occurs \cite{singh2025agentic}, effectively neutralizing the risks identified for innovative agentic applications \cite{sotiropoulos2025owasp}.


Provenance mechanisms trace data origins to counteract deepfake-induced manipulations, often within immersive environments like the metaverse. These systems leverage cryptographic metadata, blockchain for immutable audit trails, and multi-factor authentication using biometric signals \cite{awadallah2024artificial}. To combat the black box nature of agent reasoning, provenance graphs and forensic retrieval engines can be deployed to map the lineage of data from training to inference \cite{mukherjeellm, raza2025trism}. Explainable AI (XAI) enhances transparency in these verification checks \cite{capuano2022explainable}, while zero-trust frameworks utilizing blockchain and physical unclonable functions reduce risks in the hardware supply chain \cite{kulkarni2024zero}. Extensions address privacy concerns via selective disclosure mechanisms \cite{bernabe2019privacy} and federated learning architectures that protect sensitive data during model training \cite{syllaidopoulos2025comprehensive}. Cognitive security approaches integrate behavioral analysis to identify sophisticated influence operations \cite{casino2025unveiling}, with recent blockchain architectures achieving robust guarantees in data integrity and non-repudiation \cite{onami2025blockchain}. Studies also highlight the incorporation of Electric Network Frequency (ENF) as a physical anchor for authentication, providing a defense against spoofing in multimedia streams \cite{ngharamike2023enf}.

\subsubsection{Digital Watermarking}
Digital watermarking acts as a critical security layer by embedding imperceptible, algorithmic signatures into the multimodal data such as retrieval documents, images, and system prompts that MCP servers transmit to AI agents \cite{liang2024watermarking}. By employing techniques like In-Context Watermarking, organizations can inject unique identifiers directly into the prompt context without requiring access to the model's internal weights, ensuring that the generated output retains a verifiable trace of its source \cite{liu2025context}. This approach effectively secures Retrieval-Augmented Generation (RAG) workflows by acting as a knowledge watermark, allowing system administrators to detect if proprietary knowledge bases have been extracted or utilized without authorization \cite{lv2025rag}. Furthermore, some frameworks extend these protections to image-based context, ensuring that copyright claims can be algorithmically verified even after the content has been processed and regenerated by a multimodal model \cite{chen2025safeguarding}. Beyond ownership, these cryptographic signatures establish a chain of trust, enabling systems to cryptographically distinguish authorized AI-generated content from external or potentially malicious human inputs \cite{christ2024undetectable, kirchenbauer2023watermark}.

\subsubsection{Blockchain}
Blockchain technology secures the MCP based AI agents in the CPS ecosystem by establishing a decentralized, immutable trust layer that ensures data integrity, verifiable identity, and accountability. While standard MCP architectures rely on traditional security mechanisms like OAuth2 and TLS for scalability \cite{patil2025model}, high-assurance CPS environments require enhanced protection against data poisoning. By integrating cryptographic identity frameworks and smart contracts, the architecture enforces strict access control, ensuring that only verified agents can invoke specific MCP tools or modify CPS states \cite{chhetri2025model, romandini2025sok}. Furthermore, the novel concept of Model Context Contracts leverages blockchain to allow LLMs to interact deterministically with smart contracts, automating governance and creating an auditable trail of agent decisions \cite{bandara2025model, karim2025ai}. This immutable history complements architectural defenses against prompt injection and unauthorized resource access identified in recent MCP security audits \cite{salih2025addressing}, providing a forensic layer that traditional firewalls cannot offer.

\subsubsection{C2PA standards}
The C2PA standard utilizes cryptographic manifests to verify media payloads and edit histories, providing assurances of authenticity for AI-generated content \cite{balan2025framework} while maintaining utility in multi-modal environments such as audio and video \cite{akhtar2024video}. To enhance resilience, recent frameworks integrate blockchain for immutable provenance \cite{ieee2024blockchain} and employ post-quantum cryptographic tools to secure verification against future threats \cite{iavich2025combating}. Privacy is further strengthened through policy enforcement mechanisms like Self-Sovereign Identity, which allow creators to prove authorship without exposing sensitive data \cite{farhan2025self}. While these technical standards support global standardization efforts \cite{bakirov2025theoretical} and aid human discernment of fabricated media \cite{groh2024human}, Chan et al. \cite{chan2025infrastructure} argue that for autonomous agents, technical provenance must be paired with Identity Binding infrastructure. This links agent actions to legal identities to establish accountability, acknowledging that while infrastructure can govern interactions, specific defenses against adversarial attacks like prompt injection remain an open challenge.

\subsubsection{Tradeoffs in Provenance \& Authentication}
Applying the Phase 3 Selection framework reveals distinct tradeoffs between performance and security depth. Digital Watermarking offers excellent computational performance and low OPEX, making it ideal for edge-compatible verification \cite{mubarak2023survey}. However, it faces integration challenges regarding black-box prompt injection risks \cite{liu2025context}. Blockchain provides the highest integrity and non-repudiation but incurs significant latency and storage costs, limiting its use to high-assurance logging rather than real-time filtering \cite{karim2025ai}. C2PA Standards balance high cryptographic certainty with minimal verification overhead, though they require significant effort to standardize headers and sanitize metadata for privacy \cite{bakirov2025theoretical, farhan2025self}.



\begin{table}[ht]
\centering
\caption{Phase 3 Selection: Provenance \& Authentication Constraints}
\label{tab:ph3_provenance}
\scriptsize
\begin{tabular}{|p{1.25cm}|p{1.75cm}|p{1.75cm}|p{1.75cm}|}
\hline
\textbf{Constraint} & \textbf{Digital Watermarking} & \textbf{Blockchain} & \textbf{C2PA Standards} \\
\hline
\textbf{Detection} & High robustness to paraphrasing \cite{lv2025rag} . & Immutable audit trail; high integrity \cite{karim2025ai}. & Cryptographic certainty \cite{farhan2025self} \\ 
\hline
\textbf{Computational} & Zero model training overhead \cite{liu2025context}. & High latency (consensus dependent) \cite{romandini2025sok,karim2025ai}. & High generation but low verification cost \cite{iavich2025combating} \\ 
\hline
\textbf{Integration} & Low; black-box prompt injection \cite{liu2025context}. & Complex; Smart Contract deployment \cite{bandara2025model}. & High \cite{balan2025framework}. \\ 
\hline
\textbf{Operational} & Low maintenance; algorithmic generation \cite{kirchenbauer2023watermark} & High storage burden for history \cite{karim2025ai}. & High \cite{chan2025infrastructure}. \\
\hline
\textbf{Privacy} & Low; imperceptible markers \cite{christ2024undetectable}. & Medium; requires hashing layers \cite{romandini2025sok}. & Low \cite{iavich2025combating} \\ 
\hline
\textbf{Cost} & Low OPEX \cite{chen2025safeguarding}. & High (Gas/Compute) \cite{karim2025ai}. & Medium \cite{farhan2025self} \\ 
\hline
\end{tabular}
\end{table}



\subsubsection{Defense Architecture Design Choices}
In the Phase 4 Architecture, these technologies are stratified to maximize defense-in-depth. C2PA and Watermarking serve as the Perimeter Tier, rejecting unauthenticated inputs before they reach agent logic \cite{daxa2025mcp}. The Detection Tier analyzes watermarks for tampering artifacts. Blockchain is reserved for the Adaptation and Response Tiers, providing an immutable forensic log for incident investigation. This ensures a verifiable trace remains for future model adaptation even if the perimeter is breached \cite{salih2025addressing}.



\begin{table}[ht]
\centering
\caption{Phase 4 Architecture: Provenance \& Authentication Tiers}
\label{tab:ph4_provenance}
\scriptsize
\begin{tabular}{|p{1.25cm}|p{1.75cm}|p{1.75cm}|p{1.75cm}|}
\hline
\textbf{Defense Tier} & \textbf{Digital Watermarking} & \textbf{Blockchain} & \textbf{C2PA Standards} \\
\hline
\textbf{Perimeter} & Block inputs lacking valid watermarks \cite{chen2025safeguarding}. & None. & Reject payloads with broken signature chains \cite{ieee2024blockchain}. \\ 
\hline
\textbf{Detection} & Analyze watermarks for tampering/stripping \cite{lv2025rag}. & Verify transaction hash against ledger \cite{karim2025ai}. & Validate provenance history \cite{farhan2025self}. \\  
\hline
\textbf{Response} & Automated content rejection  \cite{kirchenbauer2023watermark}` & Log incident data immutably for forensics \cite{romandini2025sok}. & Revoke keys of compromised publishers \cite{chan2025infrastructure}. \\ 
\hline
\textbf{Adaptation} & Update embedding algos against removal attacks \cite{liang2024watermarking}. & Audit smart contracts for logic gaps \cite{romandini2025sok}. & Update decentralized reputation scores and trust lists \cite{farhan2025self}\\ 
\hline
\end{tabular}
\end{table}

\subsection{Multi-factor \& Context Validation}
Multi-factor mechanisms integrate environmental awareness and biometric fusion to counter deepfake threats and impersonation \cite{zeeshan2025continuous}. AI-driven adaptive Multi-factor authentication models assess risk in real-time \cite{baseri2025privacy}, with cross-modal sensor checks achieving 96.3\% verification accuracy \cite{zeeshan2025continuous}. To reinforce security, event processing frameworks analyze system behavior for anomaly detection \cite{vegh2018cyber}. Privacy-preserving techniques include federated learning \cite{baseri2025privacy} and layered geolocation verification \cite{maranco2024intense}. Furthermore, temporal context factors enhance spoofing detection by approximately 30\% \cite{ali2025multilingual}, helping systems dynamically adjust encryption to address adaptive adversaries \cite{sasikumar2025enhancing}.

\subsubsection{Sensor Fusion}
Sensor fusion aggregates diverse data sources to create unified contexts, enabling the detection of inconsistencies such as mismatched Electric Network Frequency (ENF) patterns \cite{hatami2025electric}. To process these complex inputs, AI-enhanced architectures employ feature fusion strategies; for example, hybrid CNN-LSTM models effectively capture spatio-temporal anomalies \cite{tipper2024investigation}, while Multi-Graph Attention Networks integrate global and local features to detect forged traces \cite{chen2025deepfake}. In Cyber-Physical Systems (CPS), integrating environmental fingerprints like ENF has been shown to reduce false positives by over 15\% \cite{hatami2025electric}. Furthermore, proactive fusion disrupts threats via physical anchors \cite{hatami2025anchor}, while edge computing optimizations minimize inference latency for real-time verification \cite{kinnas2025reducing}. Finally, adversarial training strengthens these fusion models by enhancing generalization against unseen attacks \cite{selvaraj2025deepfake}.

\subsubsection{Out-of-band confirmation}
In Cyber-Physical Systems (CPS), the perception layer of an agent is often susceptible to sensor out-of-band (OOB) vulnerabilities. As systematized by Xiao et al., these vulnerabilities occur when signals from a different physical modality, such as electromagnetic interference, ultrasound or lasers, induce malicious measurements via out-of-range or cross-field energy conversion pathways \cite{xiao2025sok}. These physical layer exploits allow an attacker to deceive an agent about its environment, for example, by spoofing a clear path for an autonomous robot or falsifying thermal readings in industrial equipment.

To mitigate these physical spoofing risks, security architectures are increasingly adopting communication out-of-band confirmations. These mechanisms leverage cryptographic visual channels, such as dynamic QR codes, to authenticate high-risk actions independent of the compromised primary network \cite{de2024secure}. While traditional biometrics previously offered high effectiveness for user binding, recent surveys warn that Generative AI is actively eroding these zero-trust assumptions by synthesizing identities that bypass standard liveness checks \cite{xu2025erosion}. Consequently, modern defenses must integrate context-aware opportunistic sensing \cite{saideh2024opportunistic} and hardware-anchored Physical Unclonable Functions (PUFs) \cite{alhamarneh2024strengthening} to validate the physical integrity of the device itself. Furthermore, in immersive interfaces such as Extended Reality (XR), privacy-preserving architectures are being developed to bolster resilience by decoupling authentication verification from sensitive biometric user data \cite{hallal2024recent}.

\subsubsection{Tradeoffs in Multi-factor \& Context Validation}
Under Phase 3, Sensor Fusion is selected for high-criticality zones despite high Computational and Cost loads because of its superior Detection Effectiveness against environmental spoofing \cite{tipper2024investigation}.Out-of-Band Confirmation is prioritized as a fallback due to its low footprint, although its high Operational Burden due to user friction limits its use to high-privilege actions \cite{xu2025erosion}.

\begin{table}[ht]
\centering
\caption{Phase 3 Selection: Multi-factor \& Context Validation Constraints}
\label{tab:ph3_mfa}
\scriptsize
\begin{tabular}{|p{1.25cm}|p{2.5cm}|p{2.5cm}|}
\hline
\textbf{Constraint} & \textbf{Sensor Fusion} & \textbf{Out-of-Band Confirmation}\\
\hline
\textbf{Detection} & High accuracy \cite{chen2025deepfake,hatami2025electric}. & Deterministic verification \cite{de2024secure}. \\
\hline
\textbf{Computational} & High GPU/CPU load for real-time fusion \cite{tipper2024investigation}. & Low \cite{saideh2024opportunistic}.  \\
\hline
\textbf{Integration} & Complex; requires tight ENF/grid sync \cite{hatami2025electric}. & High; requires hardware modification \cite{alhamarneh2024strengthening}. \\
\hline
\textbf{Operational} & Moderate; sensor calibration needed \cite{xiao2025sok}. & High friction; user intervention \cite{xu2025erosion}. \\
\hline
\textbf{Privacy} & High; aggregates raw env. data \cite{zeeshan2025continuous}. & Medium; selective disclosure capable \cite{hallal2024recent}. \\
\hline
\textbf{Cost} & High (Hardware) \cite{xiao2025sok}. & Low \cite{saideh2024opportunistic}. \\
\hline
\end{tabular}
\end{table}

\subsubsection{Defense Architecture Design Choices}
In the Phase 4 Architecture, Sensor Fusion operates continuously in the Detection Tier, flagging inconsistencies like mismatched timestamps \cite{hatami2025electric}. Out-of-Band Confirmation serves as a circuit breaker in the Response Tier, triggered only when detection thresholds are crossed \cite{xu2025erosion}. This optimizes the system by using automated computation for monitoring while reserving high-latency human verification for escalation.

\begin{table}[ht]
\centering
\caption{Phase 4 Architecture: Multi-factor \& Context Validation Tiers}
\label{tab:ph4_mfa}
\scriptsize
\begin{tabular}{|p{1.25cm}|p{2.5cm}|p{2.5cm}|}
\hline
\textbf{Defense Tier} & \textbf{Sensor Fusion} & \textbf{Out-of-Band Confirmation} \\
\hline
\textbf{Perimeter} & None. & None.  \\
\hline
\textbf{Detection} & Detect inconsistencies (e.g., Audio vs. Video ENF) \cite{hatami2025electric}. & Verify high-privilege commands via  Visual/QR Channel \cite{de2024secure}.  \\
\hline
\textbf{Response} & Trigger failsafe if sensor mismatch exceeds threshold \cite{maranco2024intense}. & Block execution pending user approval \cite{xu2025erosion}. \\
\hline
\textbf{Adaptation} & Retrain fusion models on new spoofing patterns \cite{selvaraj2025deepfake}. & Adjust confirmation frequency based on threat level \cite{baseri2025privacy}. \\
\hline
\end{tabular}
\end{table}

\subsection{Robust Model Training}
Robust training enhances resilience in Cyber-Physical Systems (CPS) through decentralized approaches. For instance, \cite{addo2025federated} proposes a Federated Quantum Machine Learning (FQML) framework that utilizes multi-agent collaborative strategies and trimmed-mean aggregation to specifically mitigate data poisoning and Byzantine attacks. Anomaly detection strategies focus on data-centric methods \cite{odyurt2022improving} and discrete event dynamics \cite{hsieh2021robustness}. To support these strategies, techniques include specific designed Machine Learning and Deep Learning  models for Cyber-Physical systems (CPS) security constraints \cite{hasan2024review}. Furthermore, integrating cryptographic enhancements ensures privacy-preserving computation \cite{quan2025federated}, while hybrid approaches employing robust reasoning mechanisms allow autonomous systems to make valid decisions even with incomplete or contradictory data \cite{haakansson2021robust}. These methodologies optimize training to mitigate emerging threats, including the misuse of Generative AI and adversarial deepfakes, which require advanced detection frameworks to ensure information integrity \cite{syllaidopoulos2025comprehensive, zhang2025selective}.

\subsubsection{Adversarial Training}
Adversarial training enhances defense robustness by exposing AI agents to simulated deepfake perturbations during the learning process \cite{lad2024adversarial}. This approach is increasingly integrated into diagnostic behavior analysis to prevent data intrusions in CPS \cite{selvarajan2025diagnostic}. Emerging developments extend these concepts to next-generation environments, utilizing adversarial testing and reinforcement learning within AI frameworks for 7G-enabled virtual platforms \cite{jayanthiladevi2025ai}. Furthermore, recent literature categorizes these disruptive attacks \cite{alobaid2025disruptive} and highlights the role of adversarial learning in industrial data-driven innovations \cite{sangaiah2024guest}. Comprehensive reviews emphasize that deploying these adversarial defense strategies is a critical component of the cyber kill chain for mitigating AI-driven threats \cite{kazimierczak2024impact}.

\subsubsection{Environmental fingerprints}
To enhance model robustness against AI-generated deepfakes, researchers are increasingly integrating physical invariants, such as the Electric Network Frequency (ENF), into training data. Unlike purely semantic features, ENF provides a stochastic environmental fingerprint that is difficult for generative models to replicate accurately. Korgialas et al. introduced robust estimation frameworks using LAD regression to ground audio data in physical grid signals \cite{korgialas2024leveraging}. To ensure detection models remain effective even when input data is degraded, architectures like Multi-HCNet have been developed to capture high-order harmonics \cite{li2025detection}. Furthermore, integrating novel signatures like Electrical Network Voltage (ENV) allows models to learn precise location constraints, improving resilience against spoofing in IoT and Metaverse environments \cite{zhang2025feasibility, hatami2025electric}, provided that signal-to-noise ratios remain sufficient for extraction \cite{hua2024factors}.

\subsubsection{Tradeoffs in Robust Model Training}
The Phase 3 Selection highlights resource tradeoffs. Adversarial Training offers high Detection Effectiveness but incurs significant Computational and Cost penalties during offline training  \cite{lad2024adversarial}. Environment Fingerprints via ENF are selected for their efficiency, offering high detection capability with low Computational requirements by leveraging existing grid infrastructure, though they require management of Privacy risks regarding location \cite{korgialas2024leveraging}.

\begin{table}[ht]
\centering
\caption{Phase 3 Selection: Robust Model Training Constraints}
\label{tab:ph3_training}
\scriptsize
\begin{tabular}{|p{1.5cm}|p{2.5cm}| p{2.5cm}|}
\hline
\textbf{Constraint} & \textbf{Adversarial Training}  & \textbf{Environment Fingerprints} \\
\hline
\textbf{Detection} & High robustness to perturbation \cite{chen2025deepfake}. & High specificity for physical locality \cite{zhang2025feasibility} \\
\hline
\textbf{Computational} & Very high offline training cost \cite{lad2024adversarial}. & Low; signal processing only \cite{korgialas2024leveraging}.\\
\hline
\textbf{Integration} & High; Complex parameter tuning \cite{tipper2024investigation}.  & Moderate; requires grid reference DB \cite{hatami2025electric}. \\
\hline
\textbf{Operational} & High; frequent retraining required \cite{alobaid2025disruptive}. & Low; automated extraction \cite{hsu2023detection}. \\
\hline
\textbf{Privacy} & Low; with Federated Learning \cite{sangaiah2024guest}. & Medium; Location leakage risk \cite{sakaci2024conducted, hatami2025electric}. \\
\hline
\textbf{Cost} & High GPU OPEX \cite{alobaid2025disruptive}. & Low; uses existing grid \cite{zhang2025feasibility}.\\
\hline
\end{tabular}
\end{table}

\subsubsection{Defense Architecture Design Choices}
In the Phase 4 Architecture, Adversarial Training underpins the Adaptation Tier, ensuring models are hardened through continuous retraining as new techniques emerge \cite{kazimierczak2024impact}. ENF operates within the Detection Tier, providing a real-time, physics-based liveness check to validate audio-visual streams before they influence agent decisions \cite{zeng2024enfformer}. 

\begin{table}[ht]
\centering
\caption{Phase 4 Architecture: Robust Model Training Tiers}
\label{tab:ph4_training}
\scriptsize
\begin{tabular}{|p{1.5cm}|p{2.5cm}| p{2.5cm}|}
\hline
\textbf{Defense Tier} & \textbf{Adversarial Training}  & \textbf{Environment Fingerprints} \\
\hline
\textbf{Perimeter} & None. & None.  \\
\hline
\textbf{Detection} & Classifier models detect subtle adversarial noise \cite{lad2024adversarial}  & Validate media timestamp against Grid Frequency (ENF) \cite{hua2024factors}\\
\hline
\textbf{Response} & Automated RL-based incident response \cite{jayanthiladevi2025ai}. & Flag content with mismatched ENF as Synthetic \cite{hsu2023detection} \\
\hline
\textbf{Adaptation} & Incorporate new deepfake samples into training sets \cite{kazimierczak2024impact}. & Update reference databases with regional grid logs \cite{hatami2025electric}. \\
\hline
\end{tabular}
\end{table}

\subsection{Proactive Defenses}
Proactive defenses neutralize threats preemptively through strategies like identity watermarking for facial verification \cite{zhao2023proactive} and AI-driven predictive threat analysis \cite{syllaidopoulos2025comprehensive}. In the context of Multi-Agent Systems (MAS), defenses focus on prevention and resiliency mechanisms, such as trust and reputation modeling \cite{owoputi2022security}. Federated approaches distribute defenses across nodes \cite{gaba2024innovative}. Extensions include cognitive security architectures that integrate human-AI scrutiny \cite{casino2025unveiling} and blockchain-enhanced monitoring \cite{mukesh2025comprehensive}. Furthermore, system robustness against resource failures can be evaluated using discrete timed model-based optimization \cite{hsieh2021robustness}.

\subsubsection{Disrupting Deepfake creation}
Disrupting deepfake creation interferes with generative processes via training contamination \cite{sun2022faketracer}, latent space optimizations \cite{zeng2024loft}, and adversarial watermarking \cite{lin2025robust}. Techniques like sticky adversarial signals \cite{zhuang2024pgd} disrupt generation by resisting removal attempts, while learnable hidden faces \cite{li2025big} facilitate proactive exposure of tampered media. Recent advancements include differential-aware networks \cite{he2025kad}, diffusion-based watermarking \cite{sun2025diffmark}, and neural watermarking for speaker identity protection \cite{ge2025proactive}.



\subsubsection{Traceability}
Understanding the root cause of an agent's failure is critical for accountability. Mechanistic interpretability involves dissecting LLM internals to identify specific attention heads or layers responsible for harmful behavior. In multi-agent systems, this allows defenders to trace how representations, such as malicious instructions or steganographic messages, propagate between agents \cite{lee2025towards}. To operationalize this, system-level anomaly detection frameworks now employ Dynamic Execution Graphs to map multi-step interaction chains \cite{he2025sentinelagent}. These graphs enable multi-point failure attribution, a capability essential for detecting emergent collusion where agents covertly coordinate to bias collective decisions \cite{motwani2024secret}.

Technical traceability relies on establishing a verifiable link between content and its origin. Proactive methods utilize forensic watermarking for end-to-end tracking, such as the FRW-TRACE framework for biometric data \cite{qiao2025frw} and Pseudo-Zernike transform watermarking for face swapping detection \cite{lai2024novel}. Beyond watermarking, blockchain-based architectures provide immutable provenance; for instance, the VeriTrust framework integrates Self-Sovereign Identities (SSI) to bind content to creator identities \cite{farhan2025self}.

Complementing these proactive measures are reactive attribution techniques. These include source attribution via camera fingerprints \cite{shi2025deepfake} and the analysis of generative artifacts using Vision Transformers \cite{arshed2023unmasking}. To ensure these technical findings are actionable for human analysts, Explainable AI (XAI) methods are increasingly integrated to demystify black-box detection models \cite{qian2025black}. Finally, comprehensive forensic surveys highlight the need to analyze the entire lifecycle of synthetic media to counter the rising impostor bias and user cynicism regarding content authenticity \cite{amerini2025deepfake, liu2025seeing}.

\subsubsection{Tradeoffs in Proactive Defenses} 
In the Phase 3 Selection, a tradeoff analysis favors Disrupting deepfake creation for its privacy preservation capabilities, despite its higher integration costs. By optimizing the latent space of generative models, it provides a preemptive shield with moderate computational overhead \cite{zeng2024loft}. Conversely, Traceability is critical for accountability but suffers from a high operational burden due to log storage requirements and analysis latency, which limits its real-time utility in fast-acting CPS environments \cite{farhan2025self}.


\begin{table}[ht]
\centering
\caption{Phase 3 Selection: Proactive Defense Constraints}
\label{tab:ph3_proactive}
\scriptsize
\begin{tabular}{|p{1.5cm}|p{2.5cm}|p{2.5cm}|}
\hline
\textbf{Constraint} & \textbf{Disrupting Deepfake} & \textbf{Traceability} \\
\hline
\textbf{Detection} & Pre-emptive neutralization \cite{sun2022faketracer}. & Forensic attribution \cite{qiao2025frw}. \\
\hline
\textbf{Computational} & High; iterative diffusion sampling \cite{sun2025diffmark}  & High analysis latency \cite{qian2025black}. \\ 
\hline
\textbf{Integration} & High; requires source media mod \cite{lin2025robust}. & Moderate; logging infra needed \cite{shi2025deepfake}. \\ 
\hline
\textbf{Operational} & Low; automated injection \cite{zhuang2024pgd}. & High storage for logs \cite{farhan2025self}. \\
\hline
\textbf{Privacy} & High; protects biometric data \cite{li2025big}. & Low; de-anonymization risk \cite{farhan2025self}. \\
\hline
\textbf{Cost} & Low \cite{he2025kad}. & Medium \cite{farhan2025self}. \\
\hline
\end{tabular}
\end{table}

\subsubsection{Defense Architecture Design Choices}
In the Phase 4 Architecture, disrupting deepfake functions primarily at the Perimeter Tier, attacking adversarial capabilities before they penetrate the CPS boundary \cite{sun2025diffmark}. Traceability anchors the Adaptation and Response Tiers, enabling the attribution of breaches to specific actors and facilitating threat model updates based on forensic evidence \cite{lai2024novel}.


\begin{table}[ht]
\centering
\caption{Phase 4 Architecture: Proactive Defense Tiers}
\label{tab:ph4_proactive}
\scriptsize
\begin{tabular}{|p{1.5cm}|p{2.5cm}|p{2.5cm}|}
\hline
\textbf{Defense Tier} & \textbf{Disrupting Deepfake} & \textbf{Traceability} \\
\hline
\textbf{Perimeter} & Inject adversarial signal to disrupt generation \cite{zhuang2024pgd}  & None. \\ 
\hline
\textbf{Detection} & Embed learnable hidden face as tamper indicator \cite{li2025big}  & None. \\
\hline
\textbf{Response} & Diffusion-based watermarking for source tracking \cite{sun2025diffmark} & Trace attack origin via embedded forensics \cite{lai2024novel} \\
\hline
\textbf{Adaptation} & Latent space optimization for specific image defense \cite{zeng2024loft}  & Update attribution models based on new attack vectors \cite{arshed2023unmasking}. \\ 
\hline
\end{tabular}
\end{table}

\subsection{Human \& Policy Measures}
Human and policy measures complement technical mechanisms of defense in CPS through governance, education, and strategic foresight. While technical controls secure the digital-physical interface, human operators and organizational policies dictate how these tools are deployed and monitored. To improve organizational readiness against social engineering in critical infrastructure, Pedersen et al. propose the PREDICT framework, which integrates definitive policy direction with targeted employee education \cite{pedersen2025deepfake}. Beyond the organizational level, Mubarak et al. highlight the necessity for legislative action and public awareness campaigns to mitigate the societal and political destabilization caused by deepfakes \cite{mubarak2023survey}.

From a strategic perspective, reactive measures are insufficient for CPS security. Saeed et al. argue that Cyber Threat Intelligence must be leveraged to navigate the complexity of threats where human error remains a primary vulnerability \cite{saeed2023systematic}. Furthermore, Almahmoud et al. suggest shifting toward proactive defense by utilizing machine learning to forecast long-term cyber threat trends. This foresight enables organizations to make strategic investments in mitigation technologies before threats escalate to physical harm \cite{almahmoud2025forecasting}. Ultimately, preserving trust in these systems requires interdisciplinary collaboration among technologists, legislators, and psychologists \cite{domenteanu2024living}.

\subsubsection{User Training}
In safety-critical environments, humans remain both a vulnerability and a critical defense layer. Training strategies must therefore address two distinct groups: the operators who manage CPS interfaces and the developers who build them.

Operator Awareness deals with effective training for operators that must combine real-time threat recognition with strict verification procedures. As technical detection tools may lag behind generation capabilities, Romero et al. emphasize that media literacy empowers individuals as a first line of defense against deepfake-induced manipulation \cite{romero2025deepfake}. Interactive training tools, such as platforms challenging users to distinguish between real and synthetic faces, have proven effective in sharpening these analytical skills \cite{mustak2023deepfakes}. Additionally,Generative AI (GenAI) driven simulations are now employed to construct authentic attack scenarios, strengthening the response speed of cybersecurity specialists and reducing manual workloads during incident response \cite{alhashimi2025exploring}.

Under secure development practices, the integration of GenAI into the Software Development Lifecycle necessitates a shift from manual coding to AI-augmented oversight. While GenAI tools assist in code generation and debugging, they introduce risks such as logic errors and insecure code patterns. Al-Hashimi et al. note that while GenAI can automate vulnerability scanning, human-in-the-loop review processes are essential to validate AI-generated outputs against security standards like OWASP and NIST \cite{alhashimi2025exploring}. Given the arms race between attack and defense, continuous education on evolving threat landscapes is vital to maintain the integrity of the SDLC \cite{abbas2024unmasking}.

\subsubsection{Regulations}
Regulatory frameworks are evolving to mandate accountability and transparency in AI agent deployment. The landscape includes hard regulations like EU AI Act and soft law frameworks such as NIST AI RMF, ISO 42001 that collectively establish governance structures for trustworthy systems. These regulations aim to mitigate liability risks by establishing legal frameworks for content provenance and agent behavior \cite{akhtar2024video, bakirov2025theoretical}. In the context of CPS, harmonized standards are increasingly critical to enforcing privacy and data protection across interconnected devices \cite{farhan2025self, hatami2025electric}. However, human detection capabilities are often insufficient to identify sophisticated deepfakes \cite{groh2024human}. Consequently, scholars argue for adaptive legislation driven by public sentiment \cite{ccalli2025recoding} and enforced through technical cryptographic standards \cite{iavich2025combating} to safeguard the digital trust ecosystem.

\subsubsection{Standards}
Standardization ensures interoperability and security across the multi-vendor ecosystems typical of CPS. While protocols like the MCP and A2A facilitate agent interaction and collaboration, current analyses indicate a lack of native security standards, leaving systems vulnerable to naming attacks and context poisoning \cite{posta2025deepdive}. To address these gaps, the industry is moving toward robust evaluation frameworks. The OWASP Agentic Security Initiative has established a reference threat model to identify risks such as intent manipulation and memory poisoning \cite{owasp2025agentic}. 

Complementing this, the Agent Security Bench (ASB) provides a framework for benchmarking attacks—including direct prompt injection across diverse agent scenarios \cite{zhang2025agent}. Beyond software agents, standardization must extend to the physical domain. Syllaidopoulos et al. emphasize the need for international harmonization to manage high-risk AI deployments \cite{syllaidopoulos2025comprehensive}. Emerging technical frameworks also propose standardization paths, Odyurt et al. suggest behavioral passports as a baseline for anomaly detection in industrial CPS \cite{odyurt2022improving}, while Babbar et al. propose cryptographic frameworks utilizing federated learning to standardize secure data transmission in vehicular environments \cite{babbar2025federated}.

\subsection{Emerging research directions}
Emerging defense strategies increasingly leverage the intersection of physical constraints and behavioral analysis to secure AI agents. Physical Layer Authentication has evolved beyond static checks. Strategies now employ multi-modal fusion to verify device integrity \cite{wang2021survey, hasan2024review} and utilize environmental anchors, such as Electric Network Frequency (ENF), to validate digital twins against real-time grid fluctuations \cite{hatami2025electric, hatami2025anchor}. To secure the agents themselves, research is shifting toward Behavioral Fingerprinting. Odyurt et al. propose behavioral passports that model the execution phases of CPS processes, allowing for the detection of anomalies in agent behavior that traditional signature-based methods might miss \cite{odyurt2022improving}. Furthermore, Decentralized and Explainable Frameworks are gaining prominence. Hybrid models are integrating Deep Learning with blockchain to ensure data provenance \cite{selvi2024enhancing}, while Federated Learning is being adopted to train robust models without exposing sensitive edge data \cite{gaba2024innovative}. Finally, to address the black box nature of AI agents, Explainable AI (XAI) is emerging as a critical requirement for policy compliance and human-agent trust in high-stakes counter-terrorism and cybersecurity operations \cite{syllaidopoulos2025comprehensive}.

As LLM-driven agents increasingly integrate with external environments via MCP, the attack surface expands beyond simple injection to include privilege escalation and insecure tool execution. Emerging defenses are shifting toward architectural isolation and granular access control. Formal security design patterns, such as the Dual LLM architecture have been proposed that separates the untrusted planning agent from the privileged execution agent to enforce information flow control and prevent data exfiltration \cite{beurer2025design}. To address risks specific to MCP servers, Bühler et al. introduce AgentBound, a policy enforcement framework that intercepts tool calls at the protocol level to prevent unauthorized privilege escalation and resource abuse \cite{buhler2025securing}. Furthermore, dynamic evaluation frameworks are being developed to stress-test these agentic logic flows against multi-turn adaptive attacks, moving beyond static benchmarks to ensure resilience in continuous interaction environments \cite{zhan2025adaptive}.

\section{Case Study: Authenticating Smart Grid Digital Twins}
\label{sec:case}

To demonstrate how the principles surveyed in this work apply in a concrete CPS context, we present a case study based on ANCHOR-Grid, a smart grid DT authentication framework that leverages real-world environmental data to secure cyber-physical representations \cite{hatami2025anchor}. This case study shows how the SENTINEL framework can be operationalized to address deepfake threats against DTs and underscores the role of physical anchors in CPS security.

In smart grid systems, DTs are virtual replicas of physical grid components used for simulation, monitoring, and decision support. Their fidelity to the real grid is critical: attackers who manipulate DT inputs or states can disrupt operational decision-making, thereby creating reliability and safety hazards. ANCHOR-Grid was developed to confront this risk by authenticating DTs against environmental signals that are difficult for adversaries to forge. Specifically, ANCHOR-Grid uses ENF as a real-world anchor. Because the ENF reflects the dynamics of the actual power system and fluctuates unpredictably with the physical grid's behavior, it provides a robust ground truth against which DT can be verified. This approach shifts the security goal from detecting anomalous data sequences to validating the authenticity of the DT through physical signal congruence \cite{hatami2025anchor}.

\begin{figure*}[htbp!]
\centering
\includegraphics[scale=0.38]{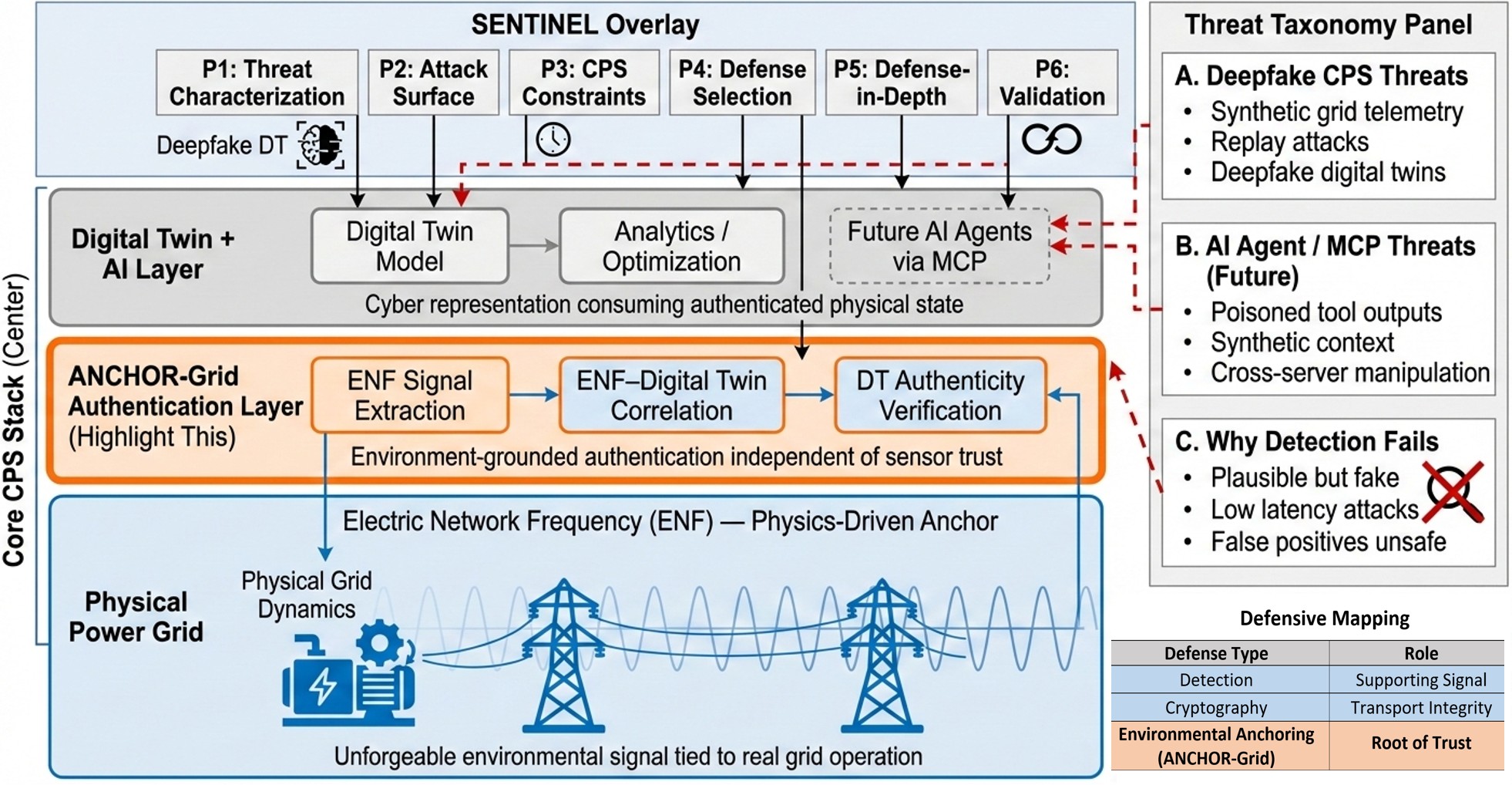}
\caption{End-to-end integration of ANCHOR-Grid with the SENTINEL framework for securing smart grid DTs against deepfake threats.}
\label{fig:anchor-grid}
\vspace{-10 pt}
\end{figure*}

Figure \ref{fig:anchor-grid} illustrates how real-world environmental anchors can be integrated with the SENTINEL framework to secure smart grid DTs against deepfake threats. A layered CPS architecture is shown in which ENF signals from the physical power grid are used by the ANCHOR-Grid authentication layer to validate DT states before they are consumed by AI-enabled analytics and future MCP-enabled agents. Deepfake and replay attacks targeting cyber representations are explicitly shown to bypass traditional detection mechanisms but to be intercepted through environment-grounded verification. The overlay of SENTINEL phases highlights how threat characterization, constraint analysis, defense selection, defense-in-depth, and continuous validation map directly onto system components, illustrating why physics-anchored trust is essential for securing AI agents in safety-critical CPS.

\subsubsection{SENTINEL Phase 1–2: Threat Characterization and Attack Surface Identification.} Applying the first two phases of SENTINEL, ANCHOR-Grid characterizes threats not merely as abstract data integrity violations, but as context-dependent risks shaped by timing, trust, and physical criticality. The attack surface extends beyond individual sensors to include the entire agent–environment interface, where digital representations of grid state influence downstream decision-making. In this setting, the adversary’s objective is not necessarily to introduce obviously anomalous data, but to generate measurements that are internally consistent and physically plausible, thereby evading detection while inducing incorrect grid state estimates. This threat model aligns closely with deepfake-style attacks surveyed earlier in this paper, where fidelity and contextual realism are leveraged to bypass trust assumptions.

\subsubsection{SENTINEL Phase 3: CPS Constraints and Feasibility Analysis.} A defining characteristic of smart grid CPS is the presence of strict timing, reliability, and safety constraints that fundamentally limit which security mechanisms are feasible in practice. In the context of DT authentication, these constraints act as a hard filter in SENTINEL Phase 3, eliminating many defenses that are effective in purely cyber or offline settings. Smart grid monitoring and control operate across multiple time scales, ranging from sub-second dynamics for frequency stability and protection functions to second-level and minute-level windows for state estimation and operational analytics. Security mechanisms introduced at the DT interface must therefore satisfy bounded latency and low false-positive requirements, as delayed or spurious alarms can trigger unnecessary mitigation actions, operator intervention, or degraded situational awareness.

ANCHOR-Grid explicitly evaluates feasibility under such CPS constraints by demonstrating that ENF-based authentication remains effective under realistic network latency and noise conditions, rather than assuming idealized communication channels. Reported results show that authentication performance is maintained under network delays on the order of $O(10^2 ms)$, a range compatible with DT validation and monitoring workflows but not with ultra-fast protection relays. This distinction is critical: ANCHOR-Grid is intentionally positioned as a trust-validation layer for cyber-physical representations, not as a replacement for real-time protection logic. Table \ref{tab:anchorgrid_constraints} summarizes representative magnitudes for key CPS constraints derived from the ANCHOR-Grid evaluation. These values demonstrate that ENF-based authentication remains effective under realistic latency, noise, and replay conditions, while maintaining acceptable detection and false positive rates. This contrasts sharply with heavyweight detection mechanisms, which typically require significantly more computation and are harder to tune under CPS timing constraints.

\begin{table*}[t]
\centering
\caption{Quantitative CPS Constraints and Design Implications in the ANCHOR-Grid Case Study (SENTINEL Phase 3)}
\label{tab:anchorgrid_constraints}
\renewcommand{\arraystretch}{1.2}
\begin{tabular}{|p{3.5cm}|p{4.5cm}|p{4.5cm}|p{4.5cm}|}
\hline
\textbf{Constraint Dimension} & \textbf{Representative Magnitude} & \textbf{Design Implication} & \textbf{Role of ANCHOR-Grid} \\
\hline
Network latency tolerance &
Authentication accuracy remains between 99.9\% and 95\% under network latencies ranging from $<5$\,ms to 200\,ms &
Security mechanisms must tolerate realistic communication delays without degrading integrity verification &
ENF-based authentication remains robust under latency levels typical of smart grid telemetry networks \\
\hline
Detection accuracy vs.\ false positives (deepfake attacks) &
99.8\% detection with 0.2\% false positives under sparse attacks; 97.5\% detection with 1.5\% false positives under high-frequency attacks &
False alarms must be tightly bounded to avoid unnecessary operator intervention or destabilizing control actions &
Physics-grounded verification maintains high accuracy while limiting false positives under increasing attack intensity \\
\hline
Replay attack resilience &
94\% detection for 5\,s-old ENF signatures; 98.5\% detection for 120\,s-old signatures &
Authentication must detect stale or replayed measurements within operationally relevant time windows &
Temporal correlation of ENF signals enables reliable replay attack detection without model retraining \\
\hline
Noise robustness &
96.5\% detection with 5\% injected noise; 88\% detection with 20\% injected noise &
Environmental noise must not invalidate authentication or cause spurious alarms &
Correlation-based ENF matching tolerates moderate noise typical of real grid measurements \\
\hline
Computational feasibility &
Lightweight signal-processing operations; no continuous deep neural inference required &
Heavyweight ML-based detection is difficult to deploy at scale under CPS resource constraints &
ENF extraction and correlation are suitable for edge or near-edge deployment \\
\hline
Operational response time &
Seconds-level validation compatible with digital twin monitoring, but not millisecond-scale relay protection &
Security layers must be positioned appropriately within CPS control hierarchies &
ANCHOR-Grid functions as a trust-validation layer for digital twins rather than a real-time protection mechanism \\
\hline
\end{tabular}
\end{table*}

From a computational perspective, ENF extraction and correlation rely on lightweight signal-processing operations whose cost scales modestly with window size, making them suitable for edge or near-edge deployment. In contrast, heavyweight deepfake detection models that require continuous inference or retraining impose computational and energy overheads that are difficult to justify under CPS resource constraints. These quantitative considerations reinforce a central insight of this survey: CPS feasibility must be evaluated alongside adversarial robustness, and defenses that ignore timing, false-positive cost, or resource limitations are unlikely to be deployable in operational settings.

\subsubsection{SENTINEL Phase 4: Defense Selection via Environmental Anchoring.} Guided by the above constraints, ANCHOR-Grid selects environmental anchoring as its primary defense mechanism. Specifically, it leverages the ENF signal, a globally observable, physics-driven characteristic of power grids, as a real-world anchor to authenticate DT inputs. Because ENF fluctuations are inherently tied to physical grid dynamics and cannot be arbitrarily forged without controlling large portions of the grid, they provide a form of provenance that is independent of sensor trustworthiness or communication integrity. This shifts the security objective from detecting malicious data to verifying consistency between digital representations and immutable physical signals, a design choice that aligns directly with the principles advocated in this survey.

\subsubsection{SENTINEL Phase 5: Defense-in-Depth Architecture.} ANCHOR-Grid integrates environmental anchoring within a broader defense-in-depth architecture. Cryptographic protections and secure communication channels serve as supporting layers, while ENF-based verification provides a grounding layer that detects desynchronization even when upstream defenses fail. This layered approach limits the blast radius of compromised components and avoids single points of failure, an essential requirement in safety-critical CPS. Importantly, the anchoring mechanism operates passively and incurs minimal computational overhead, preserving real-time performance while enhancing trustworthiness. This contrasts with detection-centric pipelines that require continuous retraining and adaptation to evolving attack strategies.

\subsubsection{SENTINEL Phase 6: Validation and Continuous Adaptation.} The final phase of SENTINEL emphasizes ongoing validation rather than one-time deployment. In ANCHOR-Grid, the continuous comparison between observed ENF signals and DT inputs enables persistent monitoring of system integrity. Deviations beyond expected tolerances signal potential compromise or drift, triggering mitigation or investigation workflows. This validation strategy highlights a key insight for securing AI-enabled CPS: resilience emerges not from perfect detection, but from continuous grounding in physical reality combined with adaptive response mechanisms.

While ANCHOR-Grid did not explicitly deploy LLM agents or MCP-mediated tools, it provides a concrete template for securing future AI agents operating DTs. As AI agents increasingly rely on MCP to access external tools, data sources, and shared contexts, the risk of synthetic yet plausible inputs grows substantially. In such settings, environmental anchors, such as those employed in ANCHOR-Grid, can serve as trust substrates that inform agent reasoning, constrain decision-making, and validate tool outputs before they influence physical actuation. This case study, therefore, demonstrates how the principles articulated in the SENTINEL framework can be realized in practice and why environment-grounded verification mechanisms are likely to be indispensable for trustworthy AI agents in CPS.

Taken together, the analyses in Sections \ref{sec:threat}-\ref{sec:mitigation} and the end-to-end CPS case study demonstrate that securing AI agents in CPS requires a shift from isolated, technique-centric defenses to lifecycle-aware system design. The SENTINEL framework integrates threat characterization and attack surface identification with feasibility filtering, defense selection, and continuous validation, ensuring that security mechanisms are evaluated not only for adversarial robustness but also for compatibility with timing, safety, and trust constraints imposed by the physical world. Across deepfake threat modalities, detection techniques, and mitigation strategies, a consistent conclusion emerges: defenses that ignore CPS constraints or lack environment-grounded provenance cannot be relied upon as decision authorities. By explicitly connecting threat analysis, quantitative feasibility, and defense-in-depth architectures, SENTINEL provides a unifying logic that explains why physics-anchored trust mechanisms, such as those illustrated in the ANCHOR-Grid case study, are indispensable for trustworthy AI agents operating in real-world CPS.


\section{Open Challenges and Future Directions}
\label{sec:future}
Open challenges in MCP for CPS encompass the need for adaptive protocols that evolve in response to advancements in deepfakes, ensuring the seamless integration of detection mechanisms without compromising tool interoperability or agent performance \cite{gaba2024innovative}. Identity fragmentation remains a persistent issue, where fragmented authentication across MCP servers heightens vulnerability to deepfake injections, necessitating unified identity management to maintain trust in agent-environment interactions \cite{odyurt2022improving}. Real-time constraints in CPS demand low-latency defenses; however, current MCP implementations struggle with computational overhead from provenance tracking, highlighting the need for optimized message formats that embed security without inflating data streams \cite{hsieh2021robustness}. Furthermore, the interplay between privacy and traceability poses dilemmas, as provenance requirements may expose sensitive context data, requiring novel encryption schemes tailored to MCP's JSON-RPC structure \cite{hasan2024review}.

Future directions for MCP in CPS include exploring hybrid architectures that fuse blockchain with protocol layers for decentralized provenance, enhancing resilience against supply-chain attacks while preserving modularity \cite{fatima2025comprehensive}. Multi-modal detection fusion within MCP could leverage cross-verification of sensor inputs, addressing gaps in single-modality defenses against sophisticated deepfakes \cite{haakansson2021robust}. Emphasis on edge-compatible implementations will drive the development of lightweight MCP variants, incorporating AI-optimized compression to enable real-time operation on resource-constrained devices \cite{syllaidopoulos2025comprehensive}. Collaborative frameworks between academia and industry should standardize MCP extensions for the ethical use of AI, fostering interoperability while mitigating risks associated with generative content \cite{wang2023robustness}. These pathways aim to fortify MCP as a secure backbone for CPS amid escalating threats.

Advancing MCP research also involves formal verification of protocol behaviors under adversarial conditions, as well as modeling deepfake scenarios to predict and preempt exploits in agent-tool exchanges \cite{gaba2024innovative}. Privacy-preserving enhancements, such as zero-knowledge proofs integrated into MCP capability descriptions, could reconcile authentication needs with data protection mandates \cite{hasan2024review}. Scalability tests on large-scale CPS simulations will inform optimizations for high-volume interactions, ensuring MCP sustains performance as generative threats proliferate \cite{selvi2024enhancing}. Ultimately, interdisciplinary efforts must prioritize user-centric designs, embedding intuitive safeguards that empower operators without hindering CPS efficiency \cite{haakansson2021robust}.

\subsection{Rapid Changing Deepfake Detection Landscape} 

The arms race between deepfake generation and detection within MCP contexts demands continuous innovation in protocol safeguards, as generative models rapidly evolve to bypass existing forensic checks in CPS data streams \cite{amerini2025deepfake}. Adversarial techniques that fine-tune GANs or diffusion models to mimic legitimate tool outputs exacerbate MCP vulnerabilities, requiring dynamic update mechanisms for server-side detectors \cite{abbasi2025comprehensive}. Multi-feature fusion approaches can counter this by analyzing inconsistencies across MCP message modalities, bolstering resistance to refined fakes \cite{patel2023deepfake}. Hybrid defenses to the escalating fidelity of synthetic content \cite{shallal2025image}.

Addressing this issue further involves proactive disruption strategies embedded in MCP, such as watermarking payloads to hinder the propagation of deepfakes during agent exchanges \cite{sohail2025deepfake}. Transfer learning frameworks enable detectors to generalize from known generative artifacts, mitigating the lag behind novel creation methods \cite{babaei2025generative}. Ensemble models that aggregate predictions from diverse architectures improve robustness, counteracting the adaptive nature of attackers \cite{akhtar2023deepfakes}. These advancements ensure MCP remains a viable conduit for secure CPS operations amid ongoing generational leaps in deepfakes \cite{patil2025navigating}.

\subsection{Challenges from Unseen Fakes}

Generalization of detectors to unseen fakes in MCP-driven CPS requires models that extrapolate beyond training distributions, incorporating self-supervised learning to identify novel artifacts in protocol interactions \cite{alrajeh2025deepfake}. Patch-wise analysis techniques dissect MCP payloads for localized inconsistencies, improving adaptability to emerging generative variants \cite{lei2025deepfake}. Cross-dataset training regimens bolster resilience, enabling detectors to handle diverse forgery types without retraining \cite{qian2025black}. Multi-attention mechanisms focus on temporal and spatial cues, thereby enhancing performance on hybrid deepfakes that blend multiple modalities \cite{alrajeh2025deepfake}.

Further enhancements involve intra-prediction frameworks that model expected MCP behaviors and flag deviations from unseen manipulations \cite{amerini2025deepfake}. Disentangled feature representations separate content from forgery traces, facilitating transfer to novel scenarios \cite{garg2023deepfake}. Gated attention architectures prioritize relevant features in dynamic CPS environments, reducing overfitting to specific fake patterns \cite{akhtar2023deepfakes}. These strategies collectively advance MCP detectors toward universal applicability against evolving unseen threats \cite{rana2024deepfakes}.

\subsection{Intelligence at the Edge}

Lightweight real-time defenses for edge devices in MCP-integrated CPS prioritize efficient architectures, such as tiny CNNs, which optimize for low-power computation while maintaining detection efficacy \cite{hatami2025electric}. MobileNet variants with quantization enable rapid inference on resource-limited hardware, embedding MCP security without excessive latency \cite{islam2024ai}. Multi-feature fusion in compact models captures essential deepfake cues while balancing accuracy and edge constraints \cite{meng2025multi}. Hybrid CNN-LSTM-Transformer designs enable the real-time analysis of streaming MCP data, making them suitable for CCTV-like applications \cite{sheng2025id}.

Advancing these defenses includes adaptive optimization techniques that fine-tune parameters on-device, enhancing responsiveness to local CPS threats \cite{sheng2025id}. End-to-end frameworks minimize overhead by integrating detection directly into MCP protocols, supporting seamless edge deployment \cite{gong2024contemporary}. Eye movement analysis in hybrid approaches adds behavioral layers, improving real-time forgery spotting with minimal resources \cite{yan2025self}. Overall, these innovations enable scalable MCP protection on edge platforms \cite{deressa2025genconvit}.

\subsection{Privacy Requirements}

Balancing privacy with provenance/authentication in MCP for CPS involves selective disclosure protocols that embed verifiable credentials without revealing full context data \cite{tahaoglu2025robust}. Blockchain-based frameworks offer immutable provenance while utilizing zero-knowledge proofs to maintain user anonymity in agent exchanges \cite{xu2025localization,jing2025zero}. Hybrid watermarking integrates post-quantum cryptography, ensuring authenticity traces resist tampering without compromising sensitive information \cite{clapten2025gated}. Multi-factor validation with human oversight adds ethical layers, mitigating over-reliance on automated provenance that could infringe privacy \cite{almuhaideb2025lightfakedetect}.

Further strategies include policy governance that enforces data minimization in MCP messages, as well as aligning authentication with regulatory standards \cite{balafrej2024enhancing}. Forensic techniques prioritize privacy-preserving artifacts, enabling detection without full data exposure \cite{ranasinghe2025privacy}. Bibliometric analyses highlight trends in ethical AI, guiding balanced implementations that integrate privacy into provenance designs \cite{karim2024mcgan}. These measures foster trust in MCP systems by harmonizing security with individual rights \cite{sohail2025deepfake}.

\subsection{Model Development}

Formal models for AI-agent security in CPS with MCP emphasize the use of applied pi-calculus for verifying protocol integrity against adversarial manipulations \cite{javed2024real}. Intrusion detection frameworks model AI behaviors formally, ensuring resilience in networked environments \cite{bansal2023real}. Multimodal fusion in formal representations captures CPS dynamics, enabling rigorous analysis of agent interactions \cite{wu2025cbam}. Blockchain-enabled models formalize trust in distributed CPS, incorporating AI optimization for secure operations \cite{yasir2025lightweight}.

Expanding these models includes adversarial learning strategies that formally defend against poisonous attacks, using GANs to simulate threats \cite{sohail2025deepfake}. Anomaly detection in power systems leverages formal AI models for real-time identification of threats. Special issue compilations on CPS security outline formal approaches integrating AI for enhanced protection \cite{islam2024ai}. These formalizations provide foundational assurances for MCP in critical infrastructures.


In summary, these open challenges underscore that securing AI agents in CPS is not primarily a modeling problem, but a systems problem governed by physical constraints, trust boundaries, and adversarial adaptability. The case study and SENTINEL-based analysis presented in this survey demonstrate that defenses ignoring timing, false-positive cost, and environmental grounding are unlikely to be deployable in safety-critical CPS, regardless of algorithmic sophistication. Progress will therefore require a shift away from detection-centric thinking toward architectures that embed provenance, physics-based anchors, and continuous validation as first-class design elements. Addressing these challenges demands coordinated advances across AI security, control and signal processing, systems engineering, and standards development, without which the next generation of autonomous, tool-using AI agents will remain fundamentally untrustworthy in the physical world.

\section{Conclusion}
\label{sec:conlusion}


In this survey, we have explored the profound vulnerabilities introduced by deepfakes at the intersection of AI agents and cyber-physical systems through the Model Context Protocol. Deepfakes, spanning visual, audio, textual, and behavioral modalities, exploit the AI-environment interface to deceive sensors, manipulate data streams, and undermine operational integrity. Within MCP, these threats manifest as context poisoning, where synthetic content infiltrates tool invocations and inter-agent communications, resulting in cascading failures in critical infrastructures such as smart grids and autonomous vehicles. The unique challenge lies in the seamless blending of digital forgery with physical actions, where a spoofed video feed or cloned voice command can trigger unsafe actions or conceal anomalies, amplifying risks beyond traditional cyberattacks. By categorizing these threats and examining real-world incidents, the survey underscores how MCP's interoperability, while enabling efficiency, inadvertently expands the attack surface for adversarial AI-generated content.

AI-agent interactions in CPS, facilitated by MCP, introduce additional layers of complexity, as agents rely on external tools and shared contexts that adversaries can hijack through deepfake-driven exploits. User-agent prompts, agent-agent trust exploitation, and agent-environment manipulations create a fertile ground for social engineering, identity spoofing, and indirect injections, where malicious actors leverage generative models to emulate legitimate behaviors or forge instructions. In MCP ecosystems, these interactions heighten the potential for supply-chain compromises and over-privileged access, as unvetted servers process deepfake-tainted data, potentially leading to unauthorized actions or data leaks. The survey highlights how such dynamics erode trustworthiness, with deepfakes not only deceiving machines but also misleading human operators, thereby challenging the core principles of safety and reliability in CPS. Addressing these threats requires recognizing their hybrid nature, combining cyber deception with physical consequences in protocol-mediated environments.

To counter these evolving dangers, an interdisciplinary collaboration is essential, drawing expertise from AI for advanced generative and detection models, cybersecurity for robust protocol designs, signal processing for forensic analysis of environmental fingerprints, and human factors for intuitive training and policy development. This convergence can foster innovative solutions, such as AI-enhanced provenance tracking informed by signal anomalies and user-centric interfaces that incorporate behavioral insights. By uniting these fields, researchers can develop holistic frameworks that anticipate deepfake adaptations within MCP, ensuring defenses evolve in tandem with threats. Collaborative efforts through joint initiatives and shared datasets will accelerate progress, bridging gaps between theoretical advancements and practical CPS deployments. Ultimately, this integration promises to create resilient systems that safeguard against the multifaceted risks posed by AI-driven manipulations.

Defense-in-depth emerges as a guiding principle, advocating layered protections that integrate provenance authentication, multi-factor validation, robust training, proactive disruptions, and human-policy measures within MCP architectures. This approach mitigates single points of failure by combining technical safeguards like digital watermarking and blockchain with procedural elements such as regulations and standards, creating a comprehensive barrier against deepfake incursions. In CPS, where failures can have physical repercussions, defense-in-depth ensures redundancy, with each layer addressing specific vulnerabilities in agent interactions and environmental interfaces. By embedding these principles into MCP specifications, systems can achieve adaptive security, dynamically responding to threats while maintaining operational continuity. This multi-tiered strategy not only deters attacks but also minimizes their impact, promoting sustained functionality in adversarial settings.

Resilience stands as the cornerstone for future MCP-enabled CPS, emphasizing systems that withstand, recover from, and adapt to deepfake threats through ongoing monitoring, formal modeling, and emerging directions like physical fingerprints and environmental anchors. This principle prioritizes long-term viability, ensuring AI agents remain trustworthy amid the arms race of generation and detection, with detectors generalizing to unseen fakes and lightweight defenses suiting edge devices. Balancing privacy with authentication further reinforces resilience, preventing over-exposure while upholding provenance. As CPS integrates more deeply with AI, embracing resilience through interdisciplinary innovation will guide the development of secure, ethical ecosystems. In closing, this survey underscores the need for sustained efforts to implement these principles, thereby paving the way for dependable AI in an increasingly interconnected world.


\bibliographystyle{IEEEtranS}
\bibliography{ref}

\begin{thebibliography}{100}
\providecommand{\url}[1]{#1}
\csname url@samestyle\endcsname
\providecommand{\newblock}{\relax}
\providecommand{\bibinfo}[2]{#2}
\providecommand{\BIBentrySTDinterwordspacing}{\spaceskip=0pt\relax}
\providecommand{\BIBentryALTinterwordstretchfactor}{4}
\providecommand{\BIBentryALTinterwordspacing}{\spaceskip=\fontdimen2\font plus
\BIBentryALTinterwordstretchfactor\fontdimen3\font minus \fontdimen4\font\relax}
\providecommand{\BIBforeignlanguage}[2]{{%
\expandafter\ifx\csname l@#1\endcsname\relax
\typeout{** WARNING: IEEEtranS.bst: No hyphenation pattern has been}%
\typeout{** loaded for the language `#1'. Using the pattern for}%
\typeout{** the default language instead.}%
\else
\language=\csname l@#1\endcsname
\fi
#2}}
\providecommand{\BIBdecl}{\relax}
\BIBdecl

\bibitem{abbas2024unmasking}
F.~Abbas and A.~Taeihagh, ``Unmasking deepfakes: A systematic review of deepfake detection and generation techniques using artificial intelligence,'' \emph{Expert Systems with Applications}, vol. 252, p. 124260, 2024.

\bibitem{abbasi2025comprehensive}
M.~Abbasi, P.~V{\'a}z, J.~Silva, and P.~Martins, ``Comprehensive evaluation of deepfake detection models: Accuracy, generalization, and resilience to adversarial attacks,'' \emph{Applied Sciences}, vol.~15, no.~3, p. 1225, 2025.

\bibitem{abiodun2022data}
O.~I. Abiodun, M.~Alawida, A.~E. Omolara, and A.~Alabdulatif, ``Data provenance for cloud forensic investigations, security, challenges, solutions and future perspectives: A survey,'' \emph{Journal of King Saud University-Computer and Information Sciences}, vol.~34, no.~10, pp. 10\,217--10\,245, 2022.

\bibitem{adapala2025aegis}
S.~T.~R. Adapala and Y.~R. Alugubelly, ``The aegis protocol: A foundational security framework for autonomous ai agents,'' \emph{arXiv preprint arXiv:2508.19267}, 2025.

\bibitem{addo2025federated}
K.~Addo, M.~Kabeya, and E.~E. Ojo, ``Federated quantum machine learning for distributed cybersecurity in multi-agent energy systems,'' \emph{Energies}, vol.~18, no.~20, p. 5418, 2025.

\bibitem{ahmed2024visual}
N.~U.~R. Ahmed, A.~Badshah, H.~Adeel, A.~Tajammul, A.~Duad, and T.~Alsahfi, ``Visual deepfake detection: Review of techniques, tools, limitations, and future prospects,'' \emph{IEEE Access}, 2024.

\bibitem{akhtar2023deepfakes}
Z.~Akhtar, ``Deepfakes generation and detection: A short survey,'' \emph{Journal of Imaging}, vol.~9, no.~1, p.~18, 2023.

\bibitem{akhtar2024video}
Z.~Akhtar, T.~L. Pendyala, and V.~S. Athmakuri, ``Video and audio deepfake datasets and open issues in deepfake technology: being ahead of the curve,'' \emph{Forensic Sciences}, vol.~4, no.~3, pp. 289--377, 2024.

\bibitem{al2025deeplasd}
H.~Al-Tairi, A.~Javed, T.~Khan, and A.~K.~J. Saudagar, ``Deeplasd countermeasure for logical access audio spoofing,'' \emph{Scientific Reports}, vol.~15, no.~1, p. 20839, 2025.

\bibitem{alanazi2024exploring}
S.~Alanazi and S.~Asif, ``Exploring deepfake technology: creation, consequences and countermeasures,'' \emph{Human-Intelligent Systems Integration}, vol.~6, no.~1, pp. 49--60, 2024.

\bibitem{albustami2025breaking}
A.~A. Albustami and A.~F. Taha, ``Breaking the flow and the bank: Stealthy cyberattacks on water network hydraulics,'' \emph{Water Research}, p. 123719, 2025.

\bibitem{alhamarneh2024strengthening}
R.~A. Alhamarneh and M.~Mahinderjit~Singh, ``Strengthening internet of things security: Surveying physical unclonable functions for authentication, communication protocols, challenges, and applications,'' \emph{Applied Sciences}, vol.~14, no.~5, p. 1700, 2024.

\bibitem{alhashimi2025exploring}
H.~A. Alhashimi, R.~A. Khan, H.~S. Alwageed, A.~M. Algarni, S.~Ayouni, and A.~O. Almagrabi, ``Exploring the role of generative ai in enhancing cybersecurity in software development life cycle,'' \emph{Array}, p. 100509, 2025.

\bibitem{ali2025multilingual}
H.~Ali, S.~Subramani, L.~Bollinani, N.~S. Adupa, S.~El-Loh, and H.~Malik, ``Multilingual dataset integration strategies for robust audio deepfake detection: A safe challenge system,'' \emph{arXiv preprint arXiv:2508.20983}, 2025.

\bibitem{ali2025novel}
Z.~Ali, T.~Hussain, C.-L. Su, G.~Parise, K.~Sayler, M.~Sadiq, and S.~H. Rouhani, ``A novel intelligent intrusion detection and prevention framework for shore-ship hybrid ac/dc microgrids under power quality disturbances,'' in \emph{2025 IEEE Industry Applications Society Annual Meeting (IAS)}.\hskip 1em plus 0.5em minus 0.4em\relax IEEE, 2025, pp. 1--7.

\bibitem{almahmoud2025forecasting}
Z.~Almahmoud, P.~D. Yoo, E.~Damiani, K.-K.~R. Choo, and C.~Y. Yeun, ``Forecasting cyber threats and pertinent mitigation technologies,'' \emph{Technological Forecasting and Social Change}, vol. 210, p. 123836, 2025.

\bibitem{almuhaideb2025lightfakedetect}
S.~AlMuhaideb, H.~Alshaya, L.~Almutairi, D.~Alomran, and S.~T. Alhamed, ``Lightfakedetect: A lightweight model for deepfake detection in videos that focuses on facial regions,'' \emph{Mathematics}, vol.~13, no.~19, p. 3088, 2025.

\bibitem{almutairi2022review}
Z.~Almutairi and H.~Elgibreen, ``A review of modern audio deepfake detection methods: challenges and future directions,'' \emph{Algorithms}, vol.~15, no.~5, p. 155, 2022.

\bibitem{alobaid2025disruptive}
A.~Alobaid, T.~Bonny, and M.~Alrahhal, ``Disruptive attacks on artificial neural networks: A systematic review of attack techniques, detection methods, and protection strategies,'' \emph{Intelligent Systems with Applications}, p. 200529, 2025.

\bibitem{alrajeh2025deepfake}
M.~Alrajeh and A.~Al-Samawi, ``Deepfake image classification using decision (binary) tree deep learning,'' \emph{Journal of Sensor and Actuator Networks}, vol.~14, no.~2, p.~40, 2025.

\bibitem{alrashoud2025deepfake}
M.~Alrashoud, ``Deepfake video detection methods, approaches, and challenges,'' \emph{Alexandria Engineering Journal}, vol. 125, pp. 265--277, 2025.

\bibitem{alshehri2024audio}
A.~Alshehri, D.~Almalki, E.~Alharbi, and S.~Albaradei, ``Audio deep fake detection with sonic sleuth model,'' \emph{Computers}, vol.~13, no.~10, p. 256, 2024.

\bibitem{amayuelas2024multiagent}
A.~Amayuelas, X.~Yang, A.~Antoniades, W.~Hua, L.~Pan, and W.~Wang, ``Multiagent collaboration attack: Investigating adversarial attacks in large language model collaborations via debate,'' \emph{arXiv preprint arXiv:2406.14711}, 2024.

\bibitem{amer2025graphshield}
E.~Amer, S.~El-Sappagh, T.~Abuhamad, B.~A.~S. Al-Rimy, and A.~Mohasseb, ``Graphshield: advanced dynamic graph-based malware detection using graph neural networks,'' \emph{Expert Systems with Applications}, p. 129812, 2025.

\bibitem{amerini2025deepfake}
I.~Amerini, M.~Barni, S.~Battiato, P.~Bestagini, G.~Boato, V.~Bruni, R.~Caldelli, F.~De~Natale, R.~De~Nicola, L.~Guarnera \emph{et~al.}, ``Deepfake media forensics: Status and future challenges,'' \emph{Journal of Imaging}, vol.~11, no.~3, p.~73, 2025.

\bibitem{an2025rag}
B.~An, S.~Zhang, and M.~Dredze, ``Rag llms are not safer: A safety analysis of retrieval-augmented generation for large language models,'' \emph{arXiv preprint arXiv:2504.18041}, 2025.

\bibitem{anggrainingsih2024transformer}
R.~Anggrainingsih, G.~M. Hassan, and A.~Datta, ``Transformer-based models for combating rumours on microblogging platforms: a review,'' \emph{Artificial Intelligence Review}, vol.~57, no.~8, p. 212, 2024.

\bibitem{anil2024many}
C.~Anil, E.~Durmus, N.~Panickssery, M.~Sharma, J.~Benton, S.~Kundu, J.~Batson, M.~Tong, J.~Mu, D.~Ford \emph{et~al.}, ``Many-shot jailbreaking,'' \emph{Advances in Neural Information Processing Systems}, vol.~37, pp. 129\,696--129\,742, 2024.

\bibitem{anusha2025deepfake}
T.~Anusha and A.~Srinagesh, ``Deepfake video detection: A comprehensive survey of advanced machine learning and deep learning techniques to combat synthetic video manipulation,'' in \emph{2025 International Conference on Multi-Agent Systems for Collaborative Intelligence (ICMSCI)}.\hskip 1em plus 0.5em minus 0.4em\relax IEEE, 2025, pp. 1033--1041.

\bibitem{arif2021voice}
T.~Arif, A.~Javed, M.~Alhameed, F.~Jeribi, and A.~Tahir, ``Voice spoofing countermeasure for logical access attacks detection,'' \emph{IEEE Access}, vol.~9, pp. 162\,857--162\,868, 2021.

\bibitem{arshed2023unmasking}
M.~A. Arshed, A.~Alwadain, R.~Faizan~Ali, S.~Mumtaz, M.~Ibrahim, and A.~Muneer, ``Unmasking deception: empowering deepfake detection with vision transformer network,'' \emph{Mathematics}, vol.~11, no.~17, p. 3710, 2023.

\bibitem{arya2024study}
M.~Arya, U.~Goyal, S.~Chawla \emph{et~al.}, ``A study on deep fake face detection techniques,'' in \emph{2024 3rd International Conference on Applied Artificial Intelligence and Computing (ICAAIC)}.\hskip 1em plus 0.5em minus 0.4em\relax IEEE, 2024, pp. 459--466.

\bibitem{aslam2024scrutinizing}
M.~M. Aslam, A.~Tufail, R.~A. A. H.~M. Apong, L.~C. De~Silva, and M.~T. Raza, ``Scrutinizing security in industrial control systems: An architectural vulnerabilities and communication network perspective,'' \emph{IEEE Access}, vol.~12, pp. 67\,537--67\,573, 2024.

\bibitem{atawneh2023phishing}
S.~Atawneh and H.~Aljehani, ``Phishing email detection model using deep learning,'' \emph{Electronics}, vol.~12, no.~20, p. 4261, 2023.

\bibitem{awadallah2024artificial}
A.~Awadallah, K.~Eledlebi, M.~J. Zemerly, D.~Puthal, E.~Damiani, K.~Taha, T.-Y. Kim, P.~D. Yoo, K.-K.~R. Choo, M.-S. Yim \emph{et~al.}, ``Artificial intelligence-based cybersecurity for the metaverse: Research challenges and opportunities,'' \emph{IEEE Communications Surveys \& Tutorials}, vol.~27, no.~2, pp. 1008--1052, 2024.

\bibitem{azzam2023forensic}
M.~Azzam, L.~Pasquale, G.~Provan, and B.~Nuseibeh, ``Forensic readiness of industrial control systems under stealthy attacks,'' \emph{Computers \& Security}, vol. 125, p. 103010, 2023.

\bibitem{babaei2025generative}
R.~Babaei, S.~Cheng, R.~Duan, and S.~Zhao, ``Generative artificial intelligence and the evolving challenge of deepfake detection: A systematic analysis,'' \emph{Journal of Sensor and Actuator Networks}, vol.~14, no.~1, p.~17, 2025.

\bibitem{babbar2025federated}
H.~Babbar, S.~Rani, and M.~Shabaz, ``Federated learning with enhanced cryptographic security for vehicular cyber-physical systems,'' \emph{Scientific Reports}, vol.~15, no.~1, p. 28593, 2025.

\bibitem{badhan2025enhancing}
A.~Badhan, P.~Sharma, M.~Dewangan, and H.~Singh, ``Enhancing deepfake detection through facial pattern recognition and transfer learning,'' in \emph{2025 7th International Conference on Inventive Material Science and Applications (ICIMA)}.\hskip 1em plus 0.5em minus 0.4em\relax IEEE, 2025, pp. 1286--1291.

\bibitem{bagdasaryan2023abusing}
E.~Bagdasaryan, T.-Y. Hsieh, B.~Nassi, and V.~Shmatikov, ``Abusing images and sounds for indirect instruction injection in multi-modal llms,'' in \emph{arXiv preprint arXiv:2307.10490}, 2023.

\bibitem{bakirov2025theoretical}
A.~Bakirov and I.~Suleimenov, ``Theoretical bases of methods of counteraction to modern forms of information warfare,'' \emph{Computers}, vol.~14, no.~10, p. 410, 2025.

\bibitem{balafrej2024enhancing}
I.~Balafrej and M.~Dahmane, ``Enhancing practicality and efficiency of deepfake detection,'' \emph{Scientific Reports}, vol.~14, no.~1, p. 31084, 2024.

\bibitem{balan2025framework}
K.~Balan, R.~Learney, and T.~Wood, ``A framework for cryptographic verifiability of end-to-end ai pipelines,'' in \emph{Proceedings of the 2025 ACM International Workshop on Security and Privacy Analytics}, 2025, pp. 49--59.

\bibitem{bandara2025model}
E.~Bandara, S.~Shetty, R.~Mukkamala, R.~Gore, P.~Foytik, S.~H. Bouk, A.~Rahman, X.~Liang, N.~W. Keong, K.~De~Zoysa \emph{et~al.}, ``Model context contracts-mcp-enabled framework to integrate llms with blockchain smart contracts,'' \emph{arXiv preprint arXiv:2510.19856}, 2025.

\bibitem{bansal2023real}
N.~Bansal, T.~Aljrees, D.~P. Yadav, K.~U. Singh, A.~Kumar, G.~K. Verma, and T.~Singh, ``Real-time advanced computational intelligence for deep fake video detection,'' \emph{Applied Sciences}, vol.~13, no.~5, p. 3095, 2023.

\bibitem{baseri2025privacy}
Y.~Baseri, A.~S. Hafid, D.~Makrakis, and H.~Fereidouni, ``Privacy-preserving federated learning framework for risk-based adaptive authentication,'' \emph{arXiv preprint arXiv:2508.18453}, 2025.

\bibitem{bernabe2019privacy}
J.~B. Bernabe, J.~L. Canovas, J.~L. Hernandez-Ramos, R.~T. Moreno, and A.~Skarmeta, ``Privacy-preserving solutions for blockchain: Review and challenges,'' \emph{Ieee Access}, vol.~7, pp. 164\,908--164\,940, 2019.

\bibitem{beurer2025design}
L.~Beurer-Kellner, B.~Buesser, A.-M. Cretu, E.~Debenedetti, D.~Dobos, D.~Fabian, M.~Fischer, D.~Froelicher, K.~Grosse, D.~Naeff \emph{et~al.}, ``Design patterns for securing llm agents against prompt injections,'' \emph{URL https://arxiv.org/abs/2506.08837}, 2025.

\bibitem{bezzi2024large}
M.~Bezzi, ``Large language models and security,'' \emph{IEEE Security \& Privacy}, vol.~22, no.~2, pp. 60--68, 2024.

\bibitem{bk2024ensuring}
S.~BK and F.~Azam, ``Ensuring security and privacy in vanet: A comprehensive survey of authentication approaches,'' \emph{Journal of Computer Networks and Communications}, vol. 2024, no.~1, p. 1818079, 2024.

\bibitem{bohara2025detecting}
R.~Bohara and A.~K. Bairwa, ``Detecting deepfake audio using spectrogram-based machine learning approaches,'' \emph{IEEE Access}, 2025.

\bibitem{boko2024disinformation}
A.~Boko, ``Disinformation detection: Developing a categorical framework through thematic analysis,'' \emph{Journalism and Media}, vol.~5, no.~4, pp. 1914--1924, 2024.

\bibitem{borhani2024detecting}
P.~Borhani-Darian, H.~Li, P.~Wu, and P.~Closas, ``Detecting gnss spoofing using deep learning,'' \emph{EURASIP Journal on advances in signal processing}, vol. 2024, no.~1, p.~14, 2024.

\bibitem{brissett2025machine}
A.~Brissett and J.~Wall, ``Machine learning and watermarking for accurate detection of ai generated phishing emails.'' \emph{Electronics}, vol.~14, no.~13, pp. 1--21, 2025.

\bibitem{brooks2019increasing}
T.~Brooks, G.~Princess, J.~Heatley, J.~Jeremy, K.~Scott \emph{et~al.}, ``Increasing threats of deepfake identities,'' \emph{US Department of Homeland Security [online]}, 2019.

\bibitem{buhler2025securing}
C.~B{\"u}hler, M.~Biagiola, L.~Di~Grazia, and G.~Salvaneschi, ``Securing ai agent execution,'' \emph{arXiv preprint arXiv:2510.21236}, 2025.

\bibitem{ieee2024blockchain}
E.~Bureac{\u{a}} and I.~Aciob{\u{a}}niței, ``A blockchain blockchain-based framework for content provenance and authenticity,'' in \emph{2024 16th International Conference on Electronics, Computers and Artificial Intelligence (ECAI)}.\hskip 1em plus 0.5em minus 0.4em\relax IEEE, 2024, pp. 1--5.

\bibitem{calderon2024deep}
D.~Calder{\'o}n-Gonz{\'a}lez, N.~{\'A}balos, B.~Bayo, P.~C{\'a}novas, D.~Griol, C.~Mu{\~n}oz-Romero, C.~P{\'e}rez, P.~Vila, and Z.~Callejas, ``Deep speech synthesis and its implications for news verification: Lessons learned in the rtve-ugr chair,'' \emph{Applied Sciences}, vol.~14, no.~21, p. 9916, 2024.

\bibitem{ccalli2025recoding}
L.~{\c{C}}alli and B.~Alma~{\c{C}}alli, ``Recoding reality: A case study of youtube reactions to generative ai videos,'' \emph{Systems}, vol.~13, no.~10, p. 925, 2025.

\bibitem{capuano2022explainable}
N.~Capuano, G.~Fenza, V.~Loia, and C.~Stanzione, ``Explainable artificial intelligence in cybersecurity: A survey,'' \emph{Ieee Access}, vol.~10, pp. 93\,575--93\,600, 2022.

\bibitem{casino2025unveiling}
F.~Casino, ``Unveiling the multifaceted concept of cognitive security: Trends, perspectives, and future challenges,'' \emph{Technology in Society}, p. 102956, 2025.

\bibitem{chan2025infrastructure}
A.~Chan, K.~Wei, S.~Huang, N.~Rajkumar, E.~Perrier, S.~Lazar, G.~K. Hadfield, and M.~Anderljung, ``Infrastructure for ai agents,'' \emph{arXiv preprint arXiv:2501.10114}, 2025.

\bibitem{charfeddine2024chatgpt}
M.~Charfeddine, H.~M. Kammoun, B.~Hamdaoui, and M.~Guizani, ``Chatgpt’s security risks and benefits: offensive and defensive use-cases, mitigation measures, and future implications,'' \emph{IEEE Access}, vol.~12, pp. 30\,263--30\,310, 2024.

\bibitem{chen2024blockagents}
B.~Chen, G.~Li, X.~Lin, Z.~Wang, and J.~Li, ``Blockagents: Towards byzantine-robust llm-based multi-agent coordination via blockchain,'' in \emph{Proceedings of the ACM Turing Award Celebration Conference-China 2024}, 2024, pp. 187--192.

\bibitem{chen2025deepfake}
G.~Chen, C.~Du, Y.~Yu, H.~Hu, H.~Duan, and H.~Zhu, ``A deepfake image detection method based on a multi-graph attention network.'' \emph{Electronics}, vol.~14, no.~3, 2025.

\bibitem{chen2025agentguard}
J.~Chen and S.~L. Cong, ``Agentguard: Repurposing agentic orchestrator for safety evaluation of tool orchestration,'' \emph{arXiv preprint}, 2025.

\bibitem{chen2024vall}
S.~Chen, S.~Liu, L.~Zhou, Y.~Liu, X.~Tan, J.~Li, S.~Zhao, Y.~Qian, and F.~Wei, ``Vall-e 2: Neural codec language models are human parity zero-shot text to speech synthesizers,'' \emph{arXiv preprint arXiv:2406.05370}, 2024.

\bibitem{chen2025neural}
S.~Chen, C.~Wang, Y.~Wu, Z.~Zhang, L.~Zhou, S.~Liu, Z.~Chen, Y.~Liu, H.~Wang, J.~Li \emph{et~al.}, ``Neural codec language models are zero-shot text to speech synthesizers,'' \emph{IEEE Transactions on Audio, Speech and Language Processing}, 2025.

\bibitem{chen2025safeguarding}
T.~Chen, J.~Lou, and W.~Wang, ``Safeguarding multimodal knowledge copyright in the rag-as-a-service environment,'' \emph{arXiv preprint arXiv:2506.10030}, 2025.

\bibitem{chen2022learning}
Z.-C. Chen, L.-H. Tsao, C.-L. Fu, S.-F. Chen, and Y.-C.~F. Wang, ``Learning facial liveness representation for domain generalized face anti-spoofing,'' in \emph{2022 IEEE International Conference on Multimedia and Expo (ICME)}.\hskip 1em plus 0.5em minus 0.4em\relax IEEE, 2022, pp. 1--6.

\bibitem{chhetri2025model}
G.~Chhetri, S.~Somvanshi, M.~M. Islam, S.~Brotee, M.~S. Mimi, D.~Koirala, B.~Pandey, and S.~Das, ``Model context protocols in adaptive transport systems: A survey,'' \emph{arXiv preprint arXiv:2508.19239}, 2025.

\bibitem{christ2024undetectable}
M.~Christ, S.~Gunn, and O.~Zamir, ``Undetectable watermarks for language models,'' in \emph{The Thirty Seventh Annual Conference on Learning Theory}.\hskip 1em plus 0.5em minus 0.4em\relax PMLR, 2024, pp. 1125--1139.

\bibitem{chung2024collusion}
H.~Chung, T.~Roughgarden, and E.~Shi, ``Collusion-resilience in transaction fee mechanism design,'' in \emph{Proceedings of the 25th ACM Conference on Economics and Computation}, 2024, pp. 1045--1073.

\bibitem{clapten2025gated}
J.~E. Clapten and v.~Balaji, ``A gated temporal attention based intra prediction framework for robust deepfake video detection,'' \emph{Scientific Reports}, vol.~15, no.~1, p. 38540, 2025.

\bibitem{C2PA}
{Coalition for Content Provenance and Authenticity (C2PA)}, ``C2pa: Verifying media content sources,'' \url{https://c2pa.org/}, 2024.

\bibitem{croitoru2024deepfake}
F.-A. Croitoru, A.-I. Hiji, V.~Hondru, N.~C. Ristea, P.~Irofti, M.~Popescu, C.~Rusu, R.~T. Ionescu, F.~S. Khan, and M.~Shah, ``Deepfake media generation and detection in the generative ai era: a survey and outlook,'' \emph{arXiv preprint arXiv:2411.19537}, 2024.

\bibitem{crothers2023machine}
E.~N. Crothers, N.~Japkowicz, and H.~L. Viktor, ``Machine-generated text: A comprehensive survey of threat models and detection methods,'' \emph{IEEE Access}, vol.~11, pp. 70\,977--71\,002, 2023.

\bibitem{dathathri2024scalable}
S.~Dathathri, A.~See, S.~Ghaisas, P.-S. Huang, R.~McAdam, J.~Welbl, V.~Bachani, A.~Kaskasoli, R.~Stanforth, T.~Matejovicova \emph{et~al.}, ``Scalable watermarking for identifying large language model outputs,'' \emph{Nature}, vol. 634, no. 8035, pp. 818--823, 2024.

\bibitem{datta2025agentic}
S.~Datta, S.~K. Nahin, A.~Chhabra, and P.~Mohapatra, ``Agentic ai security: Threats, defenses, evaluation, and open challenges,'' \emph{arXiv preprint arXiv:2510.23883}, 2025.

\bibitem{daxa2025mcp}
{Daxa.ai}, ``{MCP} security: Securing agentic {AI} with the model context protocol,'' Daxa.ai Blog, July 2025.

\bibitem{de2024secure}
L.~P. de~Melo, D.~Macedo~Amaral, R.~de~Oliveira~Albuquerque, R.~T. de~Sousa~J{\'u}nior, A.~L. Sandoval~Orozco, and L.~J. Garc{\'\i}a~Villalba, ``A secure approach out-of-band for e-bank with visual two-factor authorization protocol,'' \emph{Cryptography}, vol.~8, no.~4, p.~51, 2024.

\bibitem{deng2025ai}
Z.~Deng, Y.~Guo, C.~Han, W.~Ma, J.~Xiong, S.~Wen, and Y.~Xiang, ``Ai agents under threat: A survey of key security challenges and future pathways,'' \emph{ACM Computing Surveys}, vol.~57, no.~7, pp. 1--36, 2025.

\bibitem{deressa2025genconvit}
D.~W. Deressa, H.~Mareen, P.~Lambert, S.~Atnafu, Z.~Akhtar, and G.~Van~Wallendael, ``Genconvit: Deepfake video detection using generative convolutional vision transformer,'' \emph{Applied Sciences}, vol.~15, no.~12, p. 6622, 2025.

\bibitem{derner2024security}
E.~Derner, K.~Batisti{\v{c}}, J.~Zah{\'a}lka, and R.~Babu{\v{s}}ka, ``A security risk taxonomy for prompt-based interaction with large language models,'' \emph{IEEE Access}, 2024.

\bibitem{dibaji2022tutorial}
S.~M. Dibaji, A.~Hussain, and H.~Ishii, ``A tutorial on security and privacy challenges in cps,'' \emph{Security and Resilience of Control Systems: Theory and Applications}, pp. 121--146, 2022.

\bibitem{diel2024human}
A.~Diel, T.~Lalgi, I.~C. Schr{\"o}ter, K.~F. MacDorman, M.~Teufel, and A.~B{\"a}uerle, ``Human performance in detecting deepfakes: A systematic review and meta-analysis of 56 papers,'' \emph{Computers in Human Behavior Reports}, vol.~16, p. 100538, 2024.

\bibitem{docker2025mcp}
{Docker}, ``Mcp horror stories: The security issues threatening ai infrastructure,'' \url{https://www.docker.com/blog/mcp-security-issues-threatening-ai-infrastructure/}, 2025.

\bibitem{domenteanu2024living}
A.~Domenteanu, G.-C. T{\u{a}}taru, L.~Cr{\u{a}}ciun, A.-G. Mol{\u{a}}nescu, L.-A. Cotfas, and C.~Delcea, ``Living in the age of deepfakes: a bibliometric exploration of trends, challenges, and detection approaches,'' \emph{Information}, vol.~15, no.~9, p. 525, 2024.

\bibitem{doss2023deepfakes}
C.~Doss, J.~Mondschein, D.~Shu, T.~Wolfson, D.~Kopecky, V.~A. Fitton-Kane, L.~Bush, and C.~Tucker, ``Deepfakes and scientific knowledge dissemination,'' \emph{Scientific reports}, vol.~13, no.~1, p. 13429, 2023.

\bibitem{dou2025ai}
W.~W. Dou, I.~Goldstein, and Y.~Ji, ``Ai-powered trading, algorithmic collusion, and price efficiency,'' \emph{Jacobs Levy Equity Management Center for Quantitative Financial Research Paper, The Wharton School Research Paper}, 2025.

\bibitem{dsouza2024deepfakes}
D.~S. Dsouza, A.~E. Hajjar, and H.~Jahankhani, ``Deepfakes in social engineering attacks,'' in \emph{Space Law Principles and Sustainable Measures}.\hskip 1em plus 0.5em minus 0.4em\relax Springer, 2024, pp. 153--183.

\bibitem{easttom2025malicious}
C.~Easttom, ``Malicious use of artificial intelligence,'' in \emph{2025 IEEE 15th Annual Computing and Communication Workshop and Conference (CCWC)}.\hskip 1em plus 0.5em minus 0.4em\relax IEEE, 2025, pp. 00\,499--00\,507.

\bibitem{errico2025securing}
H.~Errico, J.~Ngiam, and S.~Sojan, ``Securing the model context protocol (mcp): Risks, controls, and governance,'' \emph{arXiv preprint arXiv:2511.20920}, 2025.

\bibitem{evtimov2025wasp}
I.~Evtimov, A.~Zharmagambetov, A.~Grattafiori, C.~Guo, and K.~Chaudhuri, ``Wasp: Benchmarking web agent security against prompt injection attacks,'' \emph{arXiv preprint arXiv:2504.18575}, 2025.

\bibitem{farhan2025self}
M.~Farhan, U.~Butt, R.~B. Sulaiman, and M.~Alraja, ``Self-sovereign identities and content provenance: Veritrust—a blockchain-based framework for fake news detection,'' \emph{Future Internet}, vol.~17, no.~10, p. 448, 2025.

\bibitem{fatima2025comprehensive}
S.~Fatima and M.~J. Arshad, ``A comprehensive review of blockchain and machine learning integration for peer-to-peer energy trading in smart grids,'' \emph{IEEE Access}, 2025.

\bibitem{feng2025survey}
Y.~Feng, Y.~Guo, Y.~Hou, Y.~Wu, M.~Lao, T.~Yu, and G.~Liu, ``A survey of security threats in federated learning,'' \emph{Complex \& Intelligent Systems}, vol.~11, no.~2, p. 165, 2025.

\bibitem{fu2023misusing}
X.~Fu, Z.~Wang, S.~Li, R.~K. Gupta, N.~Mireshghallah, T.~Berg-Kirkpatrick, and E.~Fernandes, ``Misusing tools in large language models with visual adversarial examples,'' in \emph{arXiv preprint arXiv:2310.03185}, 2023.

\bibitem{gaba2024innovative}
S.~Gaba, I.~Budhiraja, V.~Kumar, S.~Garg, and M.~M. Hassan, ``An innovative multi-agent approach for robust cyber--physical systems using vertical federated learning,'' \emph{Ad Hoc Networks}, vol. 163, p. 103578, 2024.

\bibitem{gaire2025systematization}
S.~Gaire, S.~Gyawali, S.~Mishra, S.~Niroula, D.~Thakur, and U.~Yadav, ``Systematization of knowledge: Security and safety in the model context protocol ecosystem,'' \emph{arXiv preprint arXiv:2512.08290}, 2025.

\bibitem{gao2024texture}
J.~Gao, M.~Micheletto, G.~Orr{\`u}, S.~Concas, X.~Feng, G.~L. Marcialis, and F.~Roli, ``Texture and artifact decomposition for improving generalization in deep-learning-based deepfake detection,'' \emph{Engineering Applications of Artificial Intelligence}, vol. 133, p. 108450, 2024.

\bibitem{gao2024temporal}
Y.~Gao, X.~Wang, Y.~Zhang, P.~Zeng, and Y.~Ma, ``Temporal feature prediction in audio--visual deepfake detection,'' \emph{Electronics}, vol.~13, no.~17, p. 3433, 2024.

\bibitem{garg2023deepfake}
D.~Garg and R.~Gill, ``Deepfake generation and detection-an exploratory study,'' in \emph{2023 10th IEEE Uttar Pradesh Section International Conference on Electrical, Electronics and Computer Engineering (UPCON)}, vol.~10.\hskip 1em plus 0.5em minus 0.4em\relax IEEE, 2023, pp. 888--893.

\bibitem{ge2025proactive}
W.~Ge, X.~Wang, and J.~Yamagishi, ``Proactive detection of speaker identity manipulation with neural watermarking,'' in \emph{ICLR 2025 Workshop on Watermarking for Generative AI (WMARK@ICLR2025)}, 2025.

\bibitem{ghiuruau2024distinguishing}
D.~Ghiur{\u{a}}u and D.~E. Popescu, ``Distinguishing reality from ai: approaches for detecting synthetic content,'' \emph{Computers}, vol.~14, no.~1, p.~1, 2024.

\bibitem{giannaros2023autonomous}
A.~Giannaros, A.~Karras, L.~Theodorakopoulos, C.~Karras, P.~Kranias, N.~Schizas, G.~Kalogeratos, and D.~Tsolis, ``Autonomous vehicles: Sophisticated attacks, safety issues, challenges, open topics, blockchain, and future directions,'' \emph{Journal of Cybersecurity and Privacy}, vol.~3, no.~3, pp. 493--543, 2023.

\bibitem{gong2024contemporary}
L.~Y. Gong and X.~J. Li, ``A contemporary survey on deepfake detection: datasets, algorithms, and challenges,'' \emph{Electronics}, vol.~13, no.~3, p. 585, 2024.

\bibitem{gong2025figstep}
Y.~Gong, D.~Ran, J.~Liu, C.~Wang, T.~Cong, A.~Wang, S.~Duan, and X.~Wang, ``Figstep: Jailbreaking large vision-language models via typographic visual prompts,'' in \emph{Proceedings of the AAAI Conference on Artificial Intelligence}, vol.~39, no.~22, 2025, pp. 23\,951--23\,959.

\bibitem{goodfellow2014generative}
I.~J. Goodfellow, J.~Pouget-Abadie, M.~Mirza, B.~Xu, D.~Warde-Farley, S.~Ozair, A.~Courville, and Y.~Bengio, ``Generative adversarial nets,'' \emph{Advances in neural information processing systems}, vol.~27, 2014.

\bibitem{greshake2023not}
K.~Greshake, S.~Abdelnabi, S.~Mishra, C.~Endres, T.~Holz, and M.~Fritz, ``Not what you've signed up for: Compromising real-world llm-integrated applications with indirect prompt injection,'' in \emph{Proceedings of the 16th ACM workshop on artificial intelligence and security}, 2023, pp. 79--90.

\bibitem{grimes2025bridging}
K.~Grimes, J.~Lawler, R.~C. Garrett, E.~Mathew, M.~Christiani, S.~Kingsley, Z.~S. Wu, and N.~VanHoudnos, ``Sok: Bridging research and practice in {LLM} agent security,'' Carnegie Mellon University Software Engineering Institute, Technical Report 6414, November 2025.

\bibitem{groh2024human}
M.~Groh, A.~Sankaranarayanan, N.~Singh, D.~Y. Kim, A.~Lippman, and R.~Picard, ``Human detection of political speech deepfakes across transcripts, audio, and video,'' \emph{Nature communications}, vol.~15, no.~1, p. 7629, 2024.

\bibitem{grondin2025beyond}
S.~Grondin, A.~Charpentier, and P.~Ratz, ``Beyond human intervention: Algorithmic collusion through multi-agent learning strategies,'' \emph{arXiv preprint arXiv:2501.16935}, 2025.

\bibitem{NISTCPS2017}
\BIBentryALTinterwordspacing
C.-P. S. P.~W. Group, ``Framework for cyber-physical systems: Volume 1, overview,'' National Institute of Standards and Technology, NIST Special Publication 1500-201, June 2017. [Online]. Available: \url{https://nvlpubs.nist.gov/nistpubs/SpecialPublications/NIST.SP.1500-201.pdf}
\BIBentrySTDinterwordspacing

\bibitem{guo2025system}
J.~Guo and H.~Cai, ``System prompt poisoning: Persistent attacks on large language models beyond user injection,'' \emph{arXiv preprint arXiv:2505.06493}, 2025.

\bibitem{guo2025frontier}
W.~Guo, Y.~Potter, T.~Shi, Z.~Wang, A.~Zhang, and D.~Song, ``Frontier ai's impact on the cybersecurity landscape,'' \emph{arXiv preprint arXiv:2504.05408}, 2025.

\bibitem{guo2025systematic}
Y.~Guo, P.~Liu, W.~Ma, Z.~Deng, X.~Zhu, P.~Di, X.~Xiao, and S.~Wen, ``Systematic analysis of mcp security,'' \emph{arXiv preprint arXiv:2508.12538}, 2025.

\bibitem{gupta2023comprehensive}
G.~Gupta, K.~Raja, M.~Gupta, T.~Jan, S.~T. Whiteside, and M.~Prasad, ``A comprehensive review of deepfake detection using advanced machine learning and fusion methods,'' \emph{Electronics}, vol.~13, no.~1, p.~95, 2023.

\bibitem{gupta2024generative}
P.~Gupta, B.~Ding, C.~Guan, and D.~Ding, ``Generative ai: A systematic review using topic modelling techniques,'' \emph{Data and Information Management}, vol.~8, no.~2, p. 100066, 2024.

\bibitem{gupta2024vulnerability}
P.~Gupta, H.~A. Patil, and R.~C. Guido, ``Vulnerability issues in automatic speaker verification (asv) systems,'' \emph{EURASIP Journal on Audio, Speech, and Music Processing}, vol. 2024, no.~1, p.~10, 2024.

\bibitem{ha2025fl}
J.~Ha, A.~El~Azzaoui, and J.~H. Park, ``Fl-tenb4: A federated-learning-enhanced tiny efficientnetb4-lite approach for deepfake detection in cctv environments,'' \emph{Sensors (Basel, Switzerland)}, vol.~25, no.~3, p. 788, 2025.

\bibitem{haakansson2021robust}
A.~H{\aa}kansson, A.~Saad, A.~Anand, V.~Gj{\ae}rum, H.~Robinson, and K.~Seel, ``Robust reasoning for autonomous cyber-physical systems in dynamic environments,'' \emph{Procedia Computer Science}, vol. 192, pp. 3966--3978, 2021.

\bibitem{haley2025impact}
P.~Haley, ``The impact of biometric surveillance on reducing violent crime: Strategies for apprehending criminals while protecting the innocent,'' \emph{Sensors}, vol.~25, no.~10, p. 3160, 2025.

\bibitem{hallal2024recent}
L.~Hallal, J.~Rhinelander, and R.~Venkat, ``Recent trends of authentication methods in extended reality: A survey,'' \emph{Applied System Innovation}, vol.~7, no.~3, p.~45, 2024.

\bibitem{hancock2021social}
J.~T. Hancock and J.~N. Bailenson, ``The social impact of deepfakes,'' pp. 149--152, 2021.

\bibitem{harris2024fake}
S.~Harris, H.~J. Hadi, N.~Ahmad, and M.~A. Alshara, ``Fake news detection revisited: An extensive review of theoretical frameworks, dataset assessments, model constraints, and forward-looking research agendas,'' \emph{Technologies}, vol.~12, no.~11, p. 222, 2024.

\bibitem{hasan2024nfts}
H.~R. Hasan, K.~Salah, R.~Jayaraman, I.~Yaqoob, and M.~Omar, ``Nfts for combating deepfakes and fake metaverse digital contents,'' \emph{Internet of Things}, vol.~25, p. 101133, 2024.

\bibitem{hasan2024review}
M.~K. Hasan, R.~A. Abdulkadir, S.~Islam, T.~R. Gadekallu, and N.~Safie, ``A review on machine learning techniques for secured cyber-physical systems in smart grid networks,'' \emph{Energy Reports}, vol.~11, pp. 1268--1290, 2024.

\bibitem{hasan2025model}
M.~M. Hasan, H.~Li, E.~Fallahzadeh, G.~K. Rajbahadur, B.~Adams, and A.~E. Hassan, ``Model context protocol (mcp) at first glance: Studying the security and maintainability of mcp servers,'' \emph{arXiv preprint arXiv:2506.13538}, 2025.

\bibitem{RedHatMCPSec}
R.~Hat, ``Model context protocol (mcp): Understanding security risks and controls,'' \url{https://www.redhat.com/en/blog/model-context-protocol-mcp-understanding-security-risks-and-controls}, July 2025.

\bibitem{hatami2025electric}
M.~Hatami, L.~Dorje, X.~Li, and Y.~Chen, ``Electric network frequency as environmental fingerprint for metaverse security: A comprehensive survey,'' \emph{Computers}, vol.~14, no.~8, p. 321, 2025.

\bibitem{hatami2025anchor}
M.~Hatami, Q.~Qu, Y.~Chen, J.~Mohammadi, E.~Blasch, and E.~Ardiles-Cruz, ``Anchor-grid: Authenticating smart grid digital twins using real-world anchors,'' \emph{Sensors}, vol.~25, no.~10, p. 2969, 2025.

\bibitem{he2025identity}
S.~He, Y.~Lei, Z.~Zhang, Y.~Sun, S.~Li, C.~Zhang, and J.~Ye, ``Identity deepfake threats to biometric authentication systems: Public and expert perspectives,'' \emph{arXiv preprint arXiv:2506.06825}, 2025.

\bibitem{he2025kad}
S.~He, Y.~Diao, Y.~Li, C.~Sun, L.~Wang, and Z.~Guo, ``Kad-net: Kolmogorov-arnold and differential-aware networks for robust and sensitive proactive deepfake forensics,'' \emph{Knowledge-Based Systems}, p. 114692, 2025.

\bibitem{he2025sentinelagent}
\BIBentryALTinterwordspacing
X.~He, D.~Wu, Y.~Zhai, and K.~Sun, ``Sentinelagent: Graph-based anomaly detection in llm-based multi-agent systems,'' \emph{arXiv preprint arXiv:2505.24201}, 2025. [Online]. Available: \url{https://arxiv.org/abs/2505.24201}
\BIBentrySTDinterwordspacing

\bibitem{hou2025model}
X.~Hou, Y.~Zhao, S.~Wang, and H.~Wang, ``Model context protocol (mcp): Landscape, security threats, and future research directions,'' \emph{arXiv preprint arXiv:2503.23278}, 2025.

\bibitem{hsieh2021robustness}
F.-S. Hsieh, ``Robustness analysis of cyber-physical systems based on discrete timed cyber-physical models,'' in \emph{2021 IEEE World AI IoT Congress (AIIoT)}.\hskip 1em plus 0.5em minus 0.4em\relax IEEE, 2021, pp. 0250--0254.

\bibitem{hsu2023detection}
H.-P. Hsu, Z.-R. Jiang, L.-Y. Li, T.-C. Tsai, C.-H. Hung, S.-C. Chang, S.-S. Wang, and S.-H. Fang, ``Detection of audio tampering based on electric network frequency signal,'' \emph{Sensors}, vol.~23, no.~16, p. 7029, 2023.

\bibitem{hu2024framework}
J.~Hu, X.~Yang, and L.-X. Yang, ``A framework for detecting false data injection attacks in large-scale wireless sensor networks,'' \emph{Sensors}, vol.~24, no.~5, p. 1643, 2024.

\bibitem{hua2024factors}
G.~Hua, Q.~Wang, D.~Ye, H.~Zhang, G.~Wang, and S.~Xia, ``Factors affecting forensic electric network frequency matching--a comprehensive study,'' \emph{Digital Communications and Networks}, vol.~10, no.~4, pp. 1121--1130, 2024.

\bibitem{huang2025threat}
K.~Huang and I.~Habler, ``Threat modeling google's {A2A} protocol with the {MAESTRO} framework,'' Cloud Security Alliance (CSA) Blog, April 2025.

\bibitem{huang2025agentic}
K.~Huang and C.~Hughes, ``Agentic ai red teaming,'' in \emph{Securing AI Agents: Foundations, Frameworks, and Real-World Deployment}.\hskip 1em plus 0.5em minus 0.4em\relax Springer, 2025, pp. 207--252.

\bibitem{iavich2025combating}
M.~Iavich, ``Combating fake news with cryptography in quantum era with post-quantum verifiable image proofs,'' \emph{Journal of Cybersecurity and Privacy}, vol.~5, no.~2, p.~31, 2025.

\bibitem{invariant2025tpa}
{Invariant Labs}, ``Mcp security notification: Tool poisoning attacks,'' \url{https://invariantlabs.ai/blog/mcp-security-notification-tool-poisoning-attacks}, 2025.

\bibitem{ironscales2024deepfake}
Ironscales, ``Deepfakes:assessing organizational readiness in the face of this emerging cyber threat,'' IRONSCALES, Research Report, 2024.

\bibitem{islam2024ai}
M.~B.~E. Islam, M.~Haseeb, H.~Batool, N.~Ahtasham, and Z.~Muhammad, ``Ai threats to politics, elections, and democracy: a blockchain-based deepfake authenticity verification framework,'' \emph{Blockchains}, vol.~2, no.~4, pp. 458--481, 2024.

\bibitem{ismail2025toward}
Ismail, R.~Kurnia, Z.~A. Brata, G.~A. Nelistiani, S.~Heo, H.~Kim, and H.~Kim, ``Toward robust security orchestration and automated response in security operations centers with a hyper-automation approach using agentic artificial intelligence,'' \emph{Information}, vol.~16, no.~5, p. 365, 2025.

\bibitem{iturbe2024unleashing}
E.~Iturbe, O.~Llorente-Vazquez, A.~Rego, E.~Rios, and N.~Toledo, ``Unleashing offensive artificial intelligence: Automated attack technique code generation,'' \emph{Computers \& Security}, vol. 147, p. 104077, 2024.

\bibitem{javed2025enhancing}
M.~Javed, Z.~Zhang, F.~H. Dahri, and T.~Kumar, ``Enhancing multimodal deepfake detection with local--global feature integration and diffusion models,'' \emph{Signal, Image and Video Processing}, vol.~19, no.~5, pp. 1--9, 2025.

\bibitem{javed2024real}
M.~Javed, Z.~Zhang, F.~H. Dahri, and A.~A. Laghari, ``Real-time deepfake video detection using eye movement analysis with a hybrid deep learning approach,'' \emph{Electronics}, vol.~13, no.~15, p. 2947, 2024.

\bibitem{jayanthiladevi2025ai}
A.~Jayanthiladevi, J.~Natarajan, K.~Arjun, L.~G. Atlas, M.~Arvindhan, and D.~Arockiam, ``Ai-based cybersecurity frameworks for 7g-enabled virtual therapy platforms,'' \emph{Cyber Security and Applications}, p. 100099, 2025.

\bibitem{jeffrey2023review}
N.~Jeffrey, Q.~Tan, and J.~R. Villar, ``A review of anomaly detection strategies to detect threats to cyber-physical systems,'' \emph{Electronics}, vol.~12, no.~15, p. 3283, 2023.

\bibitem{bald2024}
R.~Jiao, S.~Xie, J.~Yue, T.~Sato, L.~Wang, Y.~Wang, Q.~A. Chen, and Q.~Zhu, ``Can we trust embodied agents? exploring backdoor attacks against embodied llm-based decision-making systems,'' in \emph{OpenReview}, 2024.

\bibitem{jin2025comprehensive}
W.~Jin, H.~Du, B.~Zhao, X.~Tian, B.~Shi, and G.~Yang, ``A comprehensive survey on multi-agent cooperative decision-making: Scenarios, approaches, challenges and perspectives,'' \emph{arXiv preprint arXiv:2503.13415}, 2025.

\bibitem{jing2025zero}
G.~Jing and H.~Qi, ``Zero-knowledge audit for internet of agents: Privacy-preserving communication verification with model context protocol,'' \emph{arXiv preprint arXiv:2512.14737}, 2025.

\bibitem{johnson2025dangers}
S.~Johnson, V.~Pham, and T.~Le, ``The dangers of indirect prompt injection attacks on llm-based autonomous web navigation agents: A demonstration,'' in \emph{Proceedings of the 2025 Conference on Empirical Methods in Natural Language Processing: System Demonstrations}, 2025, pp. 729--738.

\bibitem{ju2024flooding}
T.~Ju, Y.~Wang, X.~Ma, P.~Cheng, H.~Zhao, Y.~Wang, L.~Liu, J.~Xie, Z.~Zhang, and G.~Liu, ``Flooding spread of manipulated knowledge in llm-based multi-agent communities,'' \emph{arXiv preprint arXiv:2407.07791}, 2024.

\bibitem{kandasamy2025harnessing}
V.~Kandasamy and A.~A. Roseline, ``Harnessing advanced hybrid deep learning model for real-time detection and prevention of man-in-the-middle cyber attacks,'' \emph{Scientific Reports}, vol.~15, no.~1, p. 1697, 2025.

\bibitem{karim2025ai}
M.~M. Karim, D.~H. Van, S.~Khan, Q.~Qu, and Y.~Kholodov, ``Ai agents meet blockchain: A survey on secure and scalable collaboration for multi-agents,'' \emph{Future Internet}, vol.~17, no.~2, 2025.

\bibitem{karim2024mcgan}
S.~Karim, X.~Liu, A.~A. Khan, A.~A. Laghari, A.~Qadir, and I.~Bibi, ``Mcgan—a cutting edge approach to real time investigate of multimedia deepfake multi collaboration of deep generative adversarial networks with transfer learning,'' \emph{Scientific Reports}, vol.~14, no.~1, p. 29330, 2024.

\bibitem{kazimierczak2024impact}
M.~Kazimierczak, N.~Habib, J.~H. Chan, and T.~Thanapattheerakul, ``Impact of ai on the cyber kill chain: A systematic review,'' \emph{Heliyon}, vol.~10, no.~24, 2024.

\bibitem{kehkashan2025ai}
T.~Kehkashan, R.~A. Riaz, A.~S. Al-Shamayleh, A.~Akhunzada, N.~Ali, M.~Hamza, and F.~Akbar, ``Ai-generated text detection: A comprehensive review of methods, datasets, and applications,'' \emph{Computer Science Review}, vol.~58, p. 100793, 2025.

\bibitem{khan2025survey}
A.~A. Khan, A.~A. Laghari, S.~A. Inam, S.~Ullah, M.~Shahzad, and D.~Syed, ``A survey on multimedia-enabled deepfake detection: state-of-the-art tools and techniques, emerging trends, current challenges \& limitations, and future directions,'' \emph{Discover Computing}, vol.~28, no.~1, p.~48, 2025.

\bibitem{khan2023battling}
A.~Khan, K.~M. Malik, J.~Ryan, and M.~Saravanan, ``Battling voice spoofing: a review, comparative analysis, and generalizability evaluation of state-of-the-art voice spoofing counter measures,'' \emph{Artificial Intelligence Review}, vol.~56, no. Suppl 1, pp. 513--566, 2023.

\bibitem{kharvi2024understanding}
P.~L. Kharvi, ``Understanding the impact of ai-generated deepfakes on public opinion, political discourse, and personal security in social media,'' \emph{IEEE Security \& Privacy}, vol.~22, no.~4, pp. 115--122, 2024.

\bibitem{khatun2023machine}
M.~A. Khatun, S.~F. Memon, C.~Eising, and L.~L. Dhirani, ``Machine learning for healthcare-iot security: A review and risk mitigation,'' \emph{IEEE Access}, vol.~11, pp. 145\,869--145\,896, 2023.

\bibitem{khurshid2025securing}
K.~Khurshid, K.~Khurshid, M.~U. Hadi, M.~Al~Bataineh, and N.~Saeed, ``Securing aiot surveillance: Techniques, challenges, and solutions,'' \emph{IEEE Open Journal of the Communications Society}, 2025.

\bibitem{kilinc2023audio}
H.~H. Kilinc and F.~Kaledibi, ``Audio deepfake detection by using machine and deep learning,'' in \emph{2023 10th International Conference on Wireless Networks and Mobile Communications (WINCOM)}.\hskip 1em plus 0.5em minus 0.4em\relax IEEE, 2023, pp. 1--5.

\bibitem{kinnas2025reducing}
M.~Kinnas, J.~Violos, I.~Kompatsiaris, and S.~Papadopoulos, ``Reducing inference energy consumption using dual complementary cnns,'' \emph{Future Generation Computer Systems}, vol. 165, p. 107606, 2025.

\bibitem{kira2024non}
B.~Kira, ``When non-consensual intimate deepfakes go viral: The insufficiency of the uk online safety act,'' \emph{Computer Law \& Security Review}, vol.~54, p. 106024, 2024.

\bibitem{kirchenbauer2023watermark}
J.~Kirchenbauer, J.~Geiping, Y.~Wen, J.~Katz, I.~Miers, and T.~Goldstein, ``A watermark for large language models,'' in \emph{International Conference on Machine Learning}.\hskip 1em plus 0.5em minus 0.4em\relax PMLR, 2023, pp. 17\,061--17\,084.

\bibitem{TechRadarMCP}
\BIBentryALTinterwordspacing
E.~Kontsevoy, ``Mcp’s biggest security loophole is identity fragmentation,'' \emph{TechRadar Pro}, September 2025. [Online]. Available: \url{https://www.techradar.com/pro/mcps-biggest-security-loophole-is-identity-fragmentation}
\BIBentrySTDinterwordspacing

\bibitem{kopecky2024challenges}
S.~Kopecky, ``Challenges of deepfakes,'' in \emph{Science and Information Conference}.\hskip 1em plus 0.5em minus 0.4em\relax Springer, 2024, pp. 158--166.

\bibitem{korgialas2024leveraging}
C.~Korgialas, C.~Kotropoulos, and K.~N. Plataniotis, ``Leveraging electric network frequency estimation for audio authentication,'' \emph{IEEE Access}, vol.~12, pp. 9308--9320, 2024.

\bibitem{kpmg2024deepfake}
KPMG, \emph{Deepfake Threats to Companies: Navigating AI-Driven Fraud Risks}, 2024.

\bibitem{kulkarni2024zero}
A.~Kulkarni, N.~A. Hazari, and M.~Y. Niamat, ``A zero trust-based framework employing blockchain technology and ring oscillator physical unclonable functions for security of field programmable gate array supply chain,'' \emph{IEEE Access}, vol.~12, pp. 89\,322--89\,338, 2024.

\bibitem{kumar2025advances}
A.~Kumar, D.~Singh, R.~Jain, D.~K. Jain, C.~Gan, and X.~Zhao, ``Advances in deepfake detection algorithms: Exploring fusion techniques in single and multi-modal approach,'' \emph{Information Fusion}, p. 102993, 2025.

\bibitem{kumar2025artificial}
N.~Kumar and A.~K. Singh, ``Artificial intelligence content detection techniques using watermarking: A survey,'' \emph{Image and Vision Computing}, p. 105728, 2025.

\bibitem{kumar2025secure}
S.~N.~P. Kumar, ``A secure accountability framework for multi-modal agent systems: Detecting, mitigating, and auditing data-poisoning attacks via model context protocol (mcp) servers,'' \emph{Journal of Computer Science and Technology Studies}, vol.~7, no.~12, pp. 01--05, November 2025.

\bibitem{RedHatMCPIntro}
F.~La~Vigne, ``Model context protocol: Discover the missing link for ai integration,'' \url{https://www.redhat.com/en/blog/model-context-protocol-discover-missing-link-ai-integration}, April 2025, red Hat Blog.

\bibitem{lad2024adversarial}
S.~Lad, ``Adversarial approaches to deepfake detection: A theoretical framework for robust defense,'' \emph{Journal of Artificial Intelligence General science (JAIGS)}, vol.~6, no.~1, pp. 46--58, 2024.

\bibitem{lai2024novel}
Z.~Lai, Z.~Yao, G.~Lai, C.~Wang, and R.~Feng, ``A novel face swapping detection scheme using the pseudo zernike transform based robust watermarking,'' \emph{Electronics}, vol.~13, no.~24, p. 4955, 2024.

\bibitem{landauer2025review}
M.~Landauer, F.~Skopik, B.~Stojanovi{\'c}, A.~Flatscher, and T.~Ullrich, ``A review of time-series analysis for cyber security analytics: from intrusion detection to attack prediction,'' \emph{International Journal of Information Security}, vol.~24, no.~1, p.~3, 2025.

\bibitem{lee2016introduction}
E.~A. Lee and S.~A. Seshia, \emph{Introduction to embedded systems: A cyber-physical systems approach}.\hskip 1em plus 0.5em minus 0.4em\relax MIT press, 2016.

\bibitem{lee2025towards}
J.~H. Lee, A.~Lauscher, and S.~V. Albrecht, ``Towards ethical multi-agent systems of large language models: A mechanistic interpretability perspective,'' \emph{arXiv preprint arXiv:2512.04691}, 2025.

\bibitem{lei2025deepfake}
S.~Lei, J.~Song, F.~Feng, Z.~Yan, and A.~Wang, ``Deepfake face detection and adversarial attack defense method based on multi-feature decision fusion,'' \emph{Applied Sciences}, vol.~15, no.~12, p. 6588, 2025.

\bibitem{lewis2020deepfake}
J.~K. Lewis, I.~E. Toubal, H.~Chen, V.~Sandesera, M.~Lomnitz, Z.~Hampel-Arias, C.~Prasad, and K.~Palaniappan, ``Deepfake video detection based on spatial, spectral, and temporal inconsistencies using multimodal deep learning,'' in \emph{2020 IEEE Applied Imagery Pattern Recognition Workshop (AIPR)}.\hskip 1em plus 0.5em minus 0.4em\relax IEEE, 2020, pp. 1--9.

\bibitem{li2025big}
H.~Li, S.~Yang, R.~Xia, L.~Yuan, and X.~Gao, ``Big brother is watching: Proactive deepfake detection via learnable hidden face,'' \emph{IEEE Signal Processing Letters}, 2025.

\bibitem{li2024two}
J.~Li, G.~Sun, Q.~Wu, S.~Liang, P.~Wang, and D.~Niyato, ``Two-way aerial secure communications via distributed collaborative beamforming under eavesdropper collusion,'' in \emph{IEEE INFOCOM 2024-IEEE Conference on Computer Communications}.\hskip 1em plus 0.5em minus 0.4em\relax IEEE, 2024, pp. 331--340.

\bibitem{li2024advgps}
J.~Li, B.~Li, X.~Liu, J.~Fang, F.~Juefei-Xu, Q.~Guo, and H.~Yu, ``Advgps: Adversarial gps for multi-agent perception attack,'' in \emph{2024 IEEE International Conference on Robotics and Automation (ICRA)}.\hskip 1em plus 0.5em minus 0.4em\relax IEEE, 2024, pp. 18\,421--18\,427.

\bibitem{li2025survey}
M.~Li, Y.~Ahmadiadli, and X.-P. Zhang, ``A survey on speech deepfake detection,'' \emph{ACM Computing Surveys}, vol.~57, no.~7, pp. 1--38, 2025.

\bibitem{li2025glue}
Q.~Li and Y.~Xie, ``From glue-code to protocols: A critical analysis of a2a and mcp integration for scalable agent systems,'' \emph{arXiv preprint arXiv:2505.03864}, 2025.

\bibitem{li2024survey}
X.~Li, S.~Wang, S.~Zeng, Y.~Wu, and Y.~Yang, ``A survey on llm-based multi-agent systems: workflow, infrastructure, and challenges,'' \emph{Vicinagearth}, vol.~1, no.~1, p.~9, 2024.

\bibitem{li2025detection}
Y.~Li, T.~Lu, S.~Peng, C.~He, K.~Zhao, G.~Yang, and Y.~Chen, ``Detection of electric network frequency in audio using multi-hcnet,'' \emph{Sensors}, vol.~25, no.~12, p. 3697, 2025.

\bibitem{liang2024watermarking}
Y.~Liang, J.~Xiao, W.~Gan, and P.~S. Yu, ``Watermarking techniques for large language models: A survey,'' \emph{arXiv preprint arXiv:2409.00089}, 2024.

\bibitem{lin2025robust}
Z.~Lin, H.~Lin, L.~Lin, S.~Chen, and X.~Liu, ``Robust cross-image adversarial watermark with jpeg resistance for defending against deepfake models,'' \emph{Computer Vision and Image Understanding}, p. 104459, 2025.

\bibitem{lin2025binding}
Z.~Lin, S.~Zhang, G.~Liao, D.~Tao, and T.~Wang, ``Binding agent {ID}: Unleashing the power of {AI} agents with accountability and credibility,'' \emph{arXiv preprint arXiv:2512.17538}, 2025.

\bibitem{liu2025multi}
J.~Liu, Z.~Liu, Q.~Li, W.~Kong, and X.~Li, ``Multi-domain controversial text detection based on a machine learning and deep learning stacked ensemble,'' \emph{Mathematics}, vol.~13, no.~9, p. 1529, 2025.

\bibitem{liu2025seeing}
Q.~Liu, L.~Wang, and M.~Luo, ``When seeing is not believing: self-efficacy and cynicism in the era of intelligent media,'' \emph{Humanities and Social Sciences Communications}, vol.~12, no.~1, pp. 1--13, 2025.

\bibitem{liu2024act}
T.~Liu, W.~Yang, C.~Xu, J.~Lv, H.~Wang, Y.~Zhang, S.~Xu, and D.~Man, ``Act in collusion: A persistent distributed multi-target backdoor in federated learning,'' \emph{arXiv preprint arXiv:2411.03926}, 2024.

\bibitem{liu2023asvspoof}
X.~Liu, X.~Wang, M.~Sahidullah, J.~Patino, H.~Delgado, T.~Kinnunen, M.~Todisco, J.~Yamagishi, N.~Evans, A.~Nautsch \emph{et~al.}, ``Asvspoof 2021: Towards spoofed and deepfake speech detection in the wild,'' \emph{IEEE/ACM Transactions on Audio, Speech, and Language Processing}, vol.~31, pp. 2507--2522, 2023.

\bibitem{liu2024adversarial}
Y.~Liu, L.~Xu, S.~Yang, D.~Zhao, and X.~Li, ``Adversarial sample attacks and defenses based on lstm-ed in industrial control systems,'' \emph{Computers \& Security}, vol. 140, p. 103750, 2024.

\bibitem{liu2025context}
Y.~Liu, X.~Zhao, C.~Kruegel, D.~Song, and Y.~Bu, ``In-context watermarks for large language models,'' \emph{arXiv preprint arXiv:2505.16934}, 2025.

\bibitem{liu2023prompt}
Y.~Liu, G.~Deng, Y.~Li, K.~Wang, Z.~Wang, X.~Wang, T.~Zhang, Y.~Liu, H.~Wang, Y.~Zheng \emph{et~al.}, ``Prompt injection attack against llm-integrated applications,'' \emph{arXiv preprint arXiv:2306.05499}, 2023.

\bibitem{liu2023jailbreaking}
Y.~Liu, G.~Deng, Z.~Xu, Y.~Li, Y.~Zheng, Y.~Zhang, L.~Zhao, T.~Zhang, K.~Wang, and Y.~Liu, ``Jailbreaking chatgpt via prompt engineering: An empirical study,'' \emph{arXiv preprint arXiv:2305.13860}, 2023.

\bibitem{liu2025secure}
Y.~Liu, R.~Zhang, H.~Luo, Y.~Lin, G.~Sun, D.~Niyato, H.~Du, Z.~Xiong, Y.~Wen, A.~Jamalipour \emph{et~al.}, ``Secure multi-llm agentic ai and agentification for edge general intelligence by zero-trust: A survey,'' \emph{arXiv preprint arXiv:2508.19870}, 2025.

\bibitem{liu2024formalizing}
Y.~Liu, Y.~Jia, R.~Geng, J.~Jia, and N.~Z. Gong, ``Formalizing and benchmarking prompt injection attacks and defenses,'' in \emph{33rd USENIX Security Symposium (USENIX Security 24)}, 2024, pp. 1831--1847.

\bibitem{lopez2025securing}
A.~Lopez~Pellicer, P.~Angelov, and N.~Suri, ``Securing (vision-based) autonomous systems: taxonomy, challenges, and defense mechanisms against adversarial threats,'' \emph{Artificial Intelligence Review}, vol.~58, no.~12, pp. 1--59, 2025.

\bibitem{lu2025adversarial}
H.~Lu, J.~Liu, J.~Peng, and J.~Lu, ``Adversarial attacks based on time-series features for traffic detection,'' \emph{Computers \& Security}, vol. 148, p. 104175, 2025.

\bibitem{lu2024recovery}
P.~Lu, L.~Zhang, M.~Liu, K.~Sridhar, O.~Sokolsky, F.~Kong, and I.~Lee, ``Recovery from adversarial attacks in cyber-physical systems: Shallow, deep, and exploratory works,'' \emph{ACM Computing Surveys}, vol.~56, no.~8, pp. 1--31, 2024.

\bibitem{lundberg2025potential}
E.~Lundberg and P.~Mozelius, ``The potential effects of deepfakes on news media and entertainment,'' \emph{AI \& SOCIETY}, vol.~40, no.~4, pp. 2159--2170, 2025.

\bibitem{luo2025digital}
H.~Luo, L.~Li, and J.~Li, ``Digital watermarking technology for ai-generated images: A survey,'' \emph{Math}, 2025.

\bibitem{lupinacci2025dark}
M.~Lupinacci, F.~A. Pironti, F.~Blefari, F.~Romeo, L.~Arena, and A.~Furfaro, ``The dark side of llms: Agent-based attacks for complete computer takeover,'' \emph{arXiv preprint arXiv:2507.06850}, 2025.

\bibitem{lv2025rag}
P.~Lv, M.~Sun, H.~Wang, X.~Wang, S.~Zhang, Y.~Chen, K.~Chen, and L.~Sun, ``Rag-wm: An efficient black-box watermarking approach for retrieval-augmented generation of large language models,'' in \emph{Proceedings of the 2025 ACM SIGSAC Conference on Computer and Communications Security}, 2025, pp. 1709--1723.

\bibitem{mahboubi2025lurking}
A.~Mahboubi, K.~Luong, G.~Jarrad, S.~Camtepe, M.~Bewong, M.~Bahutair, and G.~Pogrebna, ``Lurking in the shadows: Unsupervised decoding of beaconing communication for enhanced cyber threat hunting,'' \emph{Journal of Network and Computer Applications}, vol. 236, p. 104127, 2025.

\bibitem{maranco2024intense}
M.~Maranco, R.~Nidhya, M.~Sivakumar \emph{et~al.}, ``Intense triad defender for end-user security in cyber physical system,'' in \emph{2024 2nd International Conference on Networking and Communications (ICNWC)}.\hskip 1em plus 0.5em minus 0.4em\relax IEEE, 2024, pp. 1--7.

\bibitem{mazzocca2025survey}
C.~Mazzocca, A.~Acar, S.~Uluagac, R.~Montanari, P.~Bellavista, and M.~Conti, ``A survey on decentralized identifiers and verifiable credentials,'' \emph{IEEE Communications Surveys \& Tutorials}, 2025.

\bibitem{meng2025multi}
S.~Meng, Q.~Tan, Q.~Zhou, and R.~Wang, ``Multi-branch network with multi-feature enhancement for improving the generalization of facial forgery detection,'' \emph{Entropy}, vol.~27, no.~5, p. 545, 2025.

\bibitem{metcalfe-pearce2025predictions}
A.~Metcalfe-Pearce, ``2026 cybersecurity predictions,'' F5 Labs, December 2025.

\bibitem{blindguard2025}
R.~Miao, Y.~Liu, Y.~Wang, X.~Shen, Y.~Tan, Y.~Dai, S.~Pan, and X.~Wang, ``Blindguard: Safeguarding llm-based multi-agent systems under unknown attacks,'' \emph{arXiv preprint arXiv:2508.08127}, 2025.

\bibitem{miculicich2025veriguard}
L.~Miculicich, M.~Parmar, H.~Palangi, K.~D. Dvijotham, M.~Montanari, T.~Pfister, and L.~T. Le, ``Veriguard: Enhancing llm agent safety via verified code generation,'' \emph{arXiv preprint arXiv:2510.05156}, 2025.

\bibitem{mikhaylenko2025stealthy}
D.~Mikhaylenko and P.~Zhang, ``Stealthy targeted local covert attacks on cyber--physical systems,'' \emph{Automatica}, vol. 173, p. 112023, 2025.

\bibitem{MCPSpec}
\BIBentryALTinterwordspacing
\emph{Specification - Model Context Protocol}, Model Context Protocol, June 2025. [Online]. Available: \url{https://modelcontextprotocol.io/specification/latest}
\BIBentrySTDinterwordspacing

\bibitem{momin2025explainable}
M.~S. Momin, A.~Sufian, D.~Barman, M.~Leo, C.~Distante, and N.~Damer, ``Explainable deepfake detection across different modalities: An overview of methods and challenges,'' \emph{Image and Vision Computing}, p. 105738, 2025.

\bibitem{monteiro2025safety}
C.~Monteiro. (2025, May) Safety and security in the model context protocol ({MCP}).

\bibitem{moti2021generative}
Z.~Moti, S.~Hashemi, H.~Karimipour, A.~Dehghantanha, A.~N. Jahromi, L.~Abdi, and F.~Alavi, ``Generative adversarial network to detect unseen internet of things malware,'' \emph{Ad Hoc Networks}, vol. 122, p. 102591, 2021.

\bibitem{motwani2024secret}
\BIBentryALTinterwordspacing
S.~R. Motwani, M.~Baranchuk, M.~Strohmeier, V.~Bolina, P.~H. Torr, L.~Hammond, and C.~Schroeder~de Witt, ``Secret collusion among ai agents: Multi-agent deception via steganography,'' in \emph{Advances in Neural Information Processing Systems (NeurIPS)}, vol.~37, 2024. [Online]. Available: \url{https://www.proceedings.com/content/079/079017-2336open.pdf}
\BIBentrySTDinterwordspacing

\bibitem{mubarak2023survey}
R.~Mubarak, T.~Alsboui, O.~Alshaikh, I.~Inuwa-Dutse, S.~Khan, and S.~Parkinson, ``A survey on the detection and impacts of deepfakes in visual, audio, and textual formats,'' \emph{Ieee Access}, vol.~11, pp. 144\,497--144\,529, 2023.

\bibitem{mukesh2025comprehensive}
V.~Mukesh, ``A comprehensive review of advanced machine learning techniques for enhancing cybersecurity in blockchain networks,'' \emph{INTERNATIONAL JOURNAL OF ARTIFICIAL INTELLIGENCE (ISCSITR-IJAI)}, vol.~5, pp. 1--6, 2025.

\bibitem{mukherjeellm}
K.~Mukherjee and M.~Kantarcioglu, ``Llm-driven provenance forensics for threat intelligence and detection,'' 2025.

\bibitem{mustak2023deepfakes}
M.~Mustak, J.~Salminen, M.~M{\"a}ntym{\"a}ki, A.~Rahman, and Y.~K. Dwivedi, ``Deepfakes: Deceptions, mitigations, and opportunities,'' \emph{Journal of Business Research}, vol. 154, p. 113368, 2023.

\bibitem{nagarajan2022iadf}
S.~M. Nagarajan, G.~G. Deverajan, A.~K. Bashir, R.~P. Mahapatra, and M.~S. Al-Numay, ``Iadf-cps: Intelligent anomaly detection framework towards cyber physical systems,'' \emph{Computer Communications}, vol. 188, pp. 81--89, 2022.

\bibitem{nagothu2019detecting}
D.~Nagothu, Y.~Chen, E.~Blasch, A.~Aved, and S.~Zhu, ``Detecting malicious false frame injection attacks on surveillance systems at the edge using electrical network frequency signals,'' \emph{Sensors}, vol.~19, no.~11, p. 2424, 2019.

\bibitem{narajala2025enterprise}
V.~S. Narajala and I.~Habler, ``Enterprise-grade security for the model context protocol (mcp): Frameworks and mitigation strategies,'' \emph{arXiv preprint arXiv:2504.08623}, 2025.

\bibitem{nctr2025cybersecurity}
{National Council on Teacher Retirement (NCTR)}, ``Artificial intelligence ({AI}) and cyber security: An update,'' July 2025, accessed: December 22, 2025.

\bibitem{neupane2023impacts}
S.~Neupane, I.~A. Fernandez, S.~Mittal, and S.~Rahimi, ``Impacts and risk of generative ai technology on cyber defense,'' \emph{arXiv preprint arXiv:2306.13033}, 2023.

\bibitem{ngharamike2023enf}
E.~Ngharamike, L.-M. Ang, K.~P. Seng, and M.~Wang, ``Enf based digital multimedia forensics: Survey, application, challenges and future work,'' \emph{IEEE Access}, vol.~11, pp. 101\,241--101\,272, 2023.

\bibitem{nguyen2024towards}
L.-H. Nguyen, V.-L. Nguyen, R.-H. Hwang, J.-J. Kuo, Y.-W. Chen, C.-C. Huang, and P.-I. Pan, ``Towards secured smart grid 2.0: exploring security threats, protection models, and challenges,'' \emph{IEEE Communications Surveys \& Tutorials}, 2024.

\bibitem{nguyenle2025survey}
H.-H. Nguyen-Le, V.-T. Tran, T.~Nguyen, and N.-A. Le-Khac, ``A survey on proactive deepfake defense: Disruption and watermarking,'' \emph{ACM Computing Surveys}, vol.~58, no.~5, pp. 1--37, 2025.

\bibitem{nunes2022bane}
M.~Nunes, P.~Burnap, P.~Reinecke, and K.~Lloyd, ``Bane or boon: Measuring the effect of evasive malware on system call classifiers,'' \emph{Journal of Information Security and Applications}, vol.~67, p. 103202, 2022.

\bibitem{obsidian2025aiagent}
{Obsidian Security Team}, ``The 2025 {AI} agent security landscape: Players, trends, and risks,'' Obsidian Security Blog, October 2025.

\bibitem{odyurt2022improving}
U.~Odyurt, A.~D. Pimentel, and I.~G. Alonso, ``Improving the robustness of industrial cyber--physical systems through machine learning-based performance anomaly identification,'' \emph{Journal of Systems Architecture}, vol. 131, p. 102716, 2022.

\bibitem{NISTTrust2018}
N.~I. of~Standards and T.~(NIST), ``Nist smart grid and cps newsletter - december 2017: The cps framework introduces the concept of trustworthiness,'' \url{https://www.nist.gov/ctl/smart-connected-systems-division/smart-grid-group/nist-smart-grid-and-cps-newsletter-december}, December 2017, published online March 16, 2018.

\bibitem{onami2025blockchain}
S.~Onami, ``Blockchain for cybersecurity: Enhancing data integrity and trust in digital transactions,'' \emph{ResearchGate}, September 2025.

\bibitem{opara2025evaluating}
C.~Opara, P.~Modesti, and L.~Golightly, ``Evaluating spam filters and stylometric detection of ai-generated phishing emails,'' \emph{Expert Systems with Applications}, vol. 276, p. 127044, 2025.

\bibitem{owoputi2022security}
R.~Owoputi and S.~Ray, ``Security of multi-agent cyber-physical systems: A survey,'' \emph{IEEE Access}, vol.~10, pp. 121\,465--121\,479, 2022.

\bibitem{pantsar2023developing}
M.~Pantsar, ``Developing artificial human-like arithmetical intelligence (and why),'' \emph{Minds and Machines}, vol.~33, no.~3, pp. 379--396, 2023.

\bibitem{park2024ai}
P.~S. Park, S.~Goldstein, A.~O’Gara, M.~Chen, and D.~Hendrycks, ``Ai deception: A survey of examples, risks, and potential solutions,'' \emph{Patterns}, vol.~5, no.~5, 2024.

\bibitem{park2024comprehensive}
S.~H. Park, S.-H. Lee, M.~Y. Lim, P.~M. Hong, and Y.~K. Lee, ``A comprehensive risk analysis method for adversarial attacks on biometric authentication systems,'' \emph{IEEE Access}, 2024.

\bibitem{parti2024legal}
K.~Parti and J.~Szab{\'o}, ``The legal challenges of realistic and ai-driven child sexual abuse material: regulatory and enforcement perspectives in europe,'' \emph{Laws}, vol.~13, no.~6, p.~67, 2024.

\bibitem{patel2023deepfake}
Y.~Patel, S.~Tanwar, R.~Gupta, P.~Bhattacharya, I.~E. Davidson, R.~Nyameko, S.~Aluvala, and V.~Vimal, ``Deepfake generation and detection: Case study and challenges,'' \emph{IEEE Access}, vol.~11, pp. 143\,296--143\,323, 2023.

\bibitem{patil2025model}
M.~D. Patil and V.~V. Lokhande, ``Model context protocol ({MCP}): Enabling scalable {AI} data integration,'' \emph{International Journal For Multidisciplinary Research (IJFMR)}, vol.~7, no.~2, April 2025.

\bibitem{patil2025navigating}
S.~Patil, A.~Bhat, N.~Jain, and V.~Javalkar, ``Navigating deepfakes with data science: A multi-modal analysis and blockchain-based detection framework,'' in \emph{2025 International Conference on Pervasive Computational Technologies (ICPCT)}.\hskip 1em plus 0.5em minus 0.4em\relax IEEE, 2025, pp. 772--777.

\bibitem{pawelec2022deepfakes}
M.~Pawelec, ``Deepfakes and democracy (theory): How synthetic audio-visual media for disinformation and hate speech threaten core democratic functions,'' \emph{Digital society}, vol.~1, no.~2, p.~19, 2022.

\bibitem{pawlicki2025meta}
M.~Pawlicki, A.~Pawlicka, R.~Kozik, and M.~Chora{\'s}, ``A meta-survey of adversarial attacks against artificial intelligence algorithms, including diffusion models,'' \emph{Neurocomputing}, p. 131231, 2025.

\bibitem{pedersen2025deepfake}
K.~T. Pedersen, L.~Pepke, T.~St{\ae}rmose, M.~Papaioannou, G.~Choudhary, and N.~Dragoni, ``Deepfake-driven social engineering: Threats, detection techniques, and defensive strategies in corporate environments,'' \emph{Journal of Cybersecurity and Privacy}, vol.~5, no.~2, p.~18, 2025.

\bibitem{pedro2023prompt}
R.~Pedro, D.~Castro, P.~Carreira, and N.~Santos, ``From prompt injections to sql injection attacks: How protected is your llm-integrated web application?'' \emph{arXiv preprint arXiv:2308.01990}, 2023.

\bibitem{pei2024deepfake}
G.~Pei, J.~Zhang, M.~Hu, Z.~Zhang, C.~Wang, Y.~Wu, G.~Zhai, J.~Yang, C.~Shen, and D.~Tao, ``Deepfake generation and detection: A benchmark and survey,'' \emph{arXiv preprint arXiv:2403.17881}, 2024.

\bibitem{piccialli2025agentai}
F.~Piccialli, D.~Chiaro, S.~Sarwar, D.~Cerciello, P.~Qi, and V.~Mele, ``Agentai: A comprehensive survey on autonomous agents in distributed ai for industry 4.0,'' \emph{Expert Systems with Applications}, p. 128404, 2025.

\bibitem{pimpason2025phishing}
N.~Pimpason, P.~Viboonsang, and S.~Kosolsombat, ``Phishing email detection model using deep learning,'' in \emph{2025 IEEE International Conference on Cybernetics and Innovations (ICCI)}.\hskip 1em plus 0.5em minus 0.4em\relax IEEE, 2025, pp. 1--5.

\bibitem{piratla2025safeguarding}
V.~S.~A. Piratla, S.~Saxena, S.~Bhatia, and N.~Kumar, ``Safeguarding the artificial pancreas: A review of security and reliability gaps and ai driven resilience,'' in \emph{2025 5th Intelligent Cybersecurity Conference (ICSC)}.\hskip 1em plus 0.5em minus 0.4em\relax IEEE, 2025, pp. 399--411.

\bibitem{pohler2024technological}
L.~P{\"o}hler, V.~Schrader, A.~Ladwein, and F.~von Keller, ``A technological perspective on misuse of available ai,'' \emph{arXiv preprint arXiv:2403.15325}, 2024.

\bibitem{posta2025deepdive}
C.~Posta. (2025, May) Deep dive mcp and a2a attack vectors for ai agents.

\bibitem{GitHubMCP}
M.~C. Protocol, ``Official github organization,'' \url{https://github.com/modelcontextprotocol}, 2025.

\bibitem{pu2023deepfake}
J.~Pu, Z.~Sarwar, S.~M. Abdullah, A.~Rehman, Y.~Kim, P.~Bhattacharya, M.~Javed, and B.~Viswanath, ``Deepfake text detection: Limitations and opportunities,'' in \emph{2023 IEEE symposium on security and privacy (SP)}.\hskip 1em plus 0.5em minus 0.4em\relax IEEE, 2023, pp. 1613--1630.

\bibitem{pulikottil2023agent}
T.~Pulikottil, L.~A. Estrada-Jimenez, H.~Ur~Rehman, F.~Mo, S.~Nikghadam-Hojjati, and J.~Barata, ``Agent-based manufacturing—review and expert evaluation,'' \emph{The International Journal of Advanced Manufacturing Technology}, vol. 127, no.~5, pp. 2151--2180, 2023.

\bibitem{qian2025black}
H.~Qian, L.~Xia, R.~Ge, Y.~Fan, Q.~Wang, and Z.~Jing, ``From black boxes to glass boxes: Explainable ai for trustworthy deepfake forensics,'' \emph{Cryptography}, vol.~9, no.~4, p.~61, 2025.

\bibitem{qiao2025frw}
S.~Qiao, Q.~Guo, M.~Wang, H.~Zhu, J.~J. Rodrigues, and Z.~Lyu, ``Frw-trace: Forensic-ready watermarking framework for tamper-resistant biometric data and attack traceability in consumer electronics,'' \emph{IEEE Transactions on Consumer Electronics}, 2025.

\bibitem{quan2025federated}
M.~K. Quan, P.~N. Pathirana, M.~Wijayasundara, S.~Setunge, D.~C. Nguyen, C.~G. Brinton, D.~J. Love, and H.~V. Poor, ``Federated learning for cyber physical systems: a comprehensive survey,'' \emph{IEEE Communications Surveys \& Tutorials}, 2025.

\bibitem{LFC2PA}
A.~Ramaswami, ``How c2pa helps combat misleading information,'' \url{https://www.linuxfoundation.org/blog/how-c2pa-helps-combat-misleading-information}, June 2024, linux Foundation Blog.

\bibitem{rana2022deepfake}
M.~S. Rana, M.~N. Nobi, B.~Murali, and A.~H. Sung, ``Deepfake detection: A systematic literature review,'' \emph{IEEE access}, vol.~10, pp. 25\,494--25\,513, 2022.

\bibitem{rana2024deepfakes}
M.~S. Rana, M.~Solaiman, C.~Gudla, and M.~F. Sohan, ``Deepfakes--reality under threat?'' in \emph{2024 IEEE 14th Annual Computing and Communication Workshop and Conference (CCWC)}.\hskip 1em plus 0.5em minus 0.4em\relax IEEE, 2024, pp. 0721--0727.

\bibitem{ranasinghe2025privacy}
N.~Ranasinghe, P.~Liyanage, and L.~Kruglova, ``Privacy preserving distributed image processing using federated learning and cnns,'' in \emph{2025 IEEE 15th Symposium on Computer Applications \& Industrial Electronics (ISCAIE)}.\hskip 1em plus 0.5em minus 0.4em\relax IEEE, 2025, pp. 138--143.

\bibitem{rashid2025evaluating}
S.~Rashid, E.~Bollis, L.~Pellicer, D.~Rabbani, R.~Palacios, A.~Gupta, and A.~Gupta, ``Evaluating prompt injection attacks with lstm-based generative adversarial networks: A lightweight alternative to large language models,'' \emph{Machine Learning and Knowledge Extraction}, vol.~7, no.~3, p.~77, 2025.

\bibitem{ray2025review}
P.~P. Ray, ``A review on agent-to-agent protocol: Concept, state-of-the-art, challenges and future directions,'' \emph{Authorea Preprints}, 2025.

\bibitem{ray2025survey}
------, ``A survey on model context protocol: Architecture, state-of-the-art, challenges and future directions,'' \emph{Authorea Preprints}, 2025.

\bibitem{raza2025trism}
S.~Raza, R.~Sapkota, M.~Karkee, and C.~Emmanouilidis, ``Trism for agentic ai: A review of trust, risk, and security management in llm-based agentic multi-agent systems,'' \emph{arXiv preprint arXiv:2506.04133}, 2025.

\bibitem{roh2025multilingual}
J.~Roh, V.~Shejwalkar, and A.~Houmansadr, ``Multilingual and multi-accent jailbreaking of audio llms,'' \emph{arXiv preprint arXiv:2504.01094}, 2025.

\bibitem{romandini2025sok}
N.~Romandini, C.~Mazzocca, K.~Otsuki, and R.~Montanari, ``Sok: Security and privacy of ai agents for blockchain,'' \emph{arXiv preprint arXiv:2509.07131}, 2025.

\bibitem{romero2024deepfake}
F.~Romero-Moreno, ``Deepfake fraud detection: Safeguarding trust in generative ai,'' \emph{Available at SSRN 5031627}, 2024.

\bibitem{romero2025deepfake}
------, ``Deepfake detection in generative ai: A legal framework proposal to protect human rights,'' \emph{Computer Law \& Security Review}, vol.~58, p. 106162, 2025.

\bibitem{rosado2022managing}
D.~G. Rosado, A.~Santos-Olmo, L.~E. S{\'a}nchez, M.~A. Serrano, C.~Blanco, H.~Mouratidis, and E.~Fern{\'a}ndez-Medina, ``Managing cybersecurity risks of cyber-physical systems: The marisma-cps pattern,'' \emph{Computers in Industry}, vol. 142, p. 103715, 2022.

\bibitem{rosca2025new}
C.-M. Rosca, A.~Stancu, and E.~M. Iovanovici, ``The new paradigm of deepfake detection at the text level,'' \emph{Applied Sciences}, vol.~15, no.~5, p. 2560, 2025.

\bibitem{VergeMCP}
\BIBentryALTinterwordspacing
E.~Roth, ``Anthropic launches tool to connect ai systems directly to datasets,'' \emph{The Verge}, November 2024. [Online]. Available: \url{https://www.theverge.com/2024/11/25/24305774/anthropic-model-context-protocol-data-sources}
\BIBentrySTDinterwordspacing

\bibitem{sadaf2023connected}
M.~Sadaf, Z.~Iqbal, A.~R. Javed, I.~Saba, M.~Krichen, S.~Majeed, and A.~Raza, ``Connected and automated vehicles: Infrastructure, applications, security, critical challenges, and future aspects,'' \emph{Technologies}, vol.~11, no.~5, p. 117, 2023.

\bibitem{saeed2023systematic}
S.~Saeed, S.~A. Suayyid, M.~S. Al-Ghamdi, H.~Al-Muhaisen, and A.~M. Almuhaideb, ``A systematic literature review on cyber threat intelligence for organizational cybersecurity resilience,'' \emph{Sensors}, vol.~23, no.~16, p. 7273, 2023.

\bibitem{saideh2024opportunistic}
M.~Saideh, J.-P. Jamont, and L.~Vercouter, ``Opportunistic sensor-based authentication factors in and for the internet of things,'' \emph{Sensors}, vol.~24, no.~14, p. 4621, 2024.

\bibitem{sakaci2024conducted}
F.~H. Sakac{\i} and T.~Y{\i}ld{\i}r{\i}m, ``Conducted emission signal-based identification and real-time hardware security with deep learning,'' \emph{Engineering Applications of Artificial Intelligence}, vol. 136, p. 109025, 2024.

\bibitem{salau2022recent}
B.~A. Salau, A.~Rawal, and D.~B. Rawat, ``Recent advances in artificial intelligence for wireless internet of things and cyber--physical systems: A comprehensive survey,'' \emph{IEEE Internet of Things Journal}, vol.~9, no.~15, pp. 12\,916--12\,930, 2022.

\bibitem{salih2025addressing}
M.~Salih, J.~Gharib, and Y.~Gahi, ``Addressing security gaps in {MCP}: Design of a resilient reference architecture,'' in \emph{2025 11th International Conference on Optimization and Applications (ICOA)}.\hskip 1em plus 0.5em minus 0.4em\relax IEEE, 2025, pp. 1--7.

\bibitem{salvi2023robust}
D.~Salvi, H.~Liu, S.~Mandelli, P.~Bestagini, W.~Zhou, W.~Zhang, and S.~Tubaro, ``A robust approach to multimodal deepfake detection,'' \emph{Journal of Imaging}, vol.~9, no.~6, p. 122, 2023.

\bibitem{sanchez2025marisma}
L.~E. S{\'a}nchez, A.~Santos-Olmo, D.~G. Rosado, C.~Blanco, M.~A. Serrano, H.~Mouratidis, and E.~Fern{\'a}ndez-Medina, ``Marisma: A modern and context-aware framework for assessing and managing information cybersecurity risks,'' \emph{Computer Standards \& Interfaces}, vol.~92, p. 103935, 2025.

\bibitem{sandoval2024threat}
M.-P. Sandoval, M.~de~Almeida~Vau, J.~Solaas, and L.~Rodrigues, ``Threat of deepfakes to the criminal justice system: a systematic review,'' \emph{Crime Science}, vol.~13, no.~1, p.~41, 2024.

\bibitem{sangaiah2024guest}
A.~K. Sangaiah, X.~Wang, M.~S. Obaidat, P.~C. Huang, and K.~Govindan, ``Guest editorial data-driven innovation and adversarial learning models for industry 5.0 toward consumer digital ecosystems,'' \emph{IEEE Transactions on Consumer Electronics}, vol.~70, no.~2, pp. 4878--4881, 2024.

\bibitem{sasikumar2025enhancing}
K.~Sasikumar and S.~Nagarajan, ``Enhancing cloud security: A multi-factor authentication and adaptive cryptography approach using machine learning techniques,'' \emph{IEEE Open Journal of the Computer Society}, 2025.

\bibitem{say2025advancing}
T.~Say, M.~Alkan, and A.~Kocak, ``Advancing gan deepfake detection: Mixed datasets and comprehensive artifact analysis,'' \emph{Applied Sciences}, vol.~15, no.~2, p. 923, 2025.

\bibitem{schmitt2024digital}
M.~Schmitt and I.~Flechais, ``Digital deception: Generative artificial intelligence in social engineering and phishing,'' \emph{Artificial Intelligence Review}, vol.~57, no.~12, p. 324, 2024.

\bibitem{selvaraj2025deepfake}
P.~Selvaraj, S.~Jagatheesaperumal, K.~Marimuthu, O.~Saravanan, B.~Alkhamees, and M.~Hassan, ``Deepfake detection using adversarial neural network,'' \emph{Computer Modeling in Engineering \& Sciences}, vol. 143, no.~2, p. 1575, 2025.

\bibitem{selvarajan2025diagnostic}
S.~Selvarajan, H.~Manoharan, M.~Abdelhaq, A.~O. Khadidos, A.~O. Khadidos, R.~Alsaqour, and M.~Uddin, ``Diagnostic behavior analysis of profuse data intrusions in cyber physical systems using adversarial learning techniques,'' \emph{Scientific Reports}, vol.~15, no.~1, p. 7287, 2025.

\bibitem{selvi2024enhancing}
K.~Selvi and G.~Dilip, ``Enhancing cyber-physical systems security: A review of deep learning and blockchain integration,'' in \emph{2024 5th International Conference on Image Processing and Capsule Networks (ICIPCN)}.\hskip 1em plus 0.5em minus 0.4em\relax IEEE, 2024, pp. 725--734.

\bibitem{shallal2025image}
I.~Shallal, L.~Rzouga~Haddada, and N.~Essoukri Ben~Amara, ``Image forgery detection with focus on copy-move: An overview, real world challenges and future directions,'' \emph{Applied Sciences}, vol.~15, no.~21, p. 11774, 2025.

\bibitem{shang2024nonlinear}
J.~Shang, J.~Zhou, and T.~Chen, ``Nonlinear stealthy attacks on remote state estimation,'' \emph{Automatica}, vol. 167, p. 111747, 2024.

\bibitem{sharma2024comprehensive}
U.~Sharma and J.~Singh, ``A comprehensive overview of fake news detection on social networks,'' \emph{Social Network Analysis and Mining}, vol.~14, no.~1, p. 120, 2024.

\bibitem{sharma2025systematic}
V.~K. Sharma, R.~Garg, and Q.~Caudron, ``A systematic literature review on deepfake detection techniques,'' \emph{Multimedia Tools and Applications}, vol.~84, no.~20, pp. 22\,187--22\,229, 2025.

\bibitem{shayegani2023jailbreak}
E.~Shayegani, Y.~Dong, and N.~Abu-Ghazaleh, ``Jailbreak in pieces: Compositional adversarial attacks on multi-modal language models,'' \emph{arXiv preprint arXiv:2307.14539}, 2023.

\bibitem{sheng2025id}
Y.~Sheng, Z.~Zou, Z.~Yu, M.~Pang, W.~Ou, and W.~Han, ``Id-insensitive deepfake detection model based on multi-attention mechanism,'' \emph{Scientific Reports}, vol.~15, no.~1, p. 11168, 2025.

\bibitem{shi2025deepfake}
C.~Shi, M.~Qiao, Z.~Li, Z.~Akhtar, B.~Wang, M.~Han, and T.~Qiao, ``Deepfake video traceability and authentication via source attribution,'' \emph{IET Biometrics}, vol. 2025, pp. 1--14, 2025.

\bibitem{shoaib2023deepfakes}
M.~R. Shoaib, Z.~Wang, M.~T. Ahvanooey, and J.~Zhao, ``Deepfakes, misinformation, and disinformation in the era of frontier ai, generative ai, and large ai models,'' in \emph{2023 international conference on computer and applications (ICCA)}.\hskip 1em plus 0.5em minus 0.4em\relax IEEE, 2023, pp. 1--7.

\bibitem{siameh2025context}
T.~Siameh, A.~A. Addobea, and C.-H. Liu, ``Context injection vulnerabilities and resource exploitation attacks in model context protocol,'' \emph{Authorea Preprints}, 2025.

\bibitem{singh2025advancements}
L.~H. Singh, P.~Charanarur, and N.~K. Chaudhary, ``Advancements in detecting deepfakes: Ai algorithms and future prospects- a review,'' \emph{Discover Internet of Things}, vol.~5, no.~1, p.~53, 2025.

\bibitem{singh2025anomaly}
L.~D. Singh and P.~Meher, ``Anomaly detection in cyber-physical electrical systems using ai-enhanced pufs,'' in \emph{2025 Fourth International Conference on Power, Control and Computing Technologies (ICPC2T)}.\hskip 1em plus 0.5em minus 0.4em\relax IEEE, 2025, pp. 1--6.

\bibitem{singh2025agentic}
R.~Singh. (2025, November) Agentic ai control fabric: The next enterprise operating system for autonomous workflows.

\bibitem{sohail2025deepfake}
S.~Sohail, S.~M. Sajjad, A.~Zafar, Z.~Iqbal, Z.~Muhammad, and M.~Kazim, ``Deepfake image forensics for privacy protection and authenticity using deep learning,'' \emph{Information}, vol.~16, no.~4, p. 270, 2025.

\bibitem{soltani2024multi}
M.~Soltani, K.~Khajavi, M.~Jafari~Siavoshani, and A.~H. Jahangir, ``A multi-agent adaptive deep learning framework for online intrusion detection,'' \emph{Cybersecurity}, vol.~7, no.~1, p.~9, 2024.

\bibitem{son2025advancing}
S.~Son and W.~Kim, ``Advancing generalization in deepfake detection: Supervised contrastive representation learning with dual stream spatio-temporal features,'' \emph{IEEE Access}, 2025.

\bibitem{song2025beyond}
H.~Song, Y.~Shen, W.~Luo, L.~Guo, T.~Chen, J.~Wang, B.~Li, X.~Zhang, and J.~Chen, ``Beyond the protocol: Unveiling attack vectors in the model context protocol ecosystem,'' \emph{arXiv preprint arXiv:2506.02040}, 2025.

\bibitem{owasp2025agentic}
J.~Sotiropoulos, R.~F. Del~Rosario, K.~Huang \emph{et~al.}, ``Agentic ai - threats and mitigations,'' OWASP Foundation, Tech. Rep., December 2025.

\bibitem{sotiropoulos2025owasp}
J.~Sotiropoulos, K.~Katz, and R.~F. Del~Rosario, ``{OWASP} top 10 for agentic applications – the benchmark for agentic security in the age of autonomous {AI},'' OWASP GenAI Security Project Blog, December 2025.

\bibitem{soudy2024deepfake}
A.~H. Soudy, O.~Sayed, H.~Tag-Elser, R.~Ragab, S.~Mohsen, T.~Mostafa, A.~A. Abohany, and S.~O. Slim, ``Deepfake detection using convolutional vision transformers and convolutional neural networks,'' \emph{Neural Computing and Applications}, vol.~36, no.~31, pp. 19\,759--19\,775, 2024.

\bibitem{south2025authenticated}
T.~South, S.~Marro, T.~Hardjono, R.~Mahari, C.~D. Whitney, D.~Greenwood, A.~Chan, and A.~Pentland, ``Authenticated delegation and authorized ai agents,'' \emph{arXiv preprint arXiv:2501.09674}, 2025.

\bibitem{stockwell2025elections}
S.~Stockwell, ``From deepfake scams to poisoned chatbots: {AI} and election security in 2025,'' CETaS Expert Analysis, November 2025.

\bibitem{sun2025diffmark}
C.~Sun, H.~Sun, Z.~Guo, Y.~Diao, L.~Wang, D.~Ma, G.~Yang, and K.~Li, ``Diffmark: Diffusion-based robust watermark against deepfakes,'' \emph{arXiv preprint arXiv:2507.01428}, 2025.

\bibitem{sun2025study}
J.~Sun and W.~Hou, ``A study of two-branch fusion network model for low-quality deepfake detection in face videos,'' \emph{Signal, Image and Video Processing}, vol.~19, no.~8, p. 631, 2025.

\bibitem{sun2022faketracer}
P.~Sun, Y.~Li, H.~Qi, and S.~Lyu, ``Faketracer: Exposing deepfakes with training data contamination,'' in \emph{2022 IEEE International Conference on Image Processing (ICIP)}.\hskip 1em plus 0.5em minus 0.4em\relax IEEE, 2022, pp. 1161--1165.

\bibitem{syllaidopoulos2025comprehensive}
I.~Syllaidopoulos, K.~Ntalianis, and I.~Salmon, ``A comprehensive survey on ai in counter-terrorism and cybersecurity: Challenges and ethical dimensions,'' \emph{IEEE Access}, 2025.

\bibitem{tahaoglu2025robust}
G.~Tahaoglu, ``Robust deepfake audio detection via an improved next-tdnn with multi-fused self-supervised learning features,'' \emph{Applied Sciences}, vol.~15, no.~17, p. 9685, 2025.

\bibitem{tan2025review}
D.~Tan, Y.~Yang, C.~Niu, S.~Li, D.~Yang, and B.~Tan, ``A review of deep learning based multimodal forgery detection for video and audio,'' \emph{Discover Applied Sciences}, vol.~7, no.~9, p. 987, 2025.

\bibitem{tao2021cloud}
L.~Tao, X.~Wang, Y.~Liu, and J.~Wu, ``Cloud-based user behavior emulation approach for space-ground integrated networks,'' \emph{Sensors}, vol.~22, no.~1, p.~44, 2021.

\bibitem{taralkar2025nft}
J.~Taralkar and S.~Narlawar, ``Nft video tokenization: A decentralized approach to verifying media authenticity,'' in \emph{2025 IEEE Cloud Summit}.\hskip 1em plus 0.5em minus 0.4em\relax IEEE, 2025, pp. 174--180.

\bibitem{tas2024blockchain}
I.~M. Tas and S.~Baktir, ``Blockchain-based caller-id authentication (bbca): A novel solution to prevent spoofing attacks in voip/sip networks,'' \emph{IEEE access}, vol.~12, pp. 60\,123--60\,137, 2024.

\bibitem{tchaptchet2025deepfakes}
E.~Tchaptchet, E.~F. Tagne, J.~Acosta, R.~Danda, and C.~Kamhoua, ``Deepfakes detection by iris analysis,'' \emph{IEEE Access}, 2025.

\bibitem{thakur2023systematic}
K.~Thakur, M.~L. Ali, M.~A. Obaidat, and A.~Kamruzzaman, ``A systematic review on deep-learning-based phishing email detection,'' \emph{Electronics}, vol.~12, no.~21, p. 4545, 2023.

\bibitem{tian2025deepphysio}
J.~Tian, L.~Guan, Y.~Liu, L.~Zhang, and Y.~Chen, ``Deepphysio: detecting deepfake with non-personalized feature of physiological signal,'' \emph{Multimedia Systems}, vol.~31, no.~2, p.~86, 2025.

\bibitem{tipper2024investigation}
S.~Tipper, H.~F. Atlam, and H.~S. Lallie, ``An investigation into the utilisation of cnn with lstm for video deepfake detection,'' \emph{Applied Sciences}, vol.~14, no.~21, p. 9754, 2024.

\bibitem{triantafyllopoulos2025vishing}
A.~Triantafyllopoulos, A.~A. Spiesberger, I.~Tsangko, X.~Jing, V.~Distler, F.~Dietz, F.~Alt, and B.~W. Schuller, ``Vishing: Detecting social engineering in spoken communication—a first survey \& urgent roadmap to address an emerging societal challenge,'' \emph{Computer Speech \& Language}, vol.~94, p. 101802, 2025.

\bibitem{uppal2024comprehensive}
S.~Uppal, V.~Banga, S.~Neeraj, and A.~Singhal, ``A comprehensive study on mitigating synthetic identity threats using deepfake detection mechanisms,'' in \emph{2024 14th International Conference on Cloud Computing, Data Science \& Engineering (Confluence)}.\hskip 1em plus 0.5em minus 0.4em\relax IEEE, 2024, pp. 750--755.

\bibitem{UpwindMCP}
Upwind, ``Unpacking the security risks of model context protocol (mcp) servers,'' \url{https://www.upwind.io/feed/unpacking-the-security-risks-of-model-context-protocol-mcp-servers}, April 2025.

\bibitem{usmani2025spatio}
S.~Usmani, S.~Kumar, and D.~Sadhya, ``Spatio-temporal knowledge distilled video vision transformer (stkd-vvit) for multimodal deepfake detection,'' \emph{Neurocomputing}, vol. 620, p. 129256, 2025.

\bibitem{vegh2018cyber}
L.~Vegh, ``Cyber-physical systems security through multi-factor authentication and data analytics,'' in \emph{2018 IEEE international conference on industrial technology (ICIT)}.\hskip 1em plus 0.5em minus 0.4em\relax IEEE, 2018, pp. 1369--1374.

\bibitem{verma2025deepfake}
K.~Verma, D.~Mittal, S.~Samanta, K.~Gulati, O.~Kulkarni, M.~A. Dar, and C.~Biji, ``Deepfake audio detection: A comparative study of advanced deep learning models,'' \emph{IEEE Access}, 2025.

\bibitem{wahba2024creating}
K.~A. Wahba, K.~A. Ahmed, M.~R. Kamel, M.~Fathy, P.~K.~H. Abdelfatah, and S.~Hatem, ``Creating a digital human twin: Cloning voice, face, and attitude,'' in \emph{2024 International Mobile, Intelligent, and Ubiquitous Computing Conference (MIUCC)}.\hskip 1em plus 0.5em minus 0.4em\relax IEEE, 2024, pp. 199--205.

\bibitem{wang2025survey}
F.~Wang, Y.~Jiang, R.~Zhang, A.~Wei, J.~Xie, and X.~Pang, ``A survey of deep anomaly detection in multivariate time series: taxonomy, applications, and directions,'' \emph{Sensors (Basel, Switzerland)}, vol.~25, no.~1, p. 190, 2025.

\bibitem{wang2023anomaly}
H.~Wang, D.~J. Miller, and G.~Kesidis, ``Anomaly detection of adversarial examples using class-conditional generative adversarial networks,'' \emph{Computers \& Security}, vol. 124, p. 102956, 2023.

\bibitem{wang2025pro2guard}
H.~Wang, C.~M. Poskitt, J.~Sun, and J.~Wei, ``Pro2guard: Proactive runtime enforcement of llm agent safety via probabilistic model checking,'' \emph{arXiv preprint arXiv:2508.00500}, 2025.

\bibitem{wang2025comprehensive}
K.~Wang, G.~Zhang, Z.~Zhou, J.~Wu, M.~Yu, S.~Zhao, C.~Yin, J.~Fu, Y.~Yan, H.~Luo \emph{et~al.}, ``A comprehensive survey in llm (-agent) full stack safety: Data, training and deployment,'' \emph{arXiv preprint arXiv:2504.15585}, 2025.

\bibitem{wang2024aarf}
L.~Wang, J.~Deng, H.~Tan, Y.~Xu, J.~Zhu, Z.~Zhang, Z.~Li, R.~Zhan, and Z.~Gu, ``Aarf: Autonomous attack response framework for honeypots to enhance interaction based on multi-agent dynamic game,'' \emph{Mathematics}, vol.~12, no.~10, p. 1508, 2024.

\bibitem{wang2025gsafeguard}
S.~Wang, G.~Zhang, M.~Yu, G.~Wan, F.~Meng, C.~Guo, K.~Wang, and Y.~Wang, ``G-safeguard: A topology-guided security lens and treatment on llm-based multi-agent systems,'' \emph{Proceedings of the 63rd Annual Meeting of the Association for Computational Linguistics (Volume 1: Long Papers)}, pp. 7261--7276, 2025.

\bibitem{wang2023robustness}
S.~Wang, X.~Gu, J.~Chen, C.~Chen, and X.~Huang, ``Robustness improvement strategy of cyber-physical systems with weak interdependency,'' \emph{Reliability Engineering \& System Safety}, vol. 229, p. 108837, 2023.

\bibitem{wang2025fractalforensics}
T.~Wang, H.~Cheng, M.-H. Liu, and M.~Kankanhalli, ``Fractalforensics: Proactive deepfake detection and localization via fractal watermarks,'' in \emph{Proceedings of the 33rd ACM International Conference on Multimedia}, 2025, pp. 7210--7219.

\bibitem{wang2023blockchain}
X.~Wang, H.~Xie, S.~Ji, L.~Liu, and D.~Huang, ``Blockchain-based fake news traceability and verification mechanism,'' \emph{Heliyon}, vol.~9, no.~7, 2023.

\bibitem{wang2025security}
Y.~Wang, Y.~Pan, S.~Guo, and Z.~Su, ``Security of internet of agents: Attacks and countermeasures,'' \emph{IEEE Open Journal of the Computer Society}, 2025.

\bibitem{wang2021survey}
Z.~Wang, W.~Xie, B.~Wang, J.~Tao, and E.~Wang, ``A survey on recent advanced research of cps security,'' \emph{Applied Sciences}, vol.~11, no.~9, p. 3751, 2021.

\bibitem{wang2025can}
Z.~Wang, G.~Xu, and M.~Ren, ``Can attention detect ai-generated text? a novel benford's law-based approach,'' \emph{Information Processing \& Management}, vol.~62, no.~4, p. 104139, 2025.

\bibitem{wang2025mcp}
Z.~Wang, Q.~Chang, H.~Patel, S.~Biju, C.-E. Wu, Q.~Liu, A.~Ding, A.~Rezazadeh, A.~Shah, Y.~Bao \emph{et~al.}, ``Mcp-bench: Benchmarking tool-using llm agents with complex real-world tasks via mcp servers,'' \emph{arXiv preprint arXiv:2508.20453}, 2025.

\bibitem{wei2023jailbroken}
A.~Wei, N.~Haghtalab, and J.~Steinhardt, ``Jailbroken: How does llm safety training fail?'' in \emph{NeurIPS 2023}, 2023.

\bibitem{williamson2024era}
S.~M. Williamson and V.~Prybutok, ``The era of artificial intelligence deception: unraveling the complexities of false realities and emerging threats of misinformation,'' \emph{Information}, vol.~15, no.~6, p. 299, 2024.

\bibitem{willison2025camel}
S.~Willison, ``Camel offers a promising new direction for mitigating prompt injection attacks,'' \url{https://simonwillison.net/2025/Apr/11/camel/}, 2025.

\bibitem{ITProMCPMal}
\BIBentryALTinterwordspacing
E.~Woollacott, ``A malicious mcp server is silently stealing user emails,'' \emph{ITPro}, September 2025. [Online]. Available: \url{https://www.itpro.com/security/a-malicious-mcp-server-is-silently-stealing-user-emails}
\BIBentrySTDinterwordspacing

\bibitem{wu2024adversarial}
C.~H. Wu, J.~Y. Koh, R.~Salakhutdinov, D.~Fried, and A.~Raghunathan, ``Adversarial attacks on multimodal agents,'' in \emph{Proceedings of ACL 2024}, 2024.

\bibitem{wu2025cbam}
Y.~Wu, H.~Huang, Z.~Li, and S.~Zhang, ``Cbam-resnet: A lightweight resnet network focusing on time domain features for end-to-end deepfake speech detection,'' \emph{Electronics}, vol.~14, no.~12, p. 2456, 2025.

\bibitem{xiao2025sok}
S.~Xiao, W.~Zhu, Y.~Jiang, K.~Wang, P.~Wang, C.~Yan, X.~Ji, and W.~Xu, ``Sok: Understanding the fundamentals and implications of sensor out-of-band vulnerabilities,'' \emph{arXiv preprint arXiv:2508.16133}, 2025.

\bibitem{embodiedai2025survey}
W.~Xing, M.~Li, M.~Li, and M.~Han, ``Towards robust and secure embodied ai: A survey on vulnerabilities and attacks,'' \emph{arXiv preprint arXiv:2502.13175}, 2025.

\bibitem{xiong2025bmnet}
D.~Xiong, Z.~Wen, C.~Zhang, D.~Ren, and W.~Li, ``Bmnet: Enhancing deepfake detection through bilstm and multi-head self-attention mechanism,'' \emph{IEEE Access}, 2025.

\bibitem{xu2025erosion}
D.~Xu, I.~Gondal, X.~Yi, T.~Susnjak, P.~Watters, and T.~R. McIntosh, ``The erosion of cybersecurity zero-trust principles through generative ai: A survey on the challenges and future directions,'' \emph{Journal of Cybersecurity and Privacy}, vol.~5, no.~4, p.~87, 2025.

\bibitem{xu2025localization}
J.~Xu, X.~Liu, W.~Lin, W.~Shang, and Y.~Wang, ``Localization and detection of deepfake videos based on self-blending method,'' \emph{Scientific Reports}, vol.~15, no.~1, p. 3927, 2025.

\bibitem{xu2021digital}
Q.~Xu, S.~Ali, and T.~Yue, ``Digital twin-based anomaly detection in cyber-physical systems,'' in \emph{2021 14th IEEE Conference on Software Testing, Verification and Validation (ICST)}.\hskip 1em plus 0.5em minus 0.4em\relax IEEE, 2021, pp. 205--216.

\bibitem{xu2025llm}
W.~Xu, C.~Huang, S.~Gao, and S.~Shang, ``Llm-based agents for tool learning: A survey,'' \emph{Data Science and Engineering}, pp. 1--31, 2025.

\bibitem{yan2025self}
B.~Yan, P.~Liu, Y.~Yang, and Y.~Guo, ``Self-supervised feature disentanglement for deepfake detection,'' \emph{Mathematics}, vol.~13, no.~12, p. 2024, 2025.

\bibitem{yang2025survey}
Y.~Yang, H.~Chai, Y.~Song, S.~Qi, M.~Wen, N.~Li, J.~Liao, H.~Hu, J.~Lin, G.~Chang \emph{et~al.}, ``A survey of ai agent protocols,'' \emph{arXiv preprint arXiv:2504.16736}, 2025.

\bibitem{yang2025watermarking}
Z.~Yang, G.~Zhao, and H.~Wu, ``Watermarking for large language models: A survey,'' \emph{Mathematics}, vol.~13, no.~9, p. 1420, 2025.

\bibitem{yasir2025lightweight}
S.~M. Yasir and H.~Kim, ``Lightweight deepfake detection based on multi-feature fusion,'' \emph{Applied Sciences}, vol.~15, no.~4, p. 1954, 2025.

\bibitem{yi2025benchmarking}
J.~Yi, Y.~Xie, B.~Zhu, E.~Kiciman, G.~Sun, X.~Xie, and F.~Wu, ``Benchmarking and defending against indirect prompt injection attacks on large language models,'' in \emph{Proceedings of the 31st ACM SIGKDD Conference on Knowledge Discovery and Data Mining V. 1}, 2025, pp. 1809--1820.

\bibitem{yin2025emulating}
J.~Yin, M.~Gao, K.~Shu, Z.~Zhao, Y.~Huang, and J.~Wang, ``Emulating reader behaviors for fake news detection,'' \emph{IEEE Transactions on Big Data}, 2025.

\bibitem{chen2024infecting}
W.~Yu, K.~Hu, T.~Pang, C.~Du, M.~Lin, and M.~Fredrikson, ``Infecting llm agents via generalizable adversarial attack,'' in \emph{Red Teaming GenAI: What Can We Learn from Adversaries?}, 2024.

\bibitem{yu2024infecting}
------, ``Infecting llm-based multi-agents via self-propagating adversarial attacks,'' in \emph{Thirty-eighth Conference on Neural Information Processing Systems (NeurIPS)}, 2024.

\bibitem{yuan2025efficient}
K.~Yuan, Z.~Dong, X.~Li, Z.~Liu, C.~Jia, and S.~Lv, ``An efficient and collusion-resistant key parameters pre-distribution system for day-ahead electricity auctions,'' \emph{IEEE Internet of Things Journal}, 2025.

\bibitem{yuan2025multi}
Q.~Yuan, Q.~Meng, J.~Tao, G.~Li, J.~Fei, B.~Lu, and Y.~Wang, ``Multi-agent for network security monitoring and warning: A generative ai solution,'' \emph{IEEE Network}, 2025.

\bibitem{zeeshan2025continuous}
N.~Zeeshan, M.~Bakyt, N.~Moradpoor, and L.~La~Spada, ``Continuous authentication in resource-constrained devices via biometric and environmental fusion,'' \emph{Sensors}, vol.~25, no.~18, p. 5711, 2025.

\bibitem{zeng2024enfformer}
C.~Zeng, K.~Li, and Z.~Wang, ``Enfformer: Long-short term representation of electric network frequency for digital audio tampering detection,'' \emph{Knowledge-Based Systems}, vol. 297, p. 111938, 2024.

\bibitem{zeng2024loft}
S.~Zeng, W.~Wang, F.~Huang, and Y.~Fang, ``Loft: Latent space optimization and generator fine-tuning for defending against deepfakes,'' in \emph{ICASSP 2024-2024 IEEE International Conference on Acoustics, Speech and Signal Processing (ICASSP)}.\hskip 1em plus 0.5em minus 0.4em\relax IEEE, 2024, pp. 4750--4754.

\bibitem{zhan2025adaptive}
Q.~Zhan, R.~Fang, H.~S. Panchal, and D.~Kang, ``Adaptive attacks break defenses against indirect prompt injection attacks on llm agents,'' \emph{arXiv preprint arXiv:2503.00061}, 2025.

\bibitem{zhang2025audio}
B.~Zhang, H.~Cui, V.~Nguyen, and M.~Whitty, ``Audio deepfake detection: What has been achieved and what lies ahead,'' \emph{Sensors (Basel, Switzerland)}, vol.~25, no.~7, p. 1989, 2025.

\bibitem{embodiedjailbreak2024}
H.~Zhang, X.~Wang, Y.~Wang, M.~Li, C.~Zhu, Z.~Zhou, L.~Xue, S.~Hu, and L.~Y. Zhang, ``The threats of embodied multimodal llms: Jailbreaking robotic manipulation in the physical world,'' \emph{arXiv preprint arXiv:2407.20242}, 2024.

\bibitem{zhang2025agent}
H.~Zhang, J.~Huang, K.~Mei, Y.~Yao, Z.~Wang, C.~Zhan, H.~Wang, and Y.~Zhang, ``Agent security bench (asb): Formalizing and benchmarking attacks and defenses in llm-based agents,'' in \emph{International Conference on Learning Representations (ICLR)}, 2025.

\bibitem{zhang2025feasibility}
M.~Zhang, C.~Sonnadara, S.~Shah, and M.~Wu, ``Feasibility study of location authentication for iot data using power grid signatures,'' \emph{IEEE Open Journal of Signal Processing}, 2025.

\bibitem{zhang2025hierarchical}
W.~Zhang, S.~Cui, Q.~Zhang, B.~Chen, H.~Zeng, and Q.~Zhong, ``Hierarchical feature fusion and enhanced attention mechanism for robust gan-generated image detection,'' \emph{Mathematics}, vol.~13, no.~9, p. 1372, 2025.

\bibitem{zhang2025curvemark}
Y.~Zhang, X.~Jiang, H.~Sun, Y.~Zhang, and D.~Tong, ``Curvemark: Detecting ai-generated text via probabilistic curvature and dynamic semantic watermarking,'' \emph{Entropy}, vol.~27, no.~8, p. 784, 2025.

\bibitem{zhang2025selective}
Z.~Zhang, W.~Hao, A.~Sankoh, W.~Lin, E.~Mendiola-Ortiz, J.~Yang, and C.~Mao, ``I can hear you: Selective robust training for deepfake audio detection,'' \emph{arXiv preprint arXiv:2411.00121}, 2024.

\bibitem{zhao2025mind}
S.~Zhao, Q.~Hou, Z.~Zhan, Y.~Wang, Y.~Xie, Y.~Guo, L.~Chen, S.~Li, and Z.~Xue, ``Mind your server: A systematic study of parasitic toolchain attacks on the mcp ecosystem,'' \emph{arXiv preprint arXiv:2509.06572}, 2025.

\bibitem{zhao2023proactive}
Y.~Zhao, B.~Liu, M.~Ding, B.~Liu, T.~Zhu, and X.~Yu, ``Proactive deepfake defence via identity watermarking,'' in \emph{Proceedings of the IEEE/CVF winter conference on applications of computer vision}, 2023, pp. 4602--4611.

\bibitem{zhao2023unlearnable}
Z.~Zhao, J.~Duan, X.~Hu, K.~Xu, C.~Wang, R.~Zhang, Z.~Du, Q.~Guo, and Y.~Chen, ``Unlearnable examples for diffusion models: Protect data from unauthorized exploitation,'' \emph{arXiv preprint arXiv:2306.01902}, 2023.

\bibitem{zhong2023towards}
B.~Zhong, S.~Liu, M.~Caccamo, and M.~Zamani, ``Towards trustworthy ai: Sandboxing ai-based unverified controllers for safe and secure cyber-physical systems,'' in \emph{2023 62nd IEEE Conference on Decision and Control (CDC)}.\hskip 1em plus 0.5em minus 0.4em\relax IEEE, 2023, pp. 1833--1840.

\bibitem{zhou2024security}
W.~Zhou, X.~Zhu, Q.-L. Han, L.~Li, X.~Chen, S.~Wen, and Y.~Xiang, ``The security of using large language models: A survey with emphasis on chatgpt,'' \emph{IEEE/CAA Journal of Automatica Sinica}, 2024.

\bibitem{zhou2024fine}
X.~Zhou, H.~Han, S.~Shan, and X.~Chen, ``Fine-grained open-set deepfake detection via unsupervised domain adaptation,'' \emph{IEEE Transactions on Information Forensics and Security}, 2024.

\bibitem{zhuang2024pgd}
Z.~Zhuang, Y.~Tomioka, J.~Shin, and Y.~Okuyama, ``Pgd-trap: Proactive deepfake defense with sticky adversarial signals and iterative latent variable refinement,'' \emph{Electronics}, vol.~13, no.~17, p. 3353, 2024.

\end{thebibliography}


 



\vfill

\end{document}